\documentclass[12pt,preprint]{aastex}

\begin{document}

\def\MSUN{\rm M_{\odot}}
\def\RSUN{\rm R_{\odot}} 
\def\MSUNYR{\rm M_{\odot}\,yr^{-1}}
\def\MDOT{\dot{M}}

\newbox\grsign \setbox\grsign=\hbox{$>$} \newdimen\grdimen \grdimen=\ht\grsign
\newbox\simlessbox \newbox\simgreatbox
\setbox\simgreatbox=\hbox{\raise.5ex\hbox{$>$}\llap
     {\lower.5ex\hbox{$\sim$}}}\ht1=\grdimen\dp1=0pt
\setbox\simlessbox=\hbox{\raise.5ex\hbox{$<$}\llap
     {\lower.5ex\hbox{$\sim$}}}\ht2=\grdimen\dp2=0pt
\def\simgreat{\mathrel{\copy\simgreatbox}}
\def\simless{\mathrel{\copy\simlessbox}}

\title{Time variability of accretion flows: 
effects of the adiabatic index and gas temperature.}

\author{Monika Moscibrodzka\altaffilmark{1,}\altaffilmark{2}
and Daniel Proga\altaffilmark{1} }

\affil{$^1$ Department of Physics, University of Nevada, Las Vegas,
NV 89154, USA, e-mail: dproga@physics.unlv.edu, monikam@physics.unlv.edu}

\affil{$^2$ N. Copernicus Astronomical Center, Bartycka 18, 00-716, Warsaw, Poland}

\begin{abstract}

We report on next phase of our study of rotating accretion flows 
onto black holes. We consider hydrodynamical (HD) accretion flows with 
a spherically symmetric density distribution at the outer boundary but 
with spherical symmetry broken by the introduction of a small, 
latitude-dependent angular momentum. We study accretion flows by means of 
numerical two-dimensional, axisymmetric, HD simulations for variety
of the adiabatic index, $\gamma$ and 
the  gas temperature at infinity, $c_\infty$. 
Our work is an extension of work done by Proga \& Begelman who
consider models for only $\gamma=5/3$.
Our main result is that the flow properties such as the topology
of the sonic surface and time behavior strongly depend on
$\gamma$ but little on $c_\infty$.  
In particular, for $1 < \gamma < 5/3$,
the mass accretion rate shows large amplitude, slow time-variability which
is a result of mixing between slow and fast rotating gas.
This temporal behavior differs significantly from that
in models with $\gamma\simless 5/3$ where the accretion
rate is relatively constant and from that in models
with $\gamma\simgreat 1$ where the accretion exhibits small 
amplitude quasi-periodic oscillations. The key parameter responsible
for the differences is the sound speed of the accretion flow
which in turn determines whether the flow is dominated 
by gas pressure, radiation pressure or  rotation. Despite
these differences the time-averaged mass accretion rate
in units of the corresponding Bondi rate is a weak
function of $\gamma$ and $c_\infty$.

\end{abstract}

\section{Introduction}
Many types of astrophysical objects are powered by black hole (BH) accretion.
Radiation energy produced by accretion can be very
high and can explain such dramatic phenomena as quasars, powerful
radio galaxies, X-ray binaries, and gamma ray bursts (GRBs). 
However, BH accretion does not always result in high radiative output. 
This is true for both stellar BH and super massive black holes (SMBH).
In particular, objects with SMBH 
appear to spend statistically
most of their time in an inactive phase.
Inactive SMBHs are not something
that one would expect, because these black holes are
embedded in the relatively dense environments of galactic
nuclei. Therefore it is natural to suppose that the gravity
due to an SMBH will draw in matter at high rates, leading
to a high system luminosity. Monitoring of X-ray binaries
with stellar BHs reveals that
these objects often exhibit large time variability in the total energy output
in the spectral energy distribution.
Then one of the main goals of any theory of BH accretion
is two explain why accretion proceeds through
very different modes.

Generally, the radiative output from accretion depends on the mass
accretion rate $\MDOT_a$ and an efficiency factor, $\eta$.
In previous papers of this series (Proga \& Begelman 2003a,b, hereafter
PB03a and PB03b, respectively), we studied how physical conditions
at large distances from SMBH affect $\MDOT_a$ in the so-called
radiatively inefficient accretion flows (RIAF). 
RIAF with very low $\eta$ and also low $\MDOT_a$
have been proposed to explain very low radiative luminosities in systems
as such Sgr~A* 
(e.g., Ichimaru 1977; Rees et al. 1982; Narayan \& Yi 1994, 1995;
Abramowicz et al. 1995; Blandford \& Begelman 1999;
Sharma et al. 2007 and reference therein).
In PB03a, we addressed the issue of how 
$\MDOT_a$ depends on the distribution of specific angular momentum, $l$
at large radii assuming that the adiabatic index, $\gamma=5/3$.  

For high $l$, corresponding to the circularization
radius larger than the last stable orbit, gas cannot directly accrete onto
a BH, unless some physical mechanism like
e.g., viscosity or magnetic fields, or both, transports
angular momentum towards.  For inviscid
accretion flow the matter with too high $l$, either flows outward 
due to gas pressure and centrifugal forces or
accumulates close the central accretor (e.g. Hawley et al. 1984a,b,; 
Clarke et al. 1985; Chen et al 1997; PB03a; Janiuk $\&$ Proga 2007).  
The simulations presented by PB03a illustrate a general flow pattern 
with an inflow in the polar funnel, and an equatorial
outflow (Fig.~1 there).  Such flow pattern is induced by the angular momentum
distribution simply because
rotating gas tend to converge toward the equator. If $l$ is high, the gas
forms a subsonic very dynamic torus with gas flowing out. 
The mass accretion rate 
can be significantly smaller compared to corresponding
the Bondi rate.

In this work, we address the problem how the properties of the
accretion flow presented in PB03a, change when different micro-physical
properties of the flow are assumed.  In particular, we explore
effects of changing $\gamma$ in the polytropic
equation of state.  In reality, accretion flows with different
$\gamma$'s, may correspond to different types of objects or 
different phases of activity.  For
example, in the weakly active galaxies like e.g. Sgr A*,  $\gamma$
is usually considered to be around 5/3, because these flows
are believed to be radiatively inefficient and gas pressure dominated.
On the other hand, the large energetic output observed in GRBs indicates
that an accretion flow must be radiation
pressure dominated (e.g., Meszaros 2006)  and a
relativistic equation of state is required with $\gamma$=4/3.  Another
example are protogalactic disks, which are sometimes considered as
being formed by isothermal accretion flow ($\gamma\approx 1$; e.g., Mo et
al. 1998).  Also, the so-called `high' and `low' accretion
states in the X-ray binary systems, may reflect physical proprieties
changing in time (for a review see Done et al. 2007).

Our work is a straightforward extension of PB03a's work, who showed
results for $\gamma$=5/3.  We extend their models by considering the
flows, for  $\gamma$ ranging between 1 and 5/3.  We
also perform additional calculations with
sound speed at infinity $c_{s,\infty}$ much smaller than that
considered by PB03a allowing us to model flows
with the Bondi radius as large as that estimated in real systems, 
for example in Sgr~A*. 
Taking advantage of faster computers, 
we perform simulations
with higher resolution in the $\theta$ direction, and with much larger
computational domain in comparison to PB03a.  
We keep other model parameters
such as angular momentum distribution as in PB03a.

For $\gamma=5/3$, our new simulations are consistent with those presented
by PB03a. However, we find that $\MDOT_a$,
the flow structure, defined by the gas density and angular momentum
distribution, and  the sonic surface topology depend on 
$\gamma$. In particular,for $\gamma=5/3$, $\MDOT_a$ is nearly constant
whereas for $\gamma=4/3$, $\MDOT_a$ shows stochastic, large-amplitude time-variability. For $\gamma=1.01$, $\MDOT_a$ shows
small-amplitude periodic changes. 
In Sec.2, we describe a general set-up of our simulations. 
In Sec.3, we present our results.
In the last section, we discuss the results, and relate them to results found
in previous studies.

\section{Method}

\subsection{Hydrodynamics}

To calculate the structure and evolution of an accreting flow, we
solve the equations of hydrodynamics
\begin{equation}
   \frac{D\rho}{Dt} + \rho \nabla \cdot {\bf v} = 0,
\label{eq:con}
\end{equation}
\begin{equation}
   \rho \frac{D{\bf v}}{Dt} = - \nabla P + \rho \nabla \Phi,
\label{eq:mom}
\end{equation}
\begin{equation}
   \rho \frac{D}{Dt}\left(\frac{e}{ \rho}\right) = -P \nabla \cdot
   {\bf v},
\label{eq:en}
\end{equation}
where $\rho$ is the mass density, $P$ is the gas pressure, ${\bf v}$
is the velocity, $e$ is the internal energy density, and $\Phi$ is
gravitational potential.  We adopt an
adiabatic equation of state $P~=~(\gamma-1)e$, where $\gamma$ is an
adiabatic index.  Our calculations are performed in spherical polar
coordinates $(r,\theta,\phi)$. We assume axial symmetry about the
rotational axis of the accretion flow ($\theta=0^\circ$ and $180^{\circ}$).

We perform simulations using the pseudo-Newtonian potential $\Phi$
introduced by Paczy\'{n}ski \& Wiita (1980)
\begin{equation}
\Phi=-\frac{G M}{r-R_S}.
\end{equation}
This potential approximates general relativistic effects in the inner
regions, for a non-rotating black hole.  In particular, the
Paczy\'{n}ski--Wiita (P--W) potential reproduces the last stable circular
orbit at $r=3 R_S$ as well as the marginally bound orbit at $r=2 R_S$.

Our standard computational domain is defined to occupy the radial
range $r_i~=~1.5~R_S \leq r \leq \ r_o~=~ 1.2~R_B$ and the angular
range $0^\circ \leq \theta \leq 180^\circ$. We consider models with
$R'_S=10^{-3}$, $R'_S=10^{-4}$, and $R'_S=10^{-5}$.  The $r-\theta$
domain is discretized into zones with 140, 180 and 220 zones (for
$R'_S=10^{-3}$, $R'_S=10^{-4}$, and $R'_S=10^{-5}$ respectively) in
the $r$ direction and 200 zones in the $\theta$ direction.  We fix
zone size ratios, $dr_{k+1}/dr_{k}=1.05$, and
$d\theta_{l}/d\theta_{l+1} =1.0$ for $0^\circ \le \theta \le
180^\circ$.

\subsection{Initial conditions and boundary conditions}

For the initial conditions of the fluid variables we follow PB03a and
adopt a Bondi accretion flow.  In particular, we adopt $v_r$ and
$\rho$ computed using the Bernoulli function and mass accretion rate
for spherically symmetric Bondi accretion with the P--W potential.  We
set $\rho_\infty$ = $2.2 \times 10^{-23}$ $\rm{g/cm^3}$ and specify
$c_{s,\infty}$ through $R'_S\equiv R_S/R_B$ (note that
$R'_S=2c^2_{s,\infty}/c^2$, and $R_B=\frac{GM}{c_{s,\infty}^2}$ is the
Bondi radius). Thus $R'_S$ characterizes the gas temperature in 
our simulations.  We specify the initial conditions by adopting a non-zero
$l$ for the outer boundary of the flow.  

We consider a case where the angular momentum at the outer radius
$r_o$ depends on the polar angle via
\begin{equation}
l(r_o,\theta)=l_0 (1-|\cos\theta|).
\end{equation}
We express the angular momentum on the equator as
\begin{equation}
l_0=\sqrt{R'_C} R_B cs_\infty,
\end{equation}
where $R'_C$ is the circularization radius on the equator in units
of $R_B$ for the Newtonian potential (i.e., $GM/r^2= v^2_\phi/r$ at
$r= R'_C R_B$).

The boundary conditions are specified as follows. At the poles, (i.e.,
$\theta=0^\circ$ and $180^\circ$), we apply an axis-of-symmetry
boundary condition. At both the inner and outer radial boundaries, we
apply an outflow boundary condition for all dynamical variables.  As
in PB03a, to represent steady conditions at the outer radial boundary,
during the evolution of each model we continue to apply the
constraints that in the last zone in the radial direction,
$v_\theta=0$, $v_\phi=l(r,\theta)/ r \sin{\theta}$, and the density
is fixed at the outer boundary at all times. Note that we allow $v_r$
to float.

To solve eqs.(~\ref{eq:con})-(~\ref{eq:en}) we use the ZEUS-2D code 
described by Stone \& Norman (1992), modified to implement 
the P--W potential.

\section{Results}

Here we present results of ten simulations specified by four different
values of $\gamma$ (i.e., $\gamma$= 5/3, 4/3, 1.2, \& 1.01), 
and three values of $R'_S$ (i.e., $R'_S = 10^{-5}$, $10^{-4}$, 
and $10^{-3}$).  The simulations for $\gamma=$1.2 were 
performed only for $R'_S = 10^{-3}$.
Summary of all runs is presented in Tab~\ref{tab:1}.  The table
columns (1)-(8) show respectively the name of the run, the numerical
radial resolution used in the simulation, the value of $R'_S$ parameter,
the value of circularization radius in units of Bondi radius $R'_C$,
$\gamma$ adiabatic index, the end time at which we stopped each
simulation $t_f$ (the time is given in units of the dynamical time at
the inner boundary $t_{dyn}$= 595 s at r=1.5 $R'_S$ for a mass of 
a black hole to be $M_{bh}=3.6\times10^6 \MSUN$), 
the maximum specific angular momentum
$l^{max}_a$ at the inner radial boundary, and a time-averaged value of
mass accretion rate onto a central object in units of Bondi accretion
rate $\MDOT_a/\MDOT_B$ at the end of the simulation.  
To keep the same resolution at small radii for all cases,
our runs with lower $R'_S$, were performed
with more grid points in radial direction.
We followed each simulation until the quasi-stationary state was
achieved, i.e., when the time averaged mass accretion rate and torus
properties settled down.

\subsection{Mass accretion rate evolution}

Figs.~\ref{fig:1a},~\ref{fig:1b}, and \ref{fig:1c} 
show the mass accretion
rate evolution, for $R'_S=10^{-3}$, $R'_S=10^{-4}$, and
$R'_S=10^{-5}$, respectively, for different $\gamma$.  In
all cases, after an episode of the spherically symmetric inflow for
which $\MDOT_a/\MDOT_B$ =1, $\MDOT_a/\MDOT_B$ decreases and starts to
fluctuates around some time-averaged level.  For $\gamma=5/3$, 
the mass accretion rate evolves in a similar way for all $R'_S$, i.e., 
it stabilizes quickly and shows no strong time-variability.  
However, for $\gamma=4/3$ and 1.2, $\MDOT_a$  
the amplitude of the fluctuations is significant. In particular,
$\MDOT_a$ can suddenly increase by a factor of $\sim 5$ and then
also suddenly decrease. 
These flares are a result of mixing of high and low angular
momentum matter (see next sections).  For
$\gamma=4/3$, the occasional flares appear for all the values of $R'_S$
we explored.
We suppose that the flares are also typical for flows
with $\gamma=1.2$ for regardless of $R'_S$.

This strong time variability for intermediate $\gamma$, 
is quite surprising given we explore a relatively simple HD case.  
Another surprising result is that the amplitude and time scale
of the variability depends on $\gamma$. In particular,
the flares found in models with $\gamma$=4/3 and 1.2
disappears in models with $\gamma=1.01$. 
Instead in these models, $\MDOT_a$
exhibits small amplitude quasi-periodic
oscillations (see next sections), for all $R'_S$ values.  

Comparing our results for a fixed $\gamma$ but various $R'_S$, we
find that the $\MDOT_a$ time dependence  is quite insensitive to $R'_S$.
This parameter determines the size of the sonic
surface and therefore can somewhat 
affect the time-averaged $\MDOT_a$ (see Table 1).

To summarize our results for $\MDOT_a$, in
Fig.~\ref{fig:6} we plot the mass accretion rate at the end of
the simulations as a function of $R'_S$ for different $\gamma$.  
The final mass accretion rate is a relatively complex function
of our two model parameters: for $\gamma=5/3$, $\MDOT_a$
decreases with increasing $R'_S$ whereas for $\gamma=1.01$ it increases
with increasing $R'_S$. For $\gamma=4/3$, $\MDOT_a$ is not a monotonic
function of $R'_S$, i.e., for $R'_S<10^{-3}$ it decreases
with increasing $R'_S$ but for $R'_S\simless10^{-3}$, it
increases with increasing $R'_S$.

As shown by PB03a, $\MDOT_a$ depends on the shape and size
of the sonic surface. We interpret the complex dependence of $\MDOT_a$
on $\gamma$ and $R'_C$
as a result of complex relation between the size of the sonic
surface that depends on $R'_S$ and $\gamma$ and
the shape of the surface that depends 
strongly on $\gamma$ but does not depend significantly on $R'_S$.
Next, we will take a closer look at detail properties
of the simulations.

\subsection{Effects of $\gamma$ and $c_{s,\infty}$ on sonic surface topology and flow properties}

Fig.~\ref{fig:2} shows the time evolution of the direction of the
poloidal velocity  and contours of the sonic surfaces
(i.e., where the Mach number for the poloidal velocity equals 1: 
$|v_p/c_s|=1$) 
for four different $\gamma$ 
(from top to bottom, $\gamma$= 5/3, 4/3, 1.2 and 1.01). 
All panels in Fig.~\ref{fig:2} correspond to models
where $R'_S=10^{-3}$, and each column in the figure corresponds to the
same time.

In this and the remaining part of the paper, we focus on the
results obtained for $R'_S=10^{-3}$ (runs G, H, I, and J),
because of relatively similar qualitatively results obtained for
$R'_S\leq 10^{-3}$.

For $\gamma$ = 5/3 (run J, the top panels in Fig.~\ref{fig:2}), the initial
spherical sonic surface, corresponding to the Bondi sonic surface, is
very small ($r_s\sim 26.8 R_S$) so that it is hard to
see it in Fig.~\ref{fig:2}. 
Thus we refer a reader to the top right panel in  Fig. 9. 
When the rotating gas reaches 
the initial sonic surface, the rotation quickly modifies
the surface shape: the sphere
turns into two lobes elongated along the rotational axis
(``figure eight'' shape). This new shape is caused by the slowdown
of the gas in the equatorial region due to centrifugal force.
In fact, the centrifugal force and gas pressure halt the inflow
and push the gas outward along the equator.
The outflowing gas creates a  shock with the gas infalling from large radii.
The shock then propagates outward and eventually leaves
the computational domain.  One can follow the evolution
of the shock looking at the panels in Fig.~\ref{fig:2} from left
to right.  In the right panel showing the end of the
simulation, the shock has already left the computational
domain.  The right panel shows also that the flow lost with
time its symmetry with respect to the equator (we will return
to this point in Section 4).

For $\gamma = 4/3$ (run I, the second row in Fig.~\ref{fig:2}), 
the sonic surface is larger than for $\gamma=5/3$ ($r_s=253~R_{\rm s}$ 
instead 26.8~$R_{\rm S}$). At the beginning of the evolution,
the initially spherical sonic surface does not
evolve much, because rotation at the sonic surface is
slow.  However, as the rotating gas reaches small radii it starts
to slow down due to increasing centrifugal
force. Consequently the second, inner sonic surface
forms inside the original one.  We refer to the initial sonic surface as
the outer surface.
The inner sonic surface forms because
high-$l$ matter cannot accrete directly onto the central BH. 
In run I, gas is more gravitationally bound than 
in run J and does not flow out but rather accumulates in the inner region
forming a subsonic turbulent torus.  
The inner flow shocks the supersonically accreting gas before the latter 
feels its centrifugal barrier.  We define a torus as an equatorial region 
with high-$l$ gas and subsonic radial velocity.
The torus evolves. Namely its density, gas pressure and size increase 
with time. 
Subsequently the inner sonic surface grows, too.  
The two sonic surfaces eventually connect as
the inner surface reaches the outer sonic surface.  When this
happens, the accumulated matter with high $l$ starts
to flow out as in run J. 
The sonic surface topology changes and eventually
the sonic surface has the figure eight shape.
We note that in run I, the figure eight shape is achieved in a
different way than in run J.  
The second row of panels from the top in Fig.~\ref{fig:2}
show that the matter
flows out, but the shock propagation is delayed in comparison run
J (the top row of panels).  At the end of run I, the flow 
is much more asymmetric than for run J
(e.g., in run I, there is a strong outflow toward the `north' direction).

For $\gamma=1.2$ (run H, the third row of panels in Fig.~\ref{fig:2}) 
the evolution of
the sonic surface topology is similar to that in run I.  The main
difference is that the inner sonic surfaces evolves on longer time scale
in run H than run I.
The second left panel of Fig.~\ref{fig:2} captures a moment
soon after the two sonic surfaces merged. A delay in 
formation of the outflow is caused by the fact
that for lower $\gamma$, the sound speed is lower and 
the radius of the initial outer sonic surface is
larger.  At the end of run H, the flow is very asymmetric.

The simulation for $\gamma = 1.01$ (run G, the bottom row of panel
in Fig.~\ref{fig:2})
shows that again two sonic surfaces form as in runs H and I. 
However, the time scale for two sonic
surfaces to merge is much longer in comparison to other $\gamma$
cases, because a large distant between the two surfaces 
(the outer surfaces is at $\sim$500 $R_S$) and small sound speed.
In run H, we do not observe the
formation of the outflow and we see no shock propagating outward.  
The time scale of the inner
sonic surface growth rate is very long, and we are not able to follow
it, with 200 points in $\theta$ direction, because of too long
simulation time.  Although not shown here, we performed a test
simulation, with 2 times lower resolution in $\theta$ direction.
In this test run we observe the sonic surface
topology change in $\gamma=1.01$ case too, but after much longer time
in comparison to other $\gamma$s (e.g., the topology in run G changes 
on a time scale 150 times longer 
than in run H!).  We see no outflow in the equatorial plane for this set
of parameters, for very long simulation time, even after the figure eight
sonic surface is formed. 

We also performed additional computations for
$\gamma$=1.01 models with the higher flow temperatures corresponding
to $R'_S=10^{-2}$ and $2\times 10^{-2}$, but keeping the
same grid parameters as for $R'_S=10^{-3}$ models. As expected,  in 
these additional runs, the flow evolves faster than in run H. In particular,
an outflow forms on a reasonable time scale.
Generally, we found that
when we suppress the gas pressure
enough by decreasing $\gamma$ or decreasing $c_{\infty}$, 
an outflow will take a very long time to develop. A delay in development
of an outflow is a stronger function of 
$\gamma$ than $c_{\infty}$.

Fig.~\ref{fig:bilans} compares the mass flux rates 
time-averaged at the end of simulations as a function
of radius for runs I, J, H, and G.
The mass flux rate at the given radius r is given by:
\begin{equation}
\MDOT= r^2 \oint_{4 \pi} \rho v_r d\Omega
\label{eq:mass}
\end{equation}
where $d\Omega = \sin(\theta) d\theta d\phi$, $\rho$ is the density,
and $v_r$ is the radial velocity.  To calculate the mass inflow rate
$\MDOT_{in}$ at a given radius, we include
only contributions  with $v_r < 0$. Whereas
to calculate the mass outflow rate
$\MDOT_{out}$ at a given radius, we include
only contributions  with $v_r > 0$. 
To compute the net mass flow flux $\MDOT_{tot}$ we include
all contributions regardless of the $v_r$ sign.  For run J, I, and H, the
$\MDOT_{in}$ is larger than $\MDOT_{out}$, and  $\MDOT_{tot}$ is
constant. The latter indicates  that
the three models are in time-averaged steady states  at the end
of simulations. We do not show it here, but we
check that the total mass in the computational domain
does not change in time that again indicates the time-averaged steady state.
On the other hand, run G does not settles into a steady state
as  matter continues to
accumulate between $5 R_S < r < 100 R_S$ and $\MDOT_{tot}$ changes
with radius.

Fig.~\ref{fig:bilans_time} compares the mass flux rates as 
functions of time at the outer boundary of the grid. We calculate
$\MDOT_{in}$ and $\MDOT_{out}$ from eq.~\ref{eq:mass} for a given time, 
in the same way as shown in Fig.~\ref{fig:bilans} but assuming that
r=$r_o$. For runs J, I, and H, the mass inflow rate is constant 
for the short time at the beginning of the simulations.
The gaps in the early time of simulations are caused 
by the shock transition through the outer boundary.
In the later times the $\MDOT_{in}$ and the $\MDOT_{out}$
oscillate and are anticorrelated.
For G model, there is no outflow and $\MDOT_{out}$ and
$\MDOT_{in}$ is constant during the whole simulation time.

Next we present and describe in more details the structure of the flow 
for different $\gamma$. 
Figs.~\ref{fig:3a},~\ref{fig:3b}, and~\ref{fig:3c}
show the 2-D structure of various quantities from 
the simulations for different $\gamma$, on different
scales (i.e., 20 $R_S$, 200 $R_S$, and 1200 $R_S$, in
Fig.~\ref{fig:3a},~\ref{fig:3b}, and~\ref{fig:3c}, respectively), at
the end  of the simulations.  Each of the figures consists of
four rows, in which we present results for $\gamma$=5/3, 4/3, 1.2 and
1.01 (panels from top to bottom, J, I, H, and G runs respectively).
From left to right the panels show the density  temperature maps,
angular momentum contours, angular velocity contours and direction
of the poloidal velocity overplotted by the sonic surface
shape.

In all four cases, characterized by different $\gamma$ index, a torus
forms.  We distinguish three kinds of tori: gas pressure dominated
torus, with $\gamma=5/3$, 'radiation pressure' dominated torus with
$\gamma=4/3$, and for $\gamma=1.01$, due to weak dependency of
pressure on density, the torus is rotation dominated.

Run J here corresponds to run B03f1a in PB03a.  
Our results are in full agreement with the previous
simulations in all respects.  The torus is subsonic, the angular
momentum distribution in the torus is nearly constant $l \sim$ 0.92
$R_S c$ (see also Fig.~\ref{fig:5} described below).  
The matter accretes through the
polar funnel, and flows out in the equatorial region.  The mass accretion rate
for $\gamma \sim 5/3$ and $R'_S=10^{-3}$ is
consistent with the previous calculations of PB03a with two times lower
angular resolution.

For $\gamma=4/3$ (run I) the radius averaged $l$ in torus is a little
higher than in $\gamma$=5/3 case, and it is around 1.1 of the critical
value.  The density gradient between polar funnels and torus is larger
than in run J.  The sonic surface is strongly asymmetrical.  There
is the above mentioned outflow in 'north' direction. Note
that in Figs.~\ref{fig:3a},~\ref{fig:3b}, and~\ref{fig:3c}, the
angular momentum contours undergo strong compression on the `north'
side of the flow.  The contours are also compressed in the `southern' part
of the grid.  This compression is due to the torus
movement in `z' direction which we will discussed in more detail 
in the next subsection.

In run H, an equatorial outflow is asymmetric and
supersonic at large radii. Fig.~\ref{fig:3c}, illustrates strong mixing 
of high- and low-$l$
matter at large distances in the `southern' part of the grid leading
to  an increase in of the  mass accretion rate seen as flares in 
Figs.~\ref{fig:1a}, \ref{fig:1b}, and \ref{fig:1c} (thick dashed lines). 

For $\gamma=1.01$ (run G), the inflow is a very symmetric,
at large radii.  On smaller scales (r$<$ 200 $R_S$), the
high-$l$ turbulent torus (red and yellow region in density map
in Fig~\ref{fig:3b}, bottom panels) is surrounded by slightly lower
density, nearly laminar inflow region (the green region in the density maps 
in the same panel).  The high-$l$ matter flows around and supplies
the matter to the turbulent torus.  The torus
expands in the `z' direction with time
(see Figs~\ref{fig:3c} and~\ref{fig:4}). 
Thus  to reach the black hole,
the low-$l$ matter has to flow around an growing in size an obstacle.
As a result,
an oblique shock, corresponding to the $v_{\theta}$ sonic surface, 
forms around torus.
The oblique shock surrounds the above mentioned region of slightly higher
density (green region in density maps).  For r $<$ 20~$R_S$ 
(bottom panels in Fig.~\ref{fig:3a}), the torus shape remains
quite flat, and is of high density and angular momentum
with the latter approaching Keplerian distribution.  
In this inner region,
the low-$l$ gas 
flows toward the equatorial plane almost parallel to
the symmetry axis (note that
the angular momentum contours outside the torus
are parallel to the rotational axis). We check that 
the stream lines of the accreting gas, 
once they reach the torus they
follow the torus shape.

For r $<$ 5 $R_S$, the low-$l$ matter forms an accretion cusp.  For some
time during at the initial evolution, the accretion cusp
expand in the `z' direction, because of the torus cusp grows in
$\theta$ direction.  This growth slightly reduces polar funnel
accretion, which dominates $\dot{M}_a$.  However soon, the size of
the cusp stabilizes.  The effect of cusp expansion can be seen in the
accretion rate evolutionary curve dip (Fig.~\ref{fig:1a}, thin solid
line) around t=$2.0 \times 10^4$.  Later the torus expansion occurs at
larger radii only.

One of the most intriguing property of all the runs is that
the angular velocity inside the turbulent torus, has cylindrical distribution
even though the radial profile of rotation depends on $\gamma$.
Fig.~\ref{fig:5} shows radial profiles of $l$  along the equator 
for models  with different $\gamma$ at
the end of the simulations.
For comparison, Keplerian $l$ distribution
and $l$ corresponding to constant angular velocity are shown too.

As found by PB03a, $l$ is constant in the torus for $\gamma=5/3$.
We find that $l$ does not change much inside the torus
also for $\gamma$=4/3 and 1.2 though the actual value
of $l$ in the torus increases with decreasing $\gamma$.
As several other properties, the rotational profile for $\gamma=1.01$ 
differs from that for other $\gamma$. Namely, in run G,
the profile is almost Keplerian for small radii (compare solid and dash-dotted
line in Fig.~\ref{fig:5}).

A detailed inspection of Fig.~\ref{fig:5} shows that
$l$ sharply decreases with radius for very small radii in runs H and I .
This decrease is due to the fact that the inner flow
is asymmetric and the gas reaching the BH at the end
of these runs is not that of the torus but is of the gas
with low $l$. The asymmetry also explains why
$l_{max}$, listed in Table~1, is higher than $l$ at $r_{\rm i}$
in Fig.~\ref{fig:5}, i.e., the maximum accreted $l$
does not have to and is not always at the equator.
The increase of $l$  with radius at large radii is caused by
the outer boundary conditions we use
with these simulations (i.e., we reset $v_{\phi}$ at $r_{\rm o}$
to its initial value at each time step).

\subsection{Time-variability and asymmetry of the flow}

We find that the temporal behavior and sonic surface topology 
of the accretion flow
depend on $\gamma$. In runs with $\gamma=$4/3, 1.2, and 1.01
two sonic surfaces can exit. The sonic surface topology evolves
in the similar way in all runs with the three $\gamma$'s
although on different time scales (the evolution slows down with decreasing
$\gamma$). During the early evolution,
gas with large angular momentum
(larger than $l_{cr}$) is locked in a turbulent torus.
The torus expands but there is no strong outflow because the gas motion is 
suppressed by supersonically infalling matter.  During this phase, an oblique
shock (seen as a jump in $v_{\theta}$ in Figs.~\ref{fig:3a} and ~\ref{fig:3b}) 
forms around the torus.  By contrast, in runs with $\gamma=5/3$, only one 
sonic surface exists and its shape turns from initially spherical to 
the figure eight like one.
The differences in the sonic surface topologies
are reflected in the mass accretion rate curve: 
for runs with one sonic surface the curves are smooth 
whereas for runs with two sonic surfaces the curves
are not smooth (compare various curves in 
Figs.~\ref{fig:1a},~\ref{fig:1b}, and \ref{fig:1c}). 

Generally, the time behavior of the flow depends mainly
on the Mach number which in our simulations depends 
on $\gamma$ and $c_\infty$. To connect the time dependence of $\MDOT_a$
with the flow properties we start with an analysis of run G
with the highest Mach number of the inner flow. In run G,
$\MDOT_a$ shows oscillations which one can attribute to oscillations
of the inner flow in the $z$ direction (see Fig.~\ref{fig:4} showing time sequence from runs G and H).

In run G (bottom panels in fig.~\ref{fig:4}), the high-$l$ gas that 
accretes supersonically in the equatorial plane from
large radii eventually tries to join the turbulent subsonic torus 
at small radii.  The torus acts in this case as an obstacle
with which the supersonic gas collides
at about 120 $R_S$ in the equatorial plane. There a strong shock forms.
At smaller radii, the post-shock gas 
splits into two streams that flow around the torus. However,
the split is not into two exactly equal halves. At the early
phase of the evolution asymmetry is very small as it is 
due to random numerical asymmetry. However with time
the asymmetry grows due to strong shocks. In our simulations
strong shocks are handled by artificial viscosity (Stone \& Norman 1982).

The asymmetry amplified by shocks propagate into other parts
of the flow. In particular, the torus becomes asymmetric. Subsequently,
a larger fraction of the equatorial inflow can 
flow in one of the hemisphere compared
to the other hemisphere (Fig.~\ref{fig:4}, left bottom panel, illustrates
a phase where more gas flows in the 'southern' hemisphere). Once
above the torus, the equatorial inflow pushes the polar inflow away
from the torus in the 'z' direction
(seen as an oblique shock that oscillates and expands
in the $z$ direction). The asymmetric interaction of the high-$l$ equatorial
and low-$l$ polar inflows can be also seen in  
the angular momentum contours 
which are more compressed on one side of
the torus.  The green colored region in Fig.~\ref{fig:4}
expands and the density in the region at radii $\simless 40 R_S$ on the
`southern' side of the torus slightly decreases (Fig.~\ref{fig:4},
middle bottom panel).  When this happens, the polar inflow of low-$l$
matter starts to push the shock back toward the equator.
The oblique shock swings and more of  the equatorial inflow
goes around the torus from the `northern' side.  The
scenario repeats (Fig.~\ref{fig:4}, right bottom panel).
As the oscillations continue, the oblique shock expands because
matter has to flow around the constantly growing torus.

For $\gamma=1.01$, the torus is confined by the supersonic inflow.
Therefore, asymmetry is strong only at small radii whereas
at large radii the flow is symmetric as it
is not affected by the presence of the asymmetric obstacle.  
Thus the low amplitude oscillations in
the mass accretion rate curve are caused by  the oscillating torus.
The relevant time scale here is the dynamical time scale
at the radii where the equatorial  inflow is split into
two streams. 
Fig.~\ref{fig:var} presents a part of the mass accretion curve
for model G in the later times of evolution to show the periodic
variability. The time period between the peaks, corresponds to the
dynamical time scale at radius of about 84 $R_S$ that corresponds
to the outer radius of the torus where the high-$l$ gas splits up into
two parts.

We also calculated a power spectrum using the mass accretion rate
curve in the later times of evolution.  The power spectrum shows broad
peak, at frequency corresponding the dynamical time scale at radius of
about 84 $R_S$. The peak is broad because the torus increases in
size with time, in particular the torus radius increases.
We computed additional models for $\gamma=1.01$ with higher resolution in
$\theta$ direction (400 and 800 points in $\theta$ angle). We found that
oscillations appear to be independent of the grid resolution.

For $\gamma>1.01$, the flow behaves in a similar ways
as that in run G. However, there is an additional complexity
due to high gas pressure and mixing. For $\gamma>1.01$, the flow
is less gravitationally bound than in run G and the torus
is not as much confined by the supersonic inflow.
Therefore, in runs with higher gas pressure the flow
is subsonic in a relatively big region where
mixing between low- and high-$l$ occurs. Because of the shock
amplified asymmetry the mixing leads to large scale
asymmetry (compare run G and H in Fig.~\ref{fig:4}). 

As we mentioned above, for $\gamma=4/3$ and $\gamma=1.2$, we also
observe occasional bursts in the $\MDOT_a$ evolution (see
e.g. Fig.~\ref{fig:1a}).  The mass accretion rate can significantly
increase, because of mixing of the high- and low-$l$ gas
at large radii.  The flaring behavior of the mass accretion rate in models for
$\gamma=4/3$ and $\gamma=1.2$ is caused by the torus movement in the
`z' direction.  Fig.~\ref{fig:3c} (panels corresponding to $\gamma$=1.2
case) captured a situation when the torus moved toward the southern
hemisphere and partially blocked an inflow of low-$l$ gas
in the polar funnel. Then this low-$l$ gas is pushed toward equator.  
When the torus later
swings toward the northern hemisphere, a channel for  the low-$l$ widens
one sees a flare in the mass accretion rate curve.
The reason we do not observe
strong flares for $\gamma=1.01$ is that there is no
mixing of gas with high- and low-$l$ at larger distances.

We find that the flow oscillates in the `z' direction and is
asymmetric with respect to the equatorial plane.  Instabilities in the
accretion disks were studied by many authors.  Pulsational instability
of quasi-adiabatic and quasi-inviscid accretion disks, was examined by
in many studies (e.g., Kato 1978
Muchotrzeb-Czerny 1986; Kato et al. 1988; Okuda et al. 1993).
Kato et al. (1978) found that the condition for pulsation
instability depend on how viscosity changes during pulsation.  
In our simulation only artificial viscosity is introduced, which is
small. Nevertheless this viscous proves to be important. 

In summary, we find that the asymmetry of the flow is initially induced by
the numerical effects, but the propagation and amplification of the
asymmetry strongly depends on the physical conditions in the
accretion flow.  In the cases where the gas pressure is reduced
because of low $\gamma$ the asymmetry is stronger.  For higher
$\gamma$, the gas pressure is larger and the information about the
perturbation propagates faster. The gas pressure works against
effects of shocks and 
tries to restore the symmetry.
We explore effects of gas pressure to confirm its role
in maintaining symmetry in the flow. To this end, we
performed additional simulations for the
same parameters as run J, but with higher $c_{\infty}$ corresponding
to models with $R'_S=10^{-2}$.  Fig.~\ref{fig:extra} shows time
evolution of velocity field of one of our exploratory runs.  
The flow with very high $c_\infty$ is symmetric as one would expect: 
the information of
perturbances in high pressure gas propagates vary fast independently of
grid effects.  This simulations supports our reasoning 
that the asymmetry growing in the flow is not caused 
by numerical effects but rather by
shocks amplifying initial seed asymmetry.

Our time dependent simulations occassionally show structures that 
resemble features of steady state disk solutions, such as backflows 
close to the equatorial plane (Kluzniak $\&$ Kita, 2000, and our 
Fig.~\ref{fig:extra}), or tori in the innermost part 
of the flow, whose angular 
momentum changes from sub- to super-Keplerian (Figs.~\ref{fig:5},~\ref{fig:4}). Such tori may be relevant to high frequency QPOs (e.g. Blaes et al., 
2007).

\section{Conclusions}

We report on next phase of our study of rotating accretion flows 
onto black holes. We consider hydrodynamical (HD) accretion flows with 
a spherically symmetric density distribution at the outer boundary but 
with spherical symmetry broken by the introduction of a small, 
latitude-dependent angular momentum. We study accretion flows by means of 
numerical two-dimensional, axisymmetric, HD simulations for variety
of the adiabatic index, $\gamma$ and 
the  gas temperature at infinity, $c_\infty$. 
Our work is an extension of work done by PB03a who
consider models for only $\gamma=5/3$. We also reran 
some models from PB03a with higher resolution. The latter
are fully consistent with lower resolution PB03a's runs.

Our main result is that the flow properties such as the topology
of the sonic surface and time behavior strongly depend on
$\gamma$ but little on $c_\infty$.  
In particular, for $1 < \gamma < 5/3$,
the mass accretion rate shows large amplitude, slow time-variability which
is a result of mixing between slow and fast rotating gas.
This temporal behavior differs significantly from that
in models with $\gamma\approx 5/3$ where the accretion
rate is relatively constant and from that in models
with $\gamma\approx 1$ where the accretion exhibits small 
amplitude quasi-periodic oscillations. The key parameter responsible
for the differences is the sound speed of the accretion flow
which in turn determines the strength of shocks and whether
the flow is dominated 
but gas pressure ($\gamma\simless 5/3$), 
radiation pressure ($1 < \gamma < 5/3$) ,  or  rotation
($\gamma\simgreat 1$). Despite
these differences, the time-averaged mass accretion rate
in units of the corresponding Bondi rate is a weak
function of $\gamma$ and $c_\infty$.

We realize that our simulations do not capture several
physical processes that may be important in real systems
(e.g., magnetic fields, radiative cooling and heating,
energy dissipation). However, our simulations are done
within a general framework so that some of these
processes can be straightforwardly added (e.g., magnetic fields
as in PB03b, or radiative processes as in Proga 2007).
Therefore, this work could serve as a good reference
point to analyze and interpret more complete and complex
simulations (e.g., we already reran some of the models
from this paper including magnetic fields, 
Moscibrodzka and Proga, in preparation).

ACKNOWLEDGMENTS: 
We thank Bozena Czerny and Marek Abramowicz for very useful comments.
We acknowledge support provided by the Chandra award TM7-8008X 
issued by the Chandra X-Ray Observatory Center, which is operated by 
the Smithsonian Astrophysical Observatory for and on behalf of NASA 
under contract NAS8-39073. 
M.M also acknowledges supported in part by grant 1P03D~008~29 of the Polish
State Committee for Scientific Research (KBN).
While D.P. acknowledges support from NASA under ATP grant NNG05GB68G.

\newpage
\section*{ REFERENCES}
 \everypar=
   {\hangafter=1 \hangindent=.5in}

{
  Abramowicz, M.A., Chen, X., Kato, S., Lasota, J.-P., Regev, O. 
  1995, ApJ, 438, L37


  





  Blandford, R.D., \& Begelman, M.C. 1999, MNRAS, 303, L1

  Blaes O. M., {\v S}r{\'a}mkov{\'a} E., Abramowicz M.A., Klu{\'z}niak W.,
 Torkelsson U.; 2007: Ap.J. vol. 665, pp. 642-653

  
 
 
  Chen, X., Taam, R.E., Abramowicz, M.A., \& Igumenshchev, I.V. 1997, MNRAS, 285, 439

   Clarke, D., Karpik, S., \& Henriksen, R.N. 1985, ApJS, 58, 81








  Done C., Gierlinski M., Kubota A., 2007, A\&ARev, temp, 3D


 


  Hawley, J.F., Smarr, L.L., \& Wilson, J.R. 1984a, ApJ, 277, 296
 
  Hawley, J.F., Smarr, L.L., \& Wilson, J.R. 1984b, ApJS, 55, 211

 
  Ichimaru, S. 1977, ApJ, 214, 840.

 
 
Janiuk A., Proga D., 2007, arXiv0708.2711

Kato S., 1978, MNRAS, 185, 629

Kato S., Honma F., Matsumoto R., 1988, MNRAS, 231, 37
Krolik J., 1999, 

Klu{\'z}niak W., Kita D., 2000, astro-ph/0006266







   

 M\'{e}sz\'{a}ros, P., 2006, Rep. Prog. Phys., 69, 2259-2322 
 
 Mo H.J., Mao S., White S. D. M, 1998, MNRAS, 295, 319 


Muchotrzeb-Czerny B., 1986, AcA, 36, M1

Proga D., Begelman M., 2003, ApJ, 582, 69 (PB03a)

Proga D., Begelman M., 2003b, ApJ, 592, 767 (PB03b)


 
  Narayan, R., \& Yi, I. 1994, ApJ, 428, L13
 
  Narayan, R., \& Yi, I. 1995, ApJ, 444, 231
 
Okuda T., Mineshinge S., 1993, Astrophysics and Space Science, 210, 361
 
 
  Paczy\'{n}ski, B., \& Wiita, J. 1980, A\&A, 88, 23

 Proga D., 2007, ApJ, 661, 693

 

Sharma, P., Quataert, E., Hammett, G., \& Stone, J. M. 2007, ApJ, 667, 714

  Rees, M.J., Begelman, M.C., Blandford, R.D., \& Phinney, E.S. 1982, 
  Nature, 295, 17

 




 
  Stone, J.M., \& Norman, M.L. 1992, ApJS, 80, 753






\newpage

\begin{table*}
\footnotesize
\begin{center}
\caption{ Summary of parameter survey.}
\begin{tabular}{l c c c c c c  l  } \\ \hline 
                 & Radial       &              &                 &            &                 &              &                      \\
Run              & resolution   & $R'_S$       & $R'_C$          & $\gamma$   & $t_f$           & $l^{max}_a$  & $\MDOT_a/\MDOT_B$    \\ 
                 &              &              &                 &            &                 &              &                       \\
              A  &   220        & $10^{-5}$    & $5\times10^{-4}$& 1.01       &$1.97 \times 10^4$& $0.77$       &   0.45                \\   
              B  &   220        & $10^{-5}$    & $5\times10^{-4}$& 4/3        &$3.10 \times 10^4$& $0.9$        &   0.25                \\   
              C  &   220        & $10^{-5}$    & $5\times10^{-4}$& 5/3        &$3.14 \times 10^4$& $0.92$       &   0.13                \\   
                 &              &              &                 &            &                 &              &                       \\
              D  &   180        & $10^{-4}$    & $5\times10^{-3}$& 1.01       &$2.80 \times 10^4$& $0.82$       &   0.40                \\   
              E  &   180        & $10^{-4}$    & $5\times10^{-3}$& 4/3        &$2.75 \times 10^4$& $0.85$       &   0.13                \\   
              F  &   180        & $10^{-4}$    & $5\times10^{-3}$& 5/3        &$3.19 \times 10^4$& $0.87$       &   0.15                \\   
                 &              &              &                 &            &                 &              &                       \\
              G  &   140        & $10^{-3}$    & $5\times10^{-2}$& 1.01       &$3.48 \times 10^4$& $0.9$        &   0.30                \\   
              H   &  140        & $10^{-3}$    & $5\times10^{-2}$& 1.2        &$3.10 \times 10^4$& $0.94$       &   0.20                \\   
              I  &   140        & $10^{-3}$    & $5\times10^{-2}$& 4/3        &$3.56 \times 10^4$& $0.9$        &   0.17                \\   
              J  &   140        & $10^{-3}$    & $5\times10^{-2}$& 5/3        &$2.67 \times 10^4$& $0.87$       &   0.30                \\   
                 &              &              &                 &            &                 &              &                       \\
\hline
\end{tabular}
\label{tab:1}
\end{center}
\normalsize
\end{table*}

\eject

\newpage

\begin{figure*}
\begin{picture}(0,600)
\put(200,0){\includegraphics{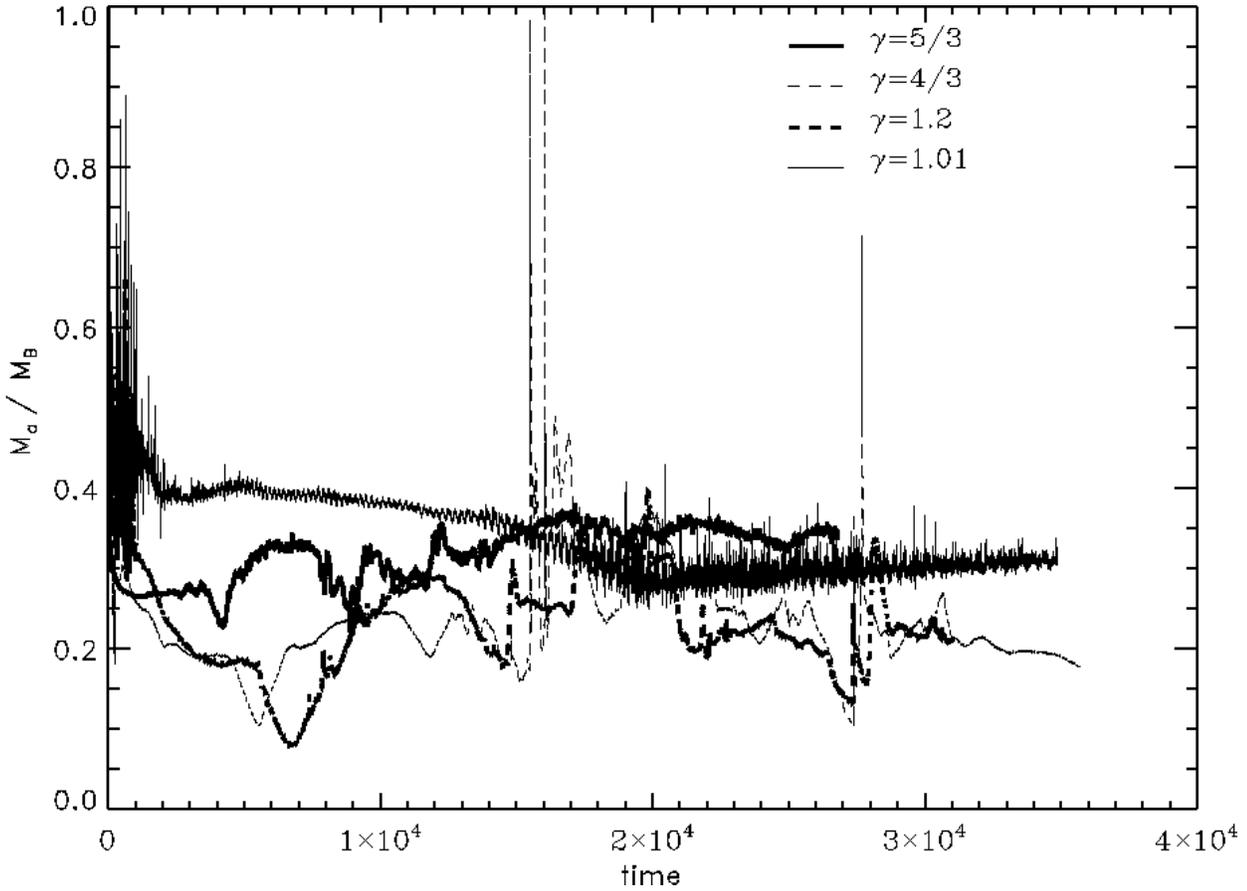}}

\end{picture}
\caption{Mass accretion rate as a function of time for $R'_S=10^{-3}$. Time is given in the units
of dynamical time at the inner boundary. Note the strong flaring behavior for $\gamma=$1.2, and 4/3. 
}\label{fig:1a}
\end{figure*}

\eject

\newpage
\begin{figure*}
\begin{picture}(0,600)

\put(200,0){\includegraphics{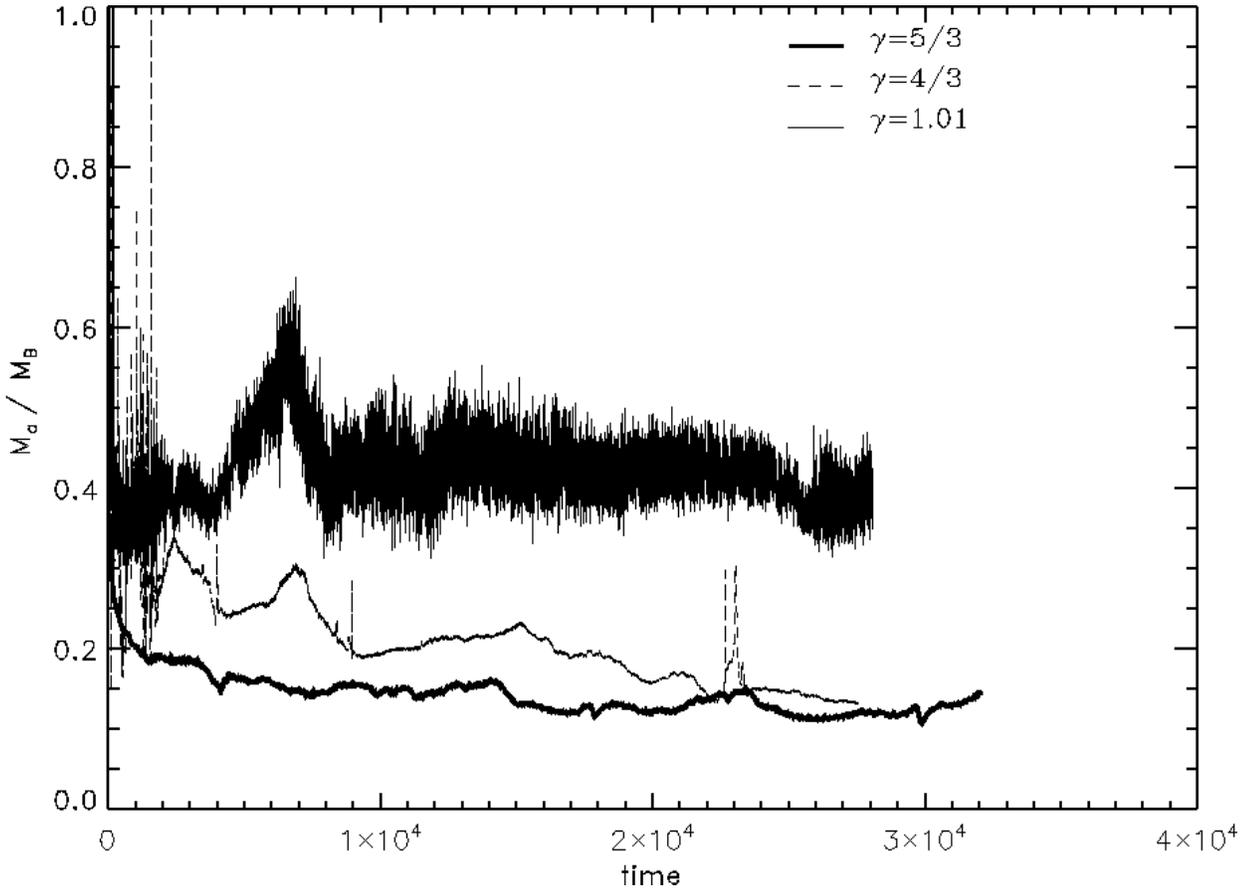}}

\end{picture}
\caption{As in Fig. 1 but for $R'_S=10^{-4}$.}
\label{fig:1b}
\end{figure*}

\eject

\newpage

\begin{figure*}
\begin{picture}(0,600)
\put(200,0){\includegraphics{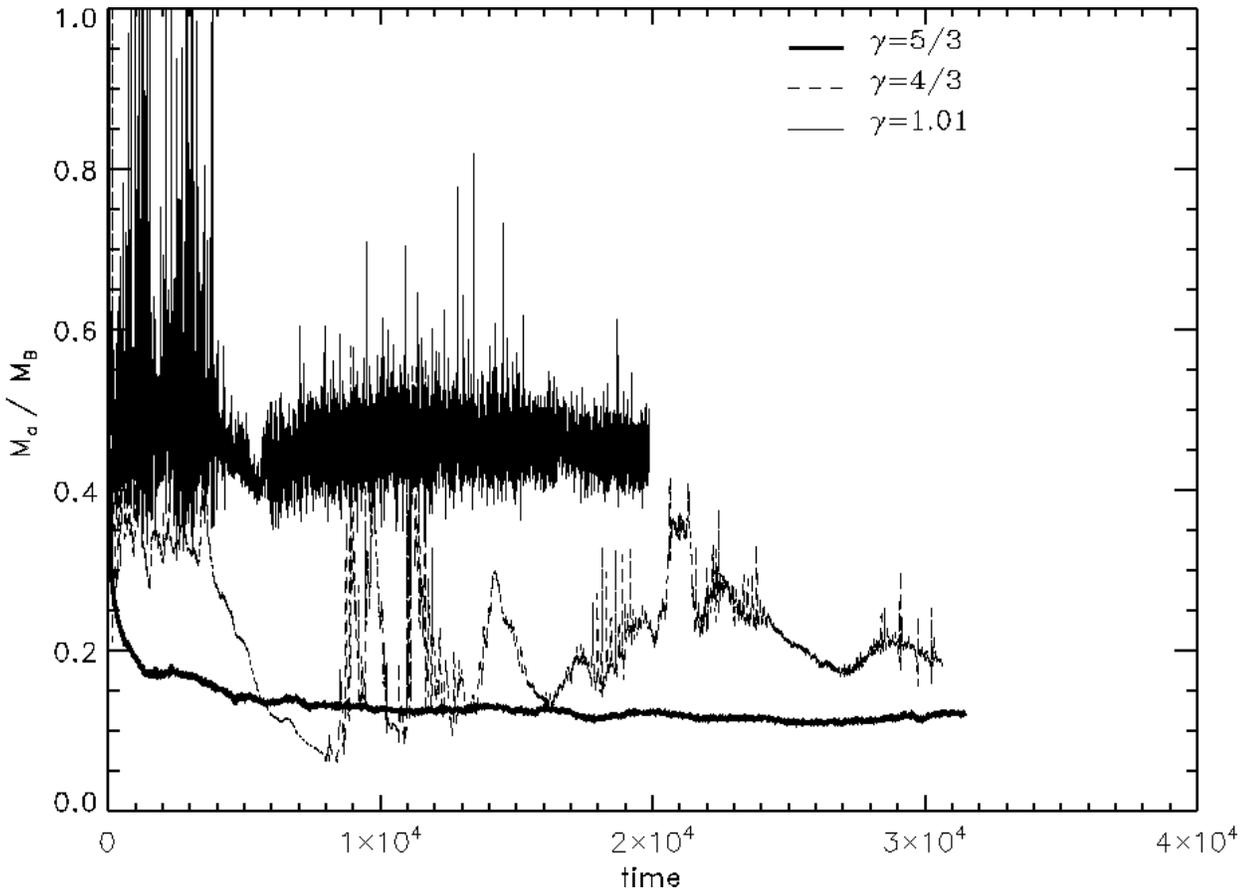}}
\end{picture}
\caption{As in Fig. 1 but $R'_S=10^{-5}$.
}\label{fig:1c}
\end{figure*}

\eject

\newpage
\begin{figure*}
\begin{picture}(0,600)
\put(350,0){\includegraphics{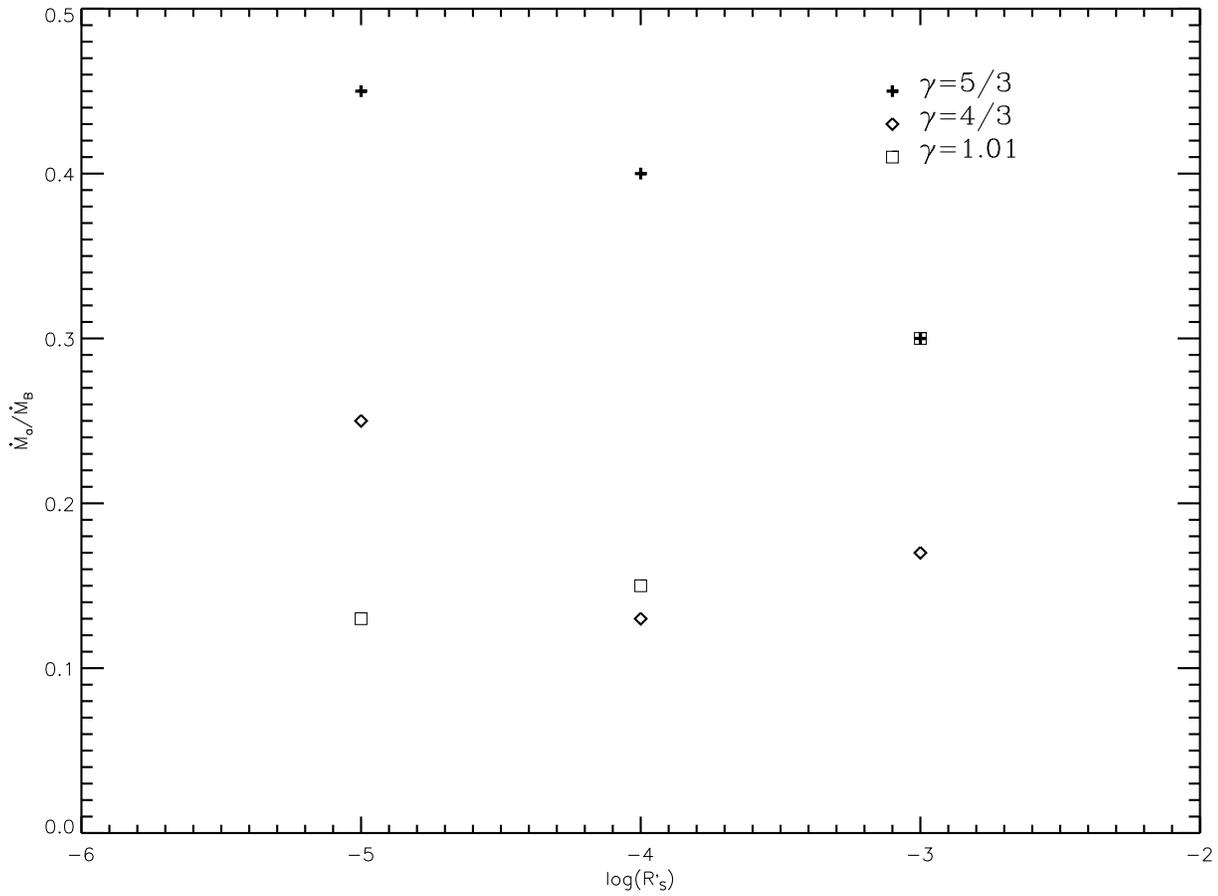}}
\end{picture}
\caption{Mass accretion rate at the end of the simulations 
as a function of $R'_S$ for different $\gamma$.
}\label{fig:6}
\end{figure*}
\eject

\newpage

\begin{figure*}
\begin{picture}(0,600)

\put(80,450){\includegraphics{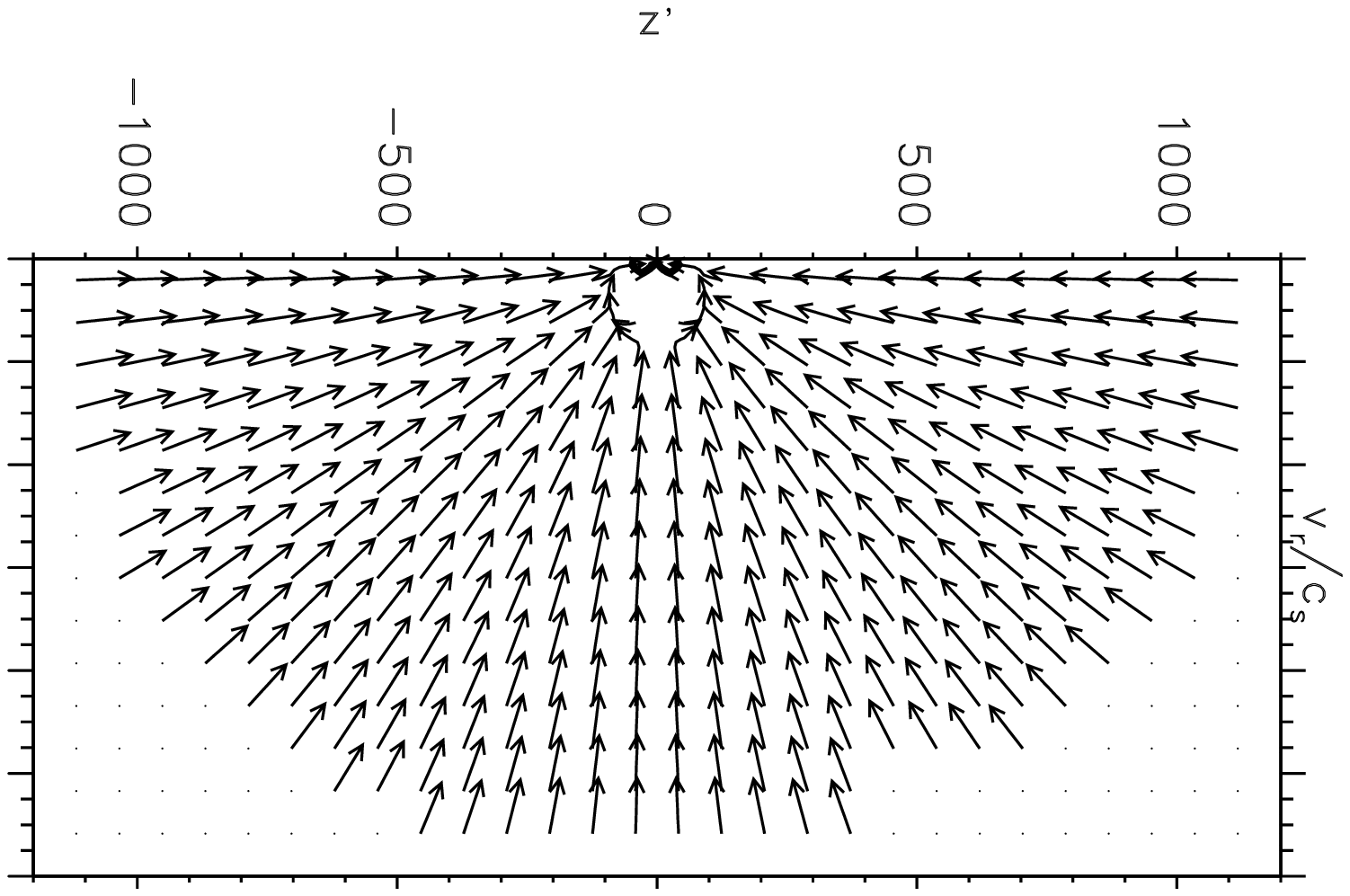}}

\put(180,450){\includegraphics{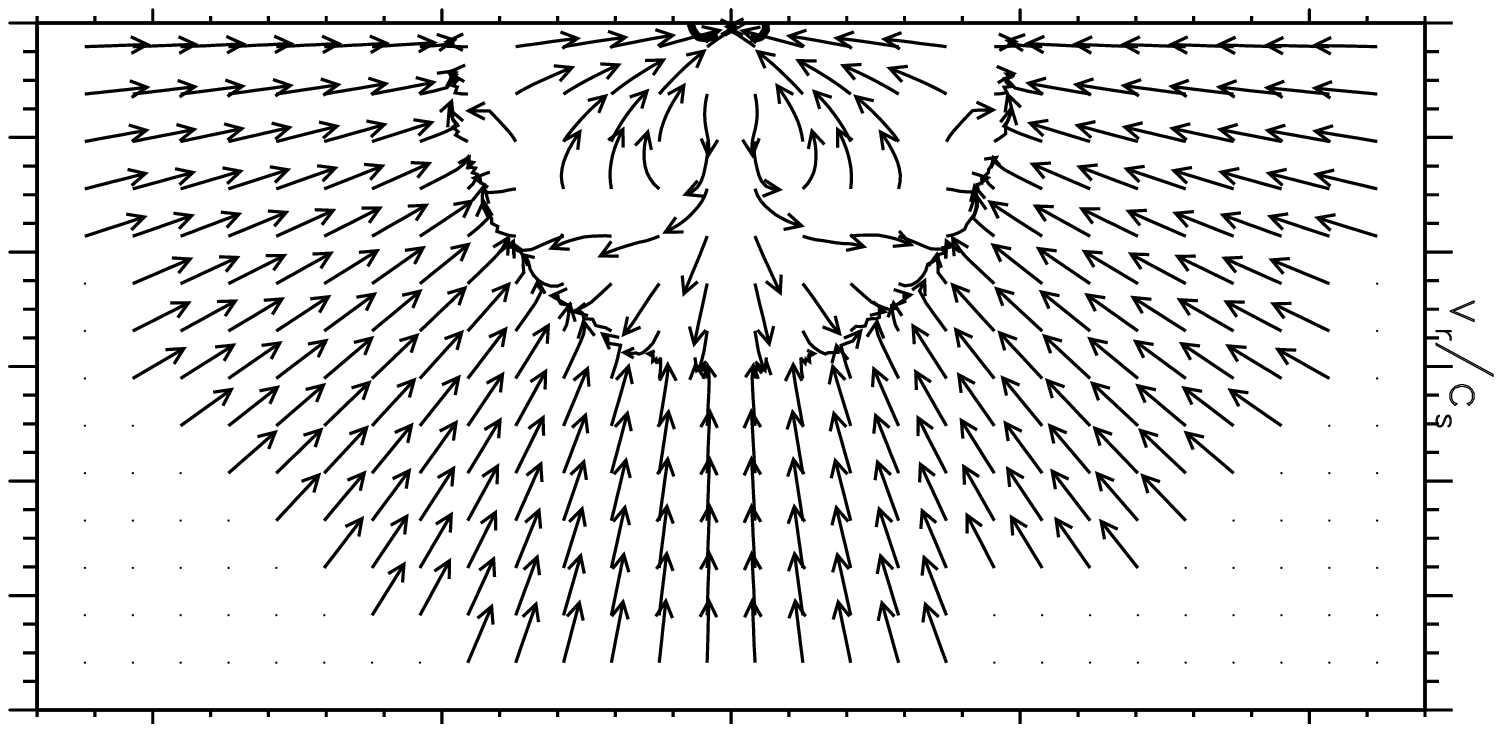}}

\put(280,450){\includegraphics{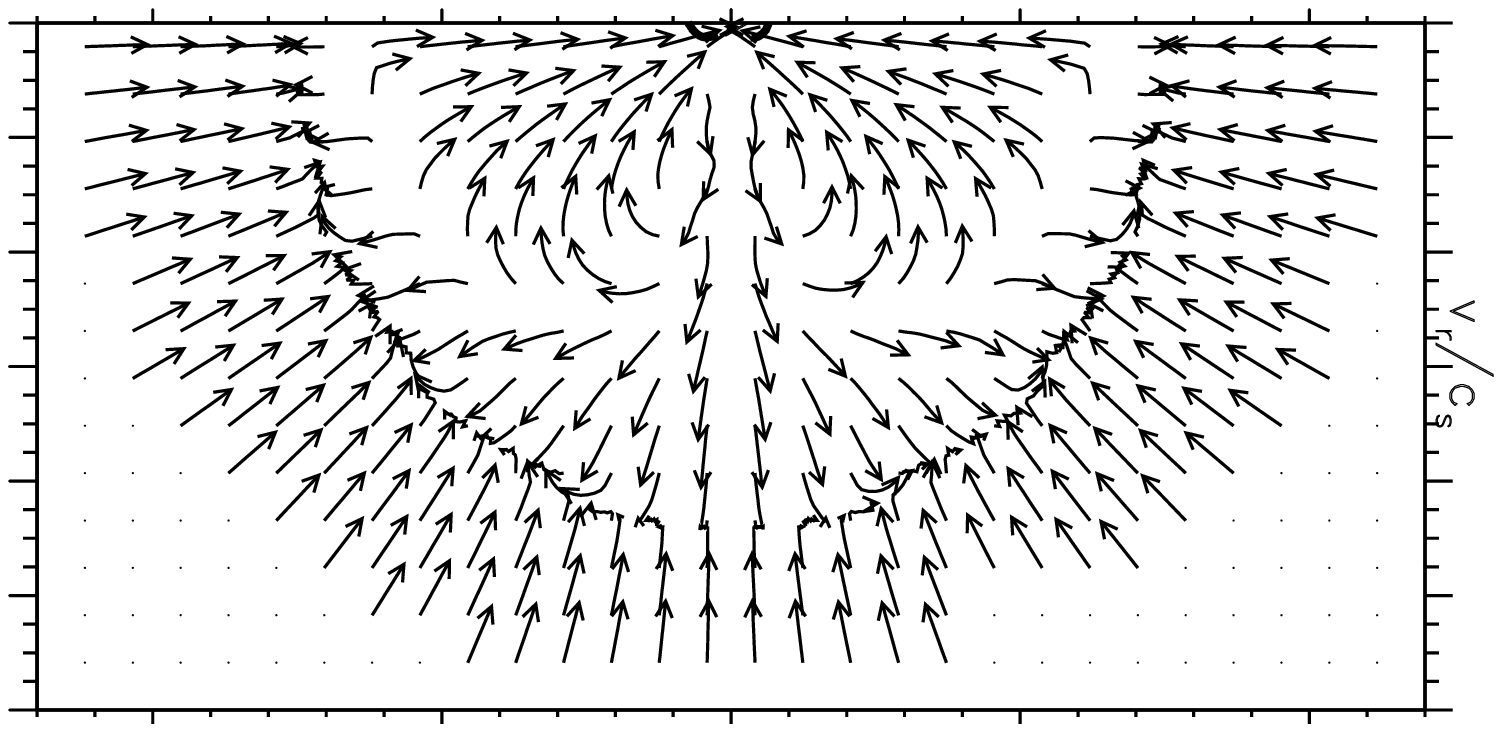}}

\put(380,450){\includegraphics{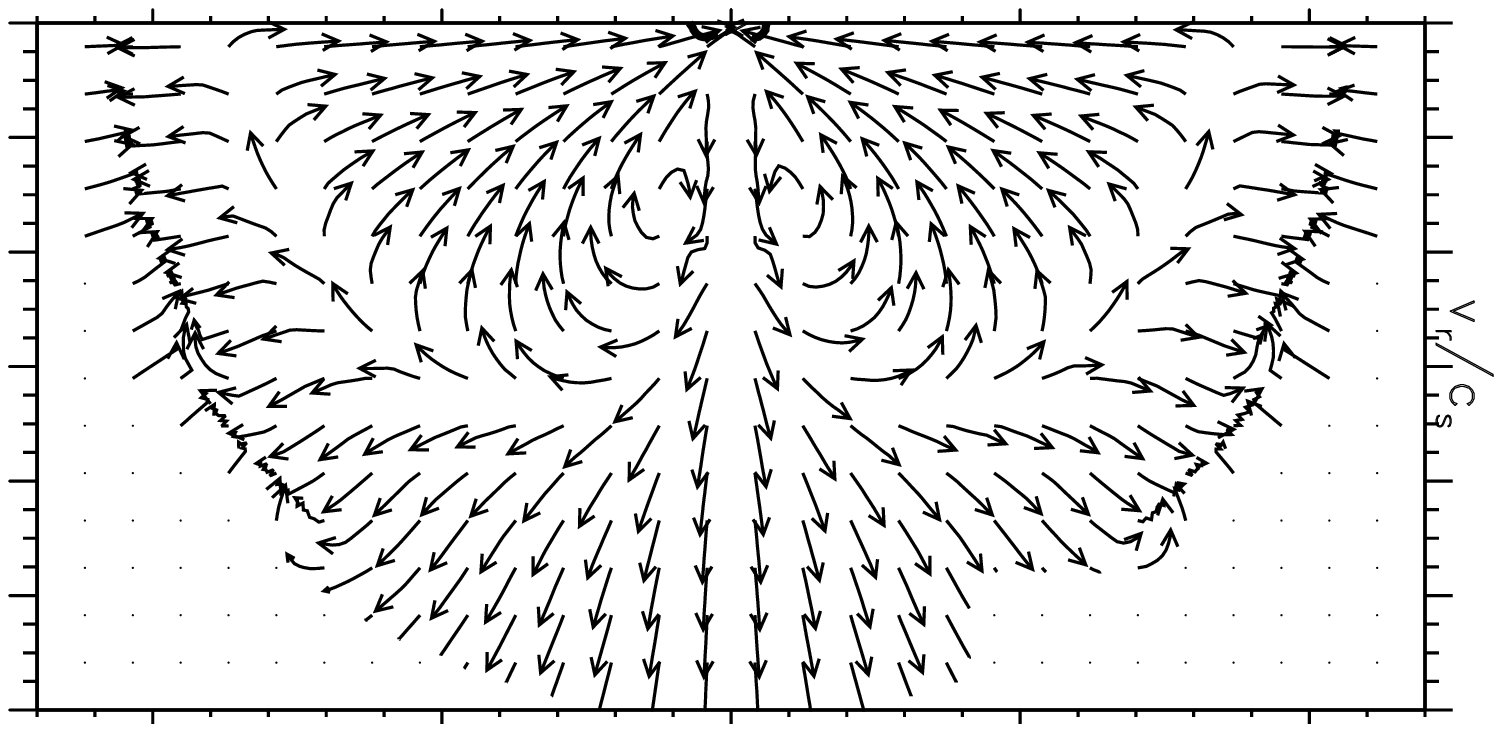}}

\put(480,450){\includegraphics{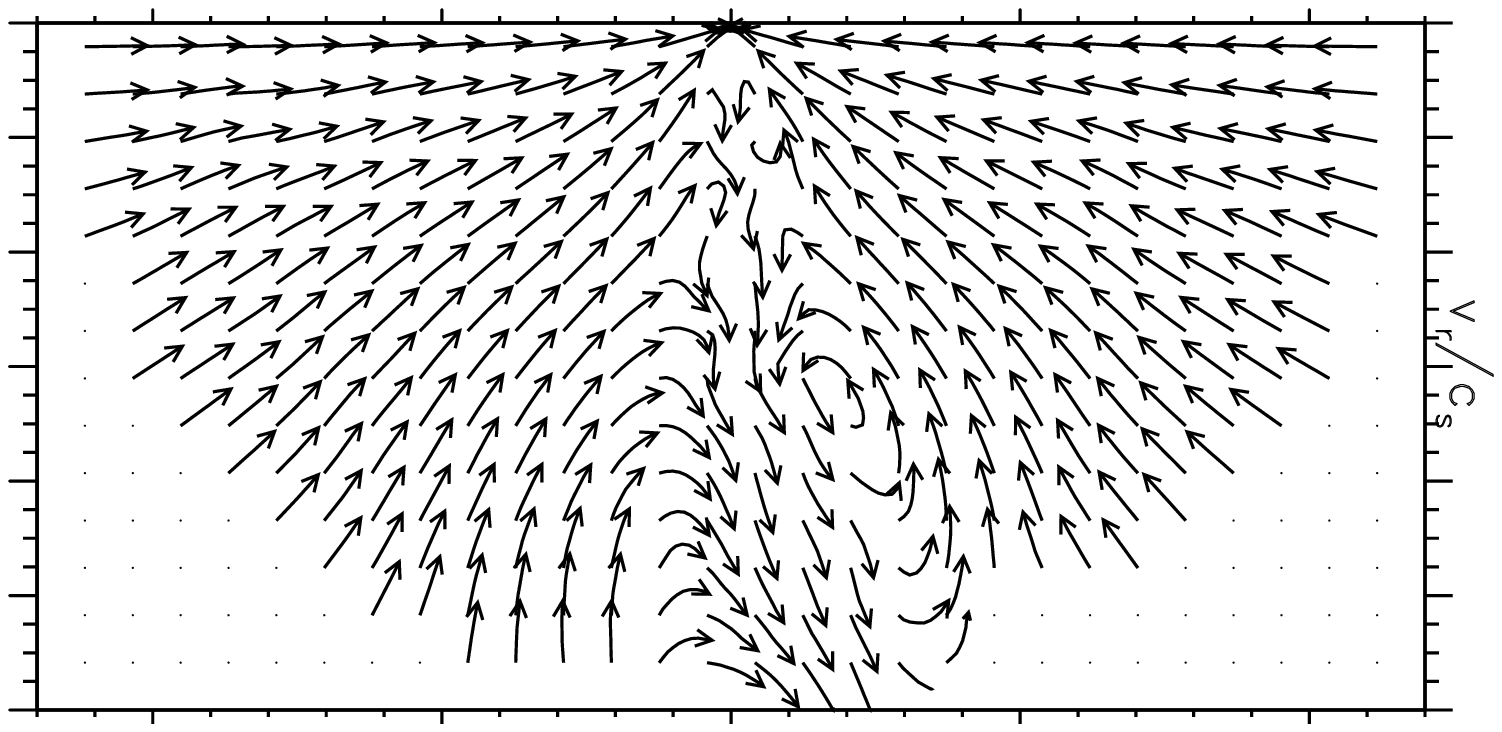}}

\put(80,300){\includegraphics{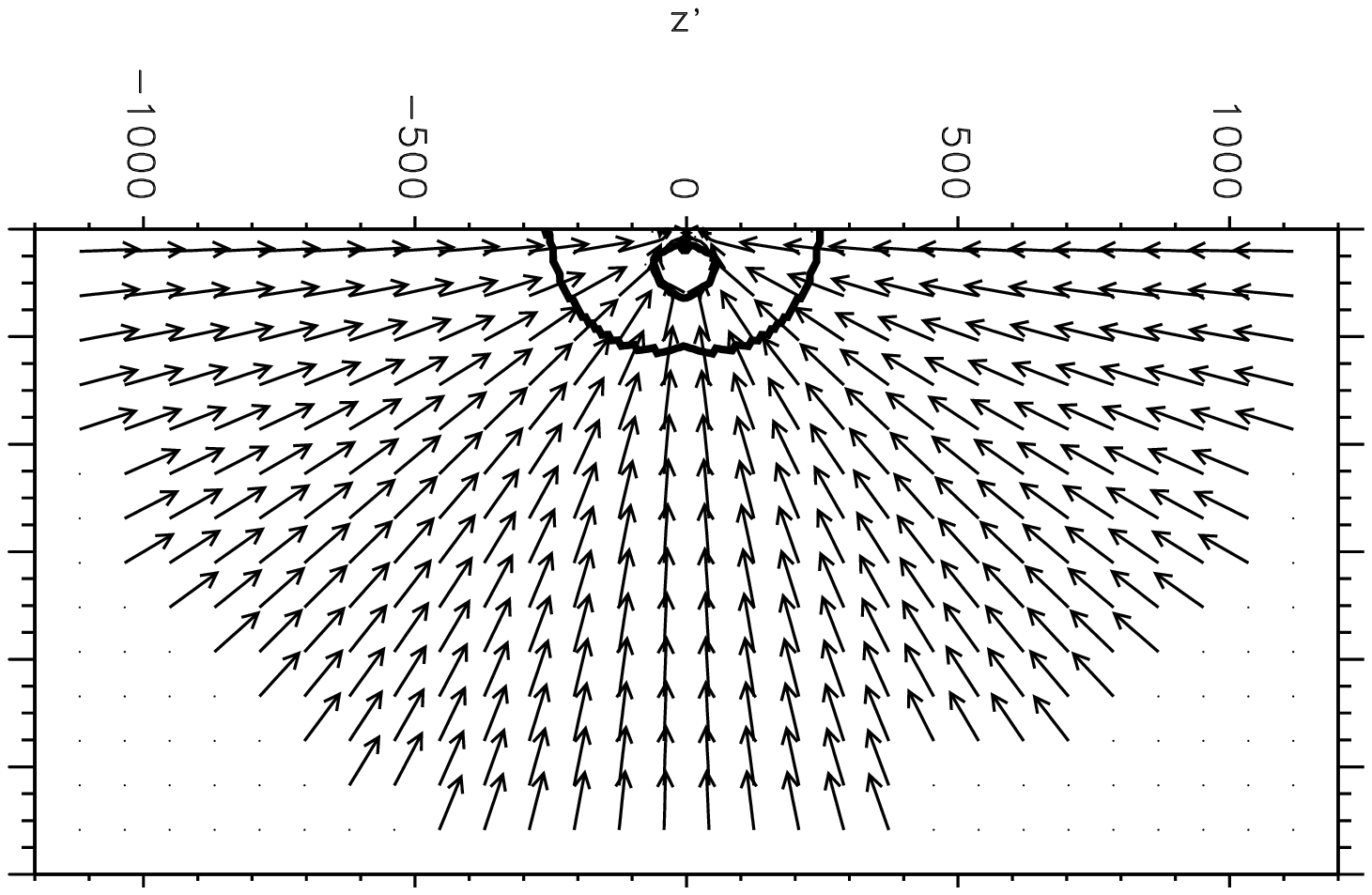}}

\put(180,300){\includegraphics{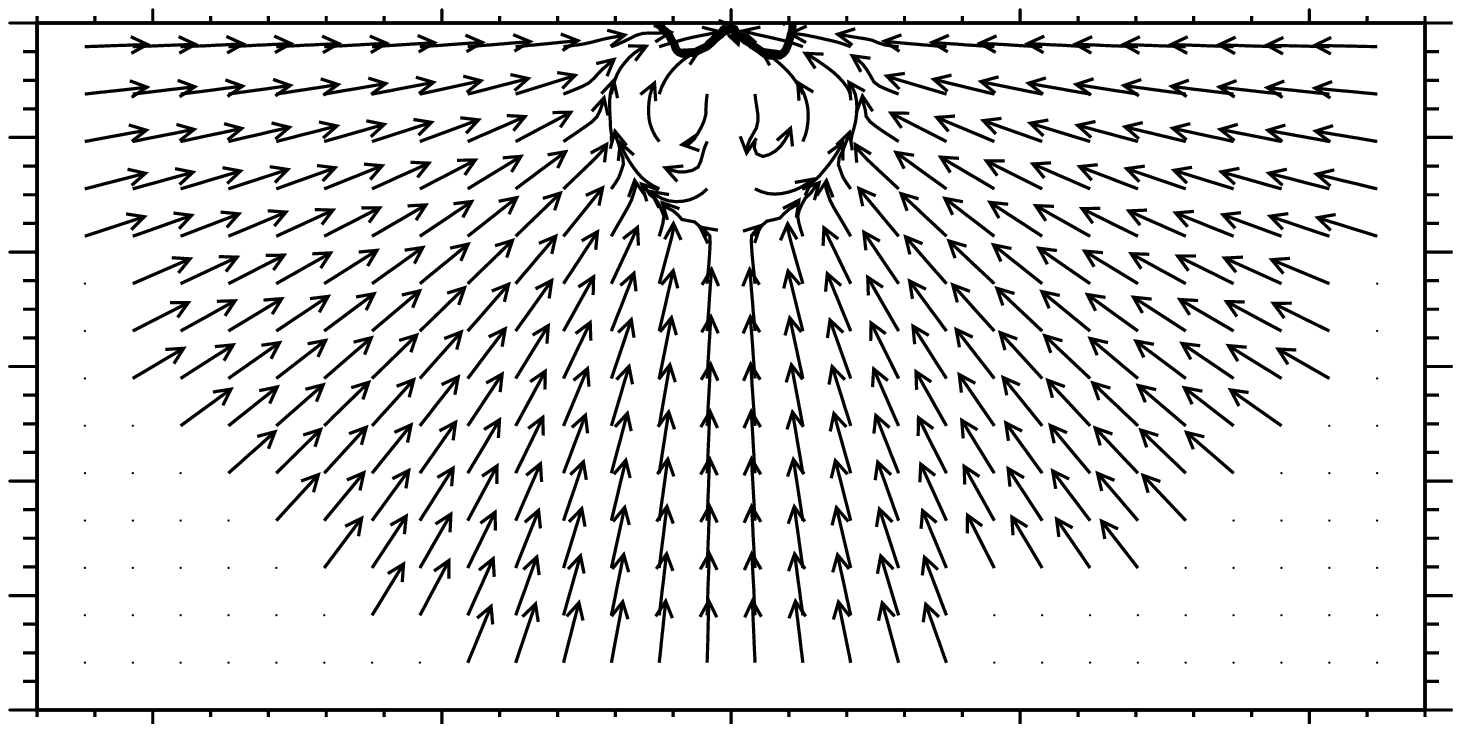}}

\put(280,300){\includegraphics{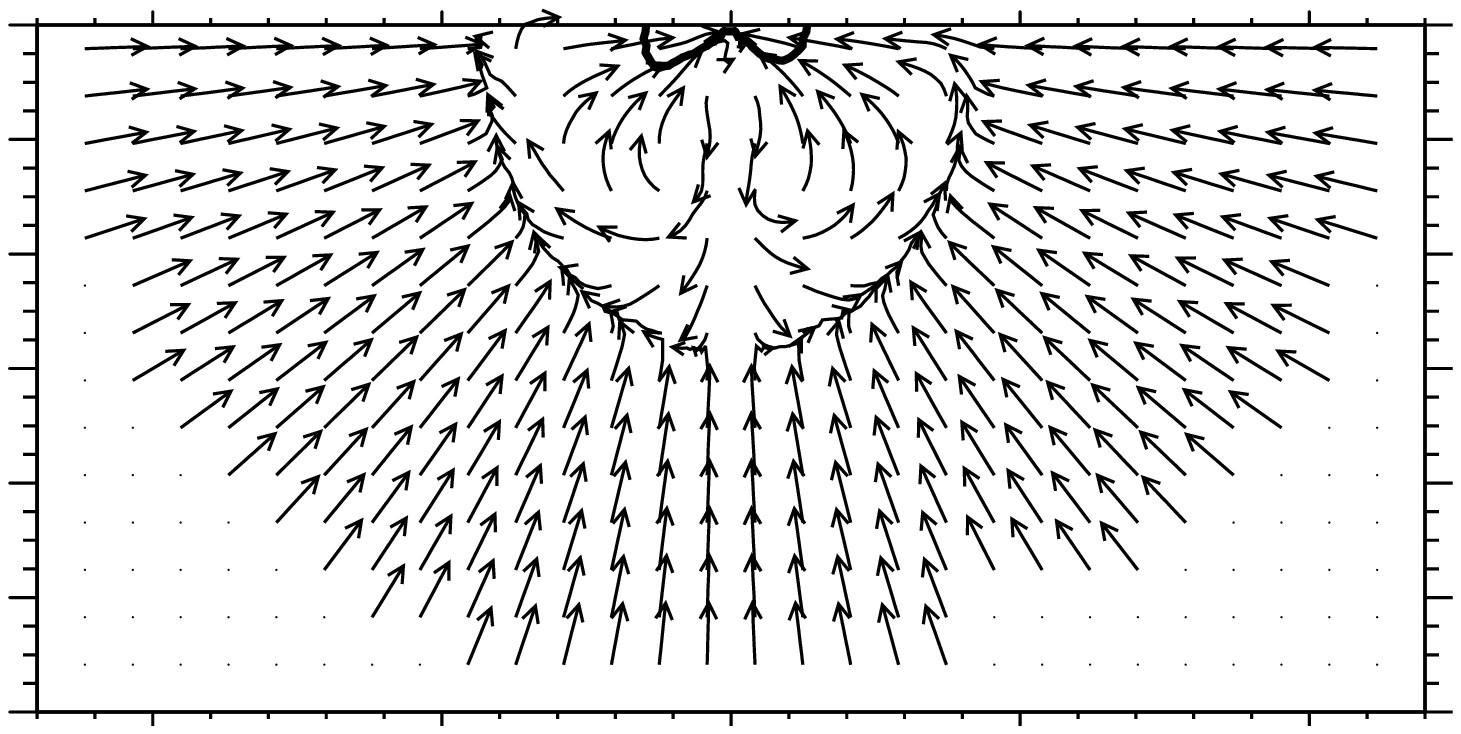}}

\put(380,300){\includegraphics{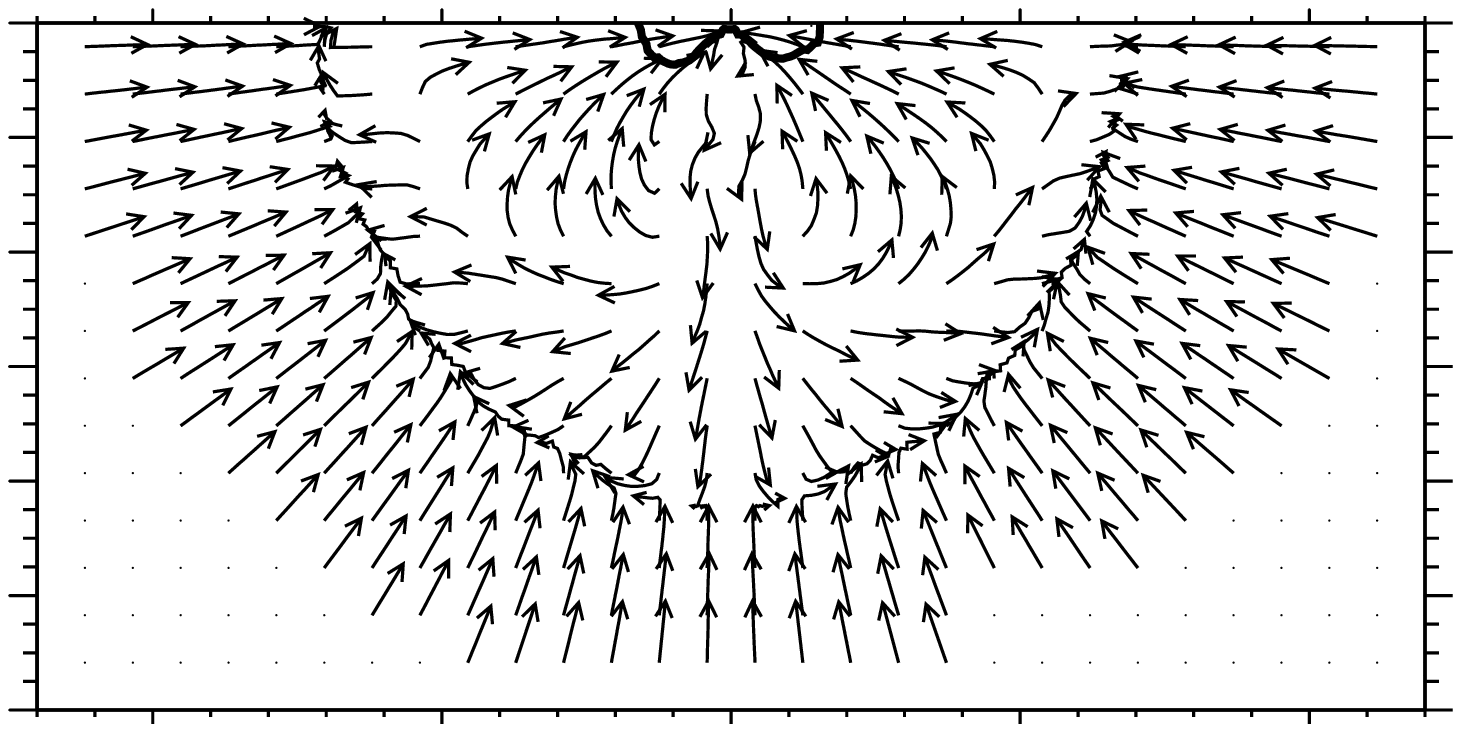}}

\put(480,300){\includegraphics{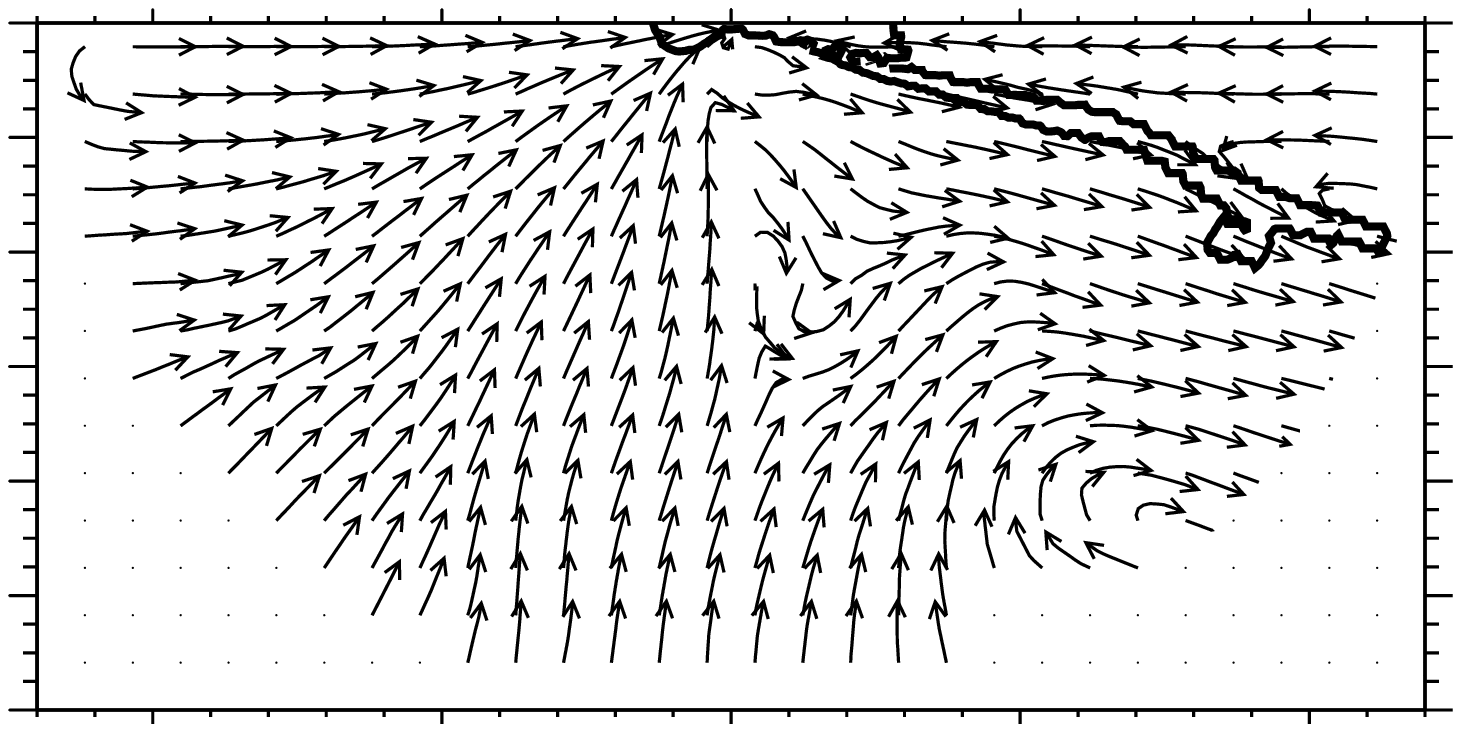}}

\put(80,150){\includegraphics{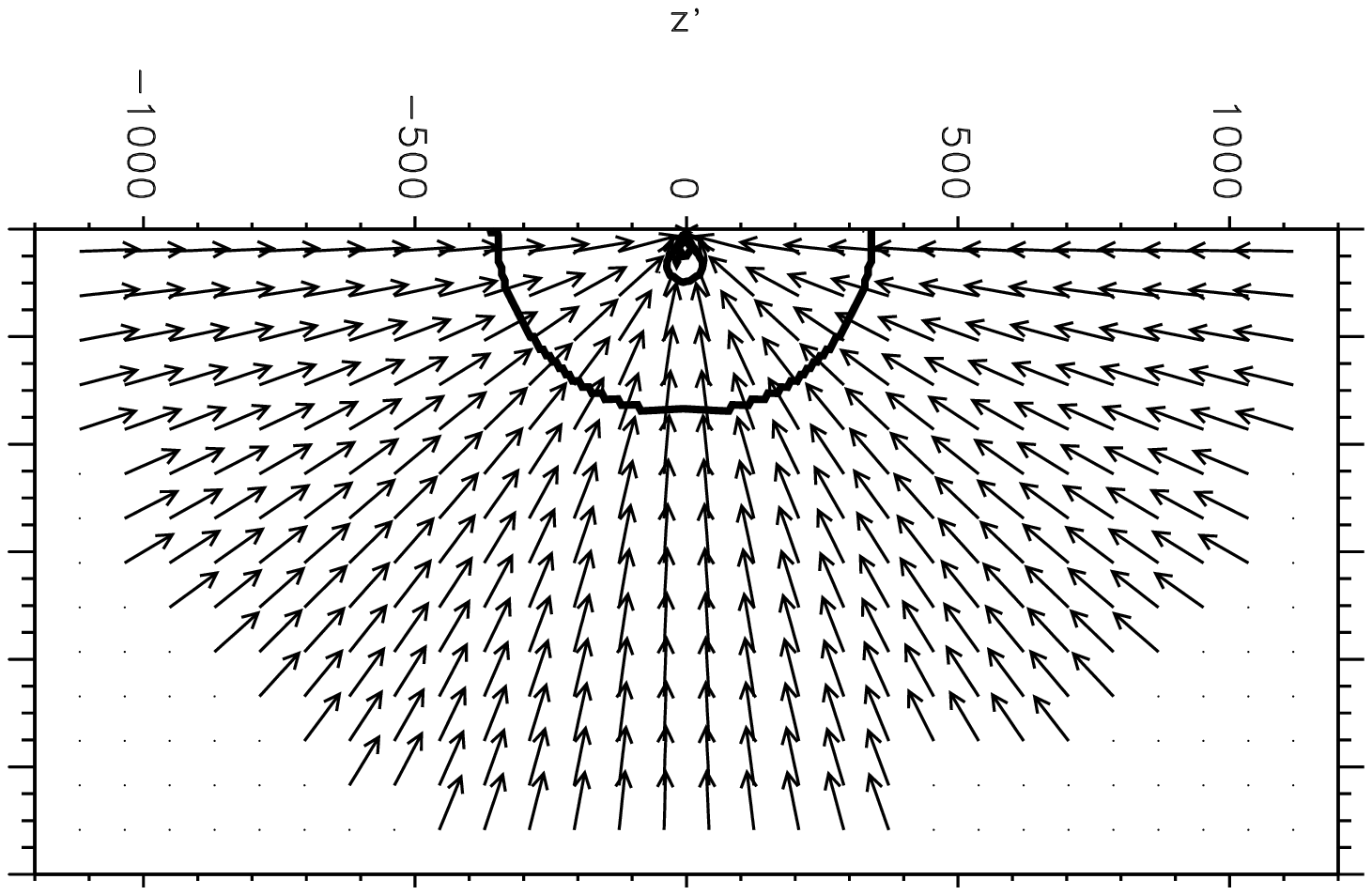}}

\put(180,150){\includegraphics{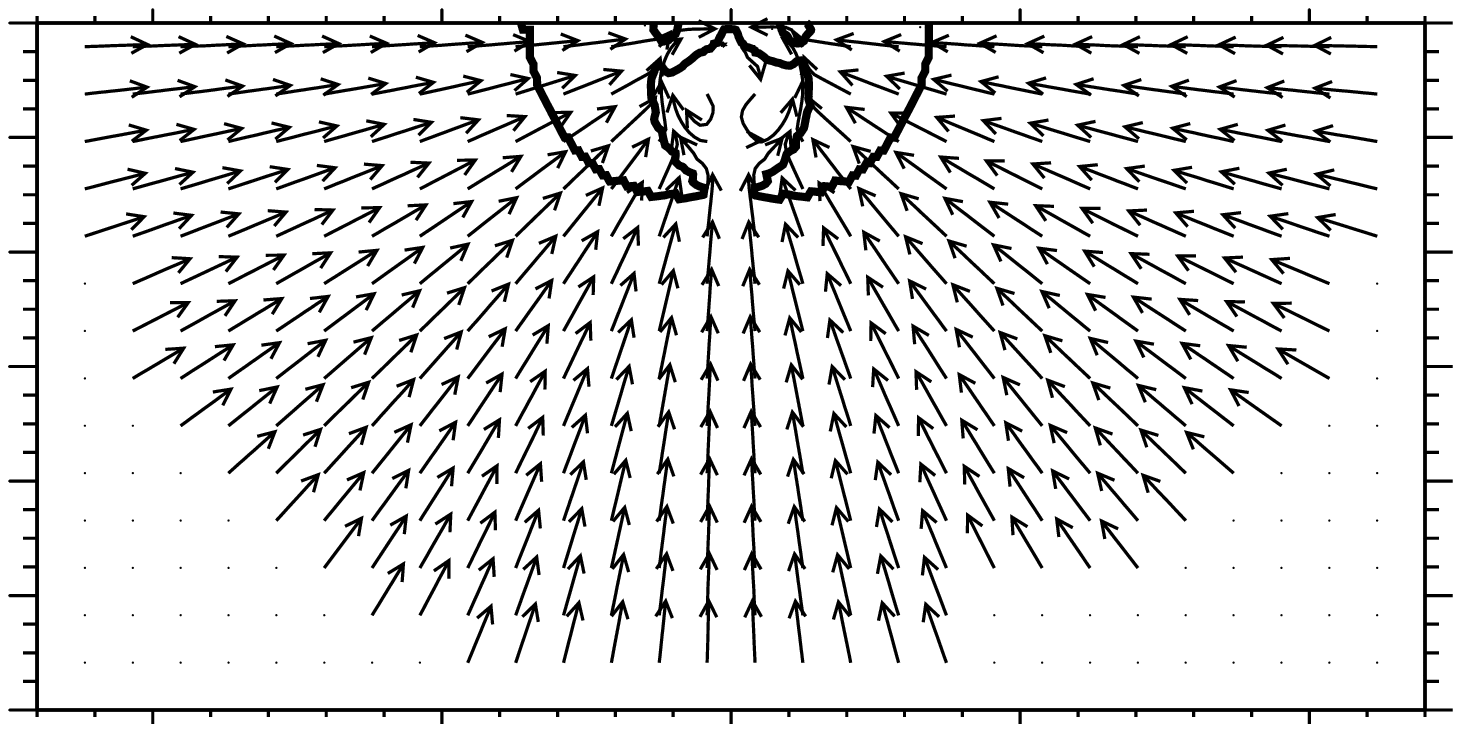}}

\put(280,150){\includegraphics{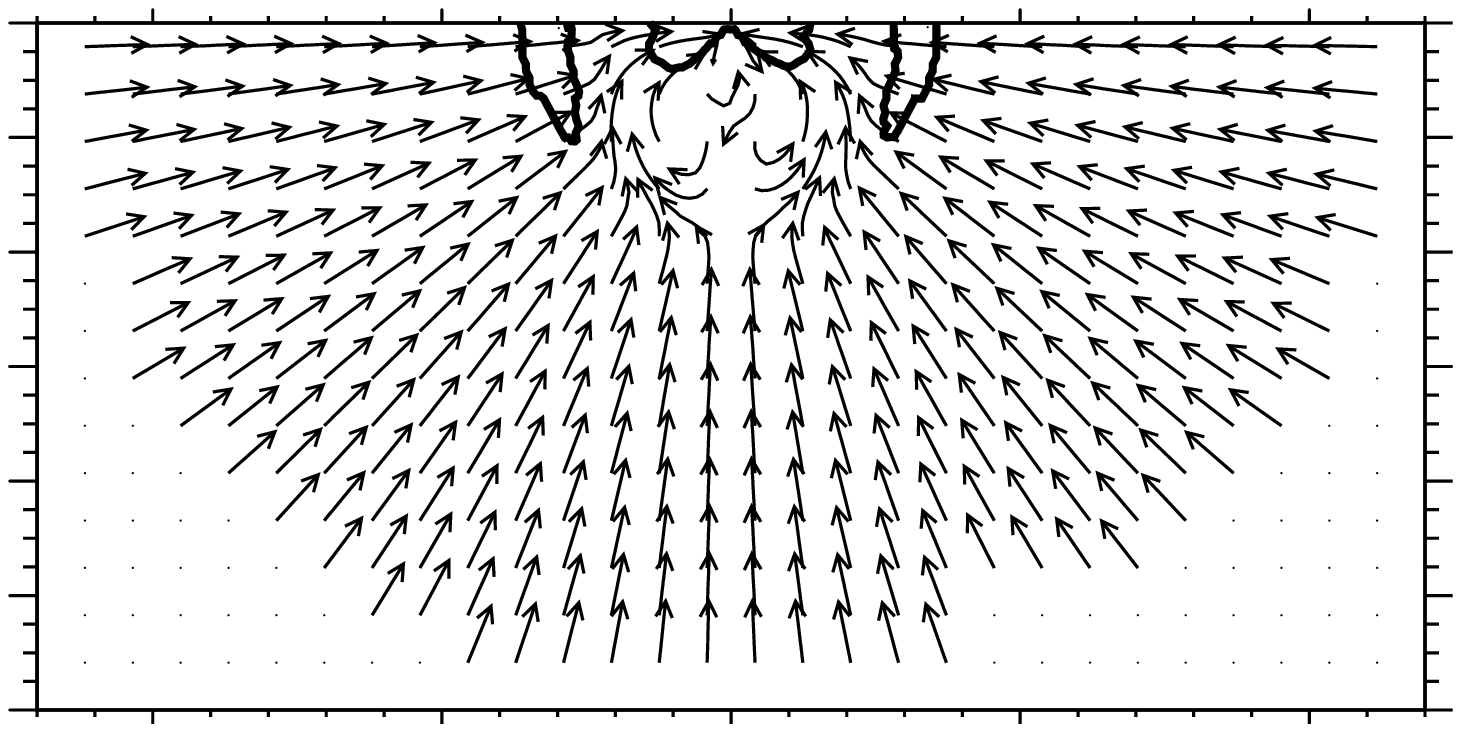}}

\put(380,150){\includegraphics{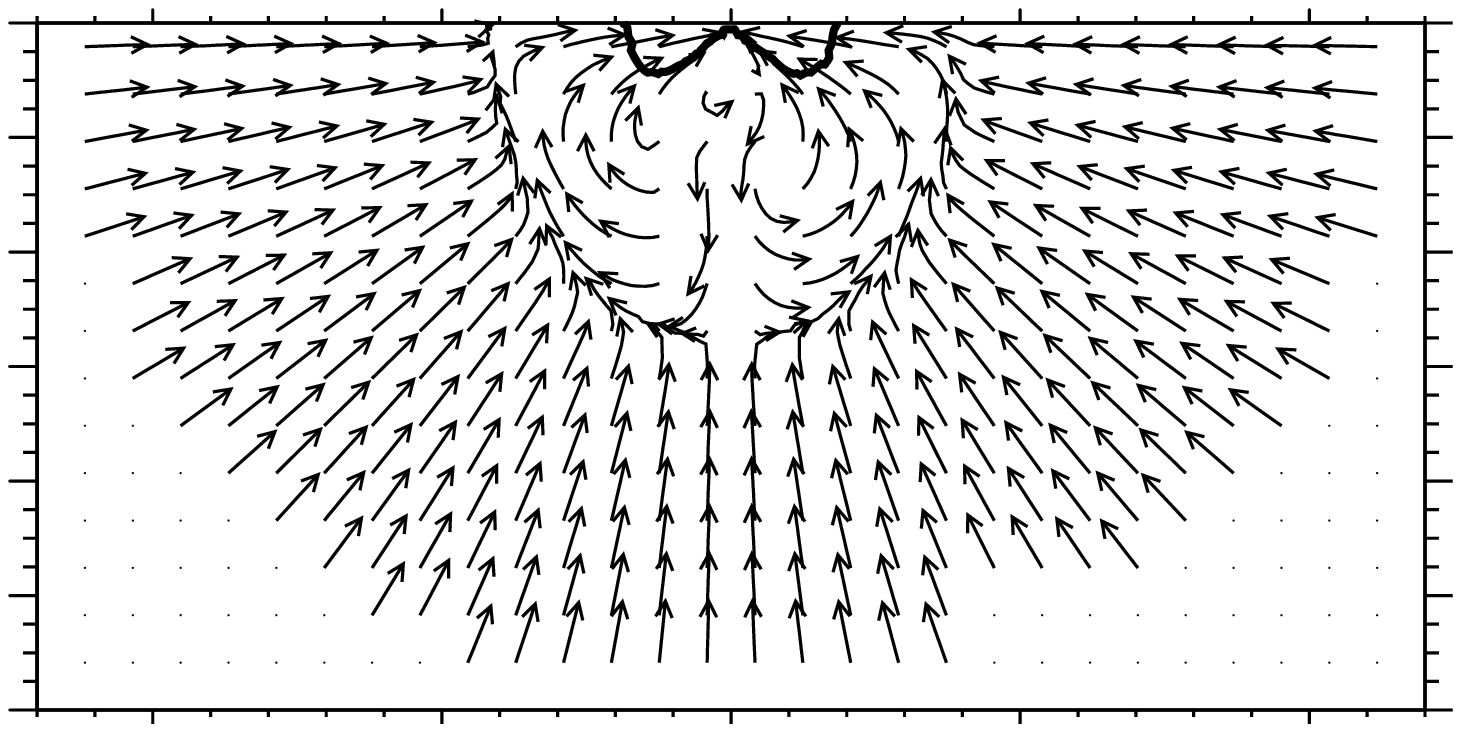}}

\put(480,150){\includegraphics{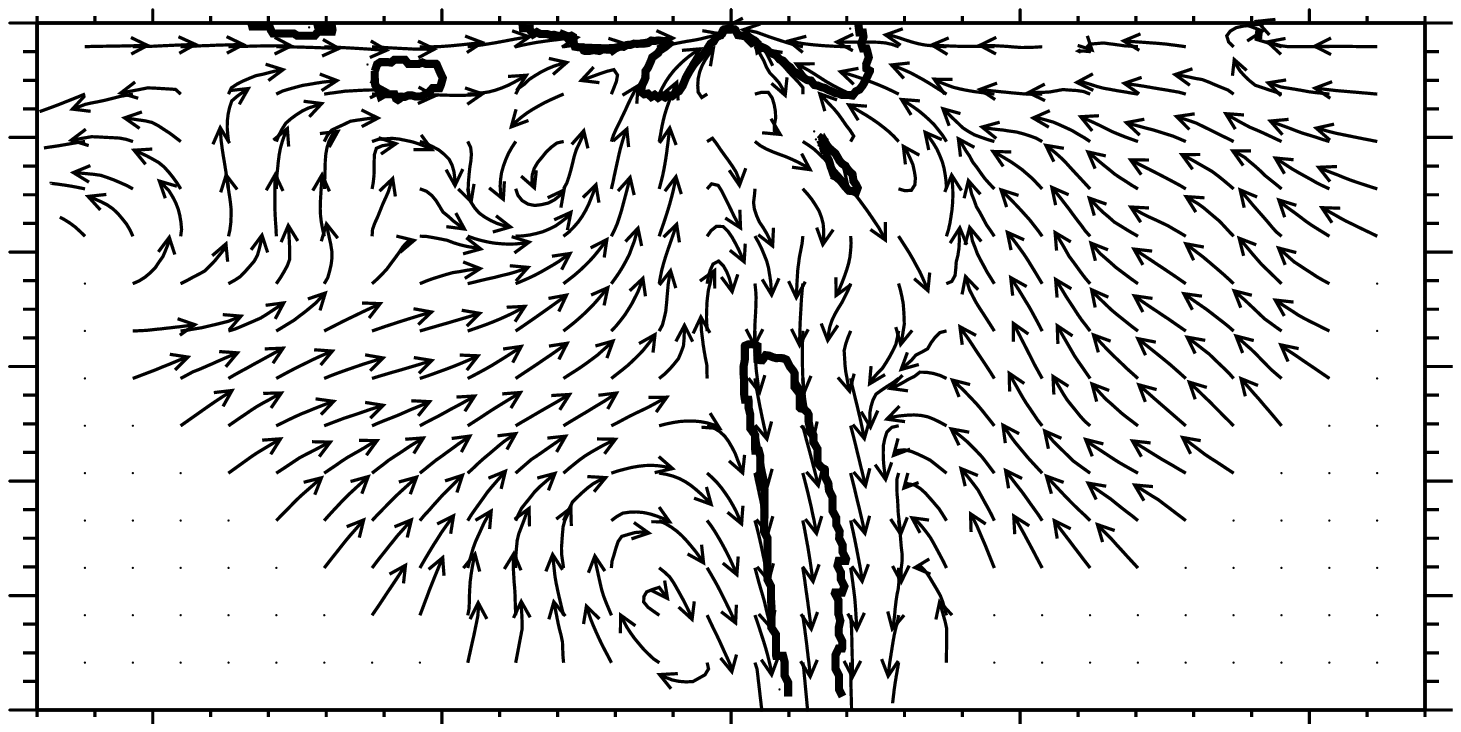}}

\put(80,0){\includegraphics{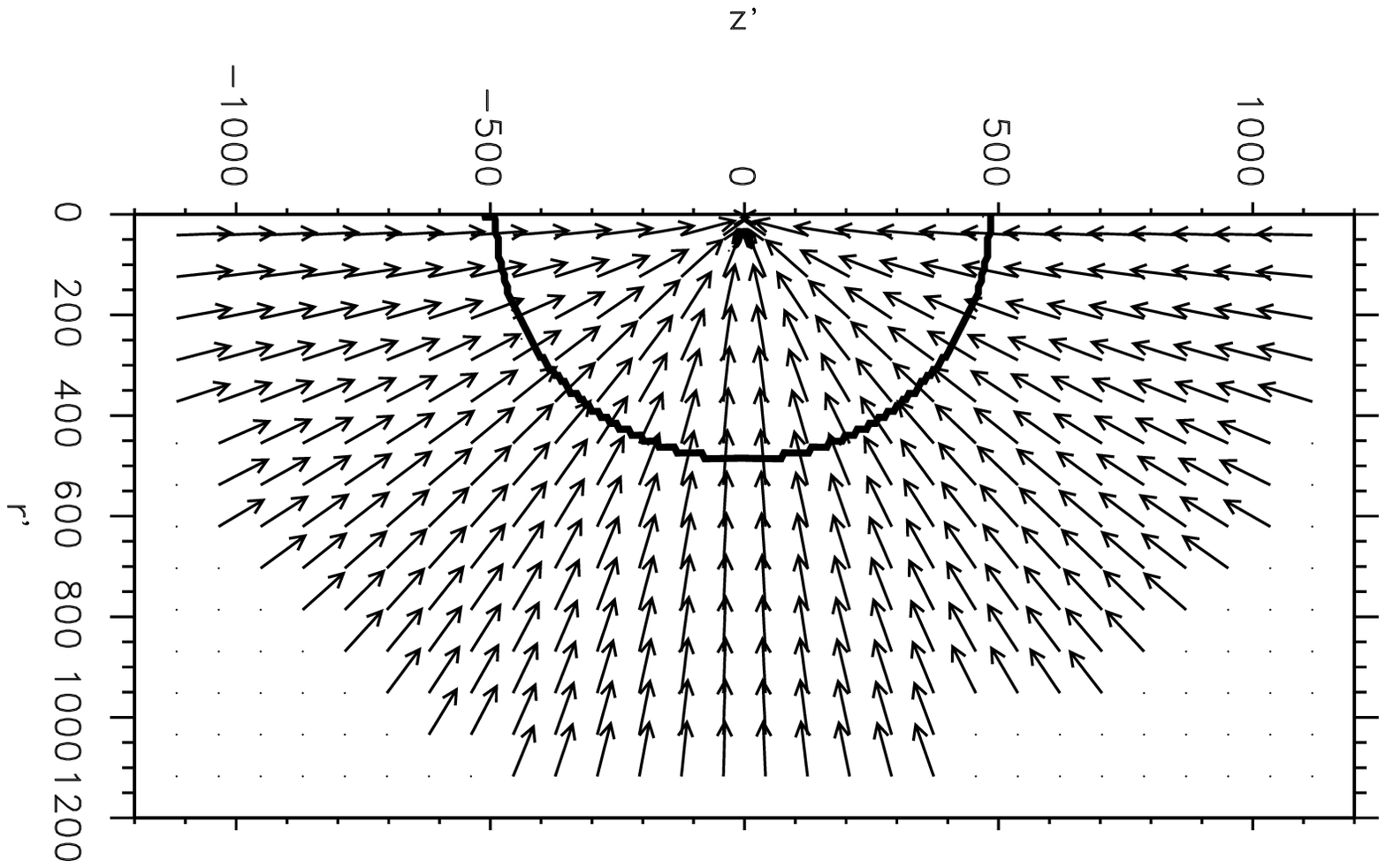}}

\put(180,0){\includegraphics{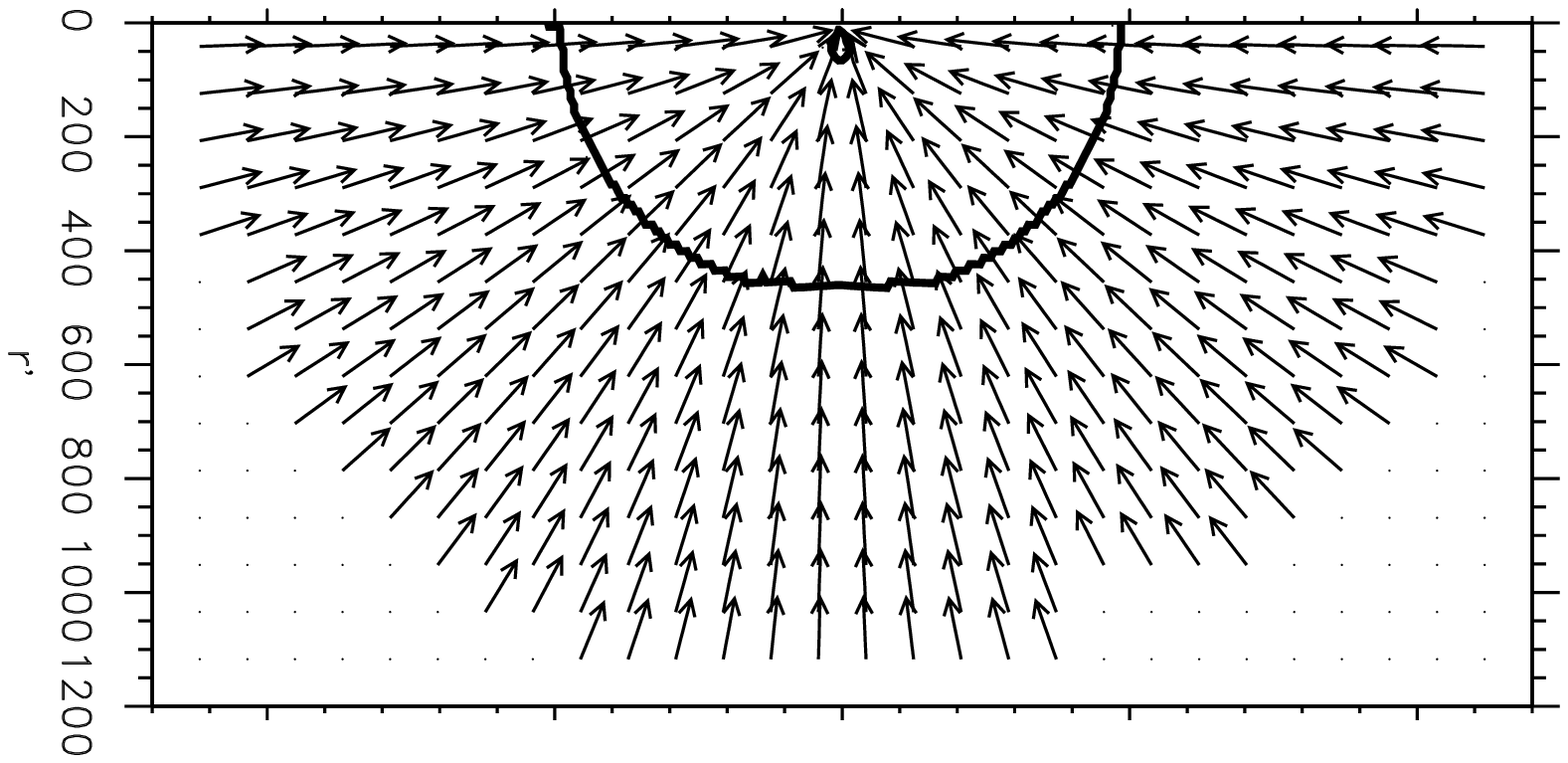}}

\put(280,0){\includegraphics{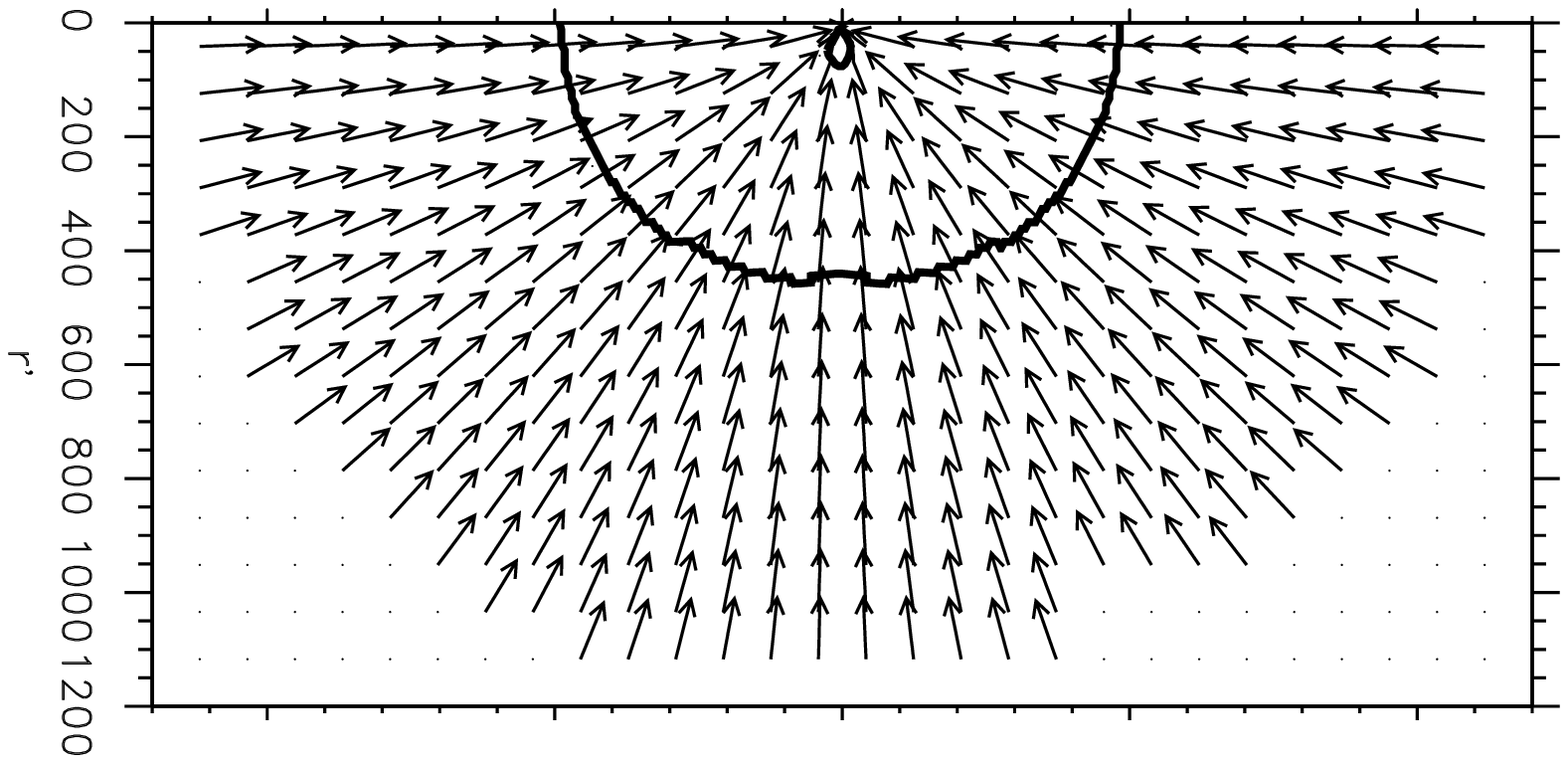}}

\put(380,0){\includegraphics{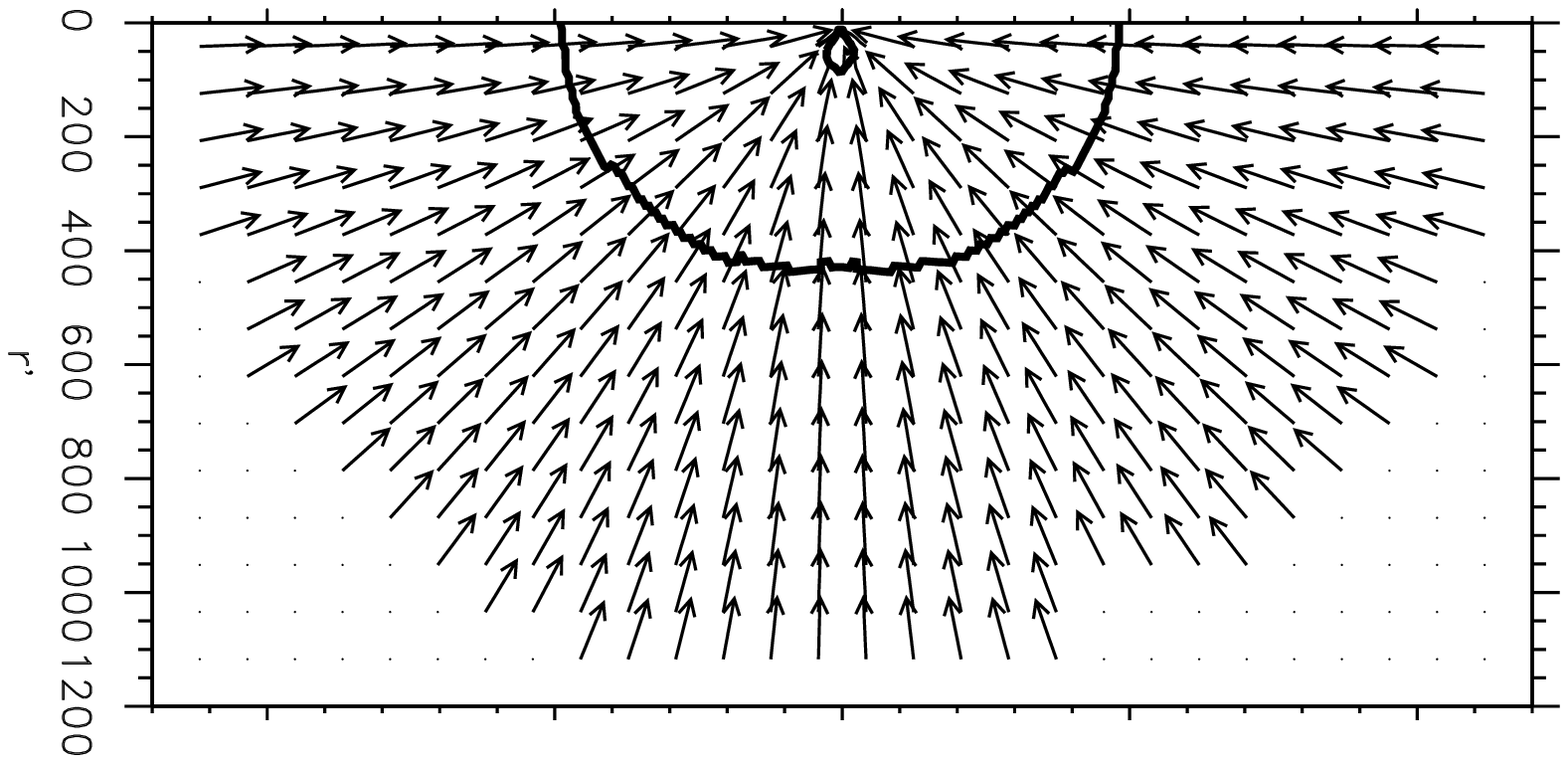}}

\put(480,0){\includegraphics{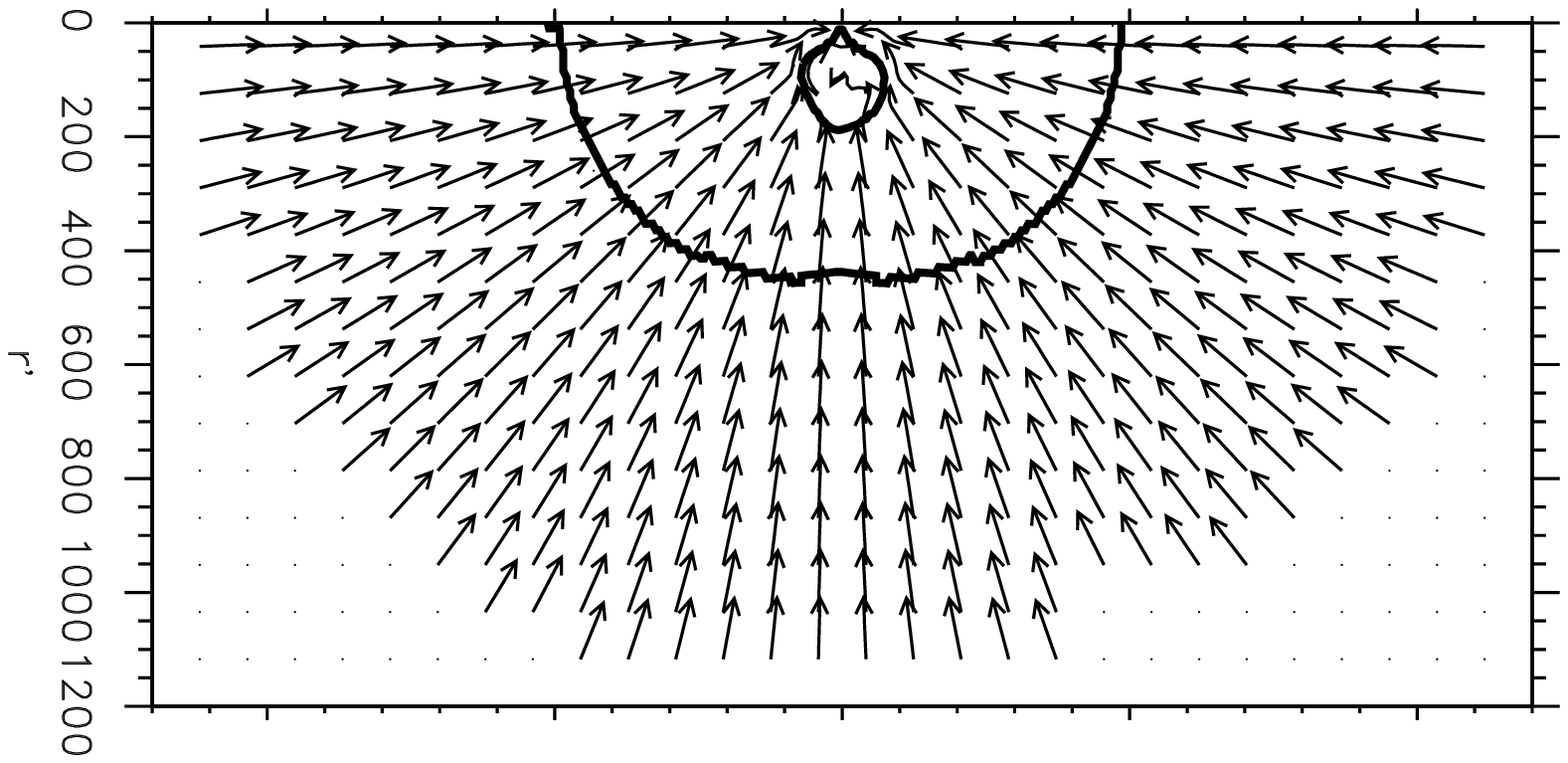}}

\end{picture}

\caption{Sequences of poloidal velocity fields, and sonic surface at
five different times (from left to right t = 177.3, 701.0, 1169.1,
1754.3, and $t_f$ respectively, in units of dynamical time at inner
radius of computational domain) for $\gamma$=5/3, 4/3, 1.2, and 1.01
(from top to bottom).
}\label{fig:2}
\end{figure*}

\newpage

\begin{figure*}
\begin{picture}(0,600)

\put(100,300){\includegraphics{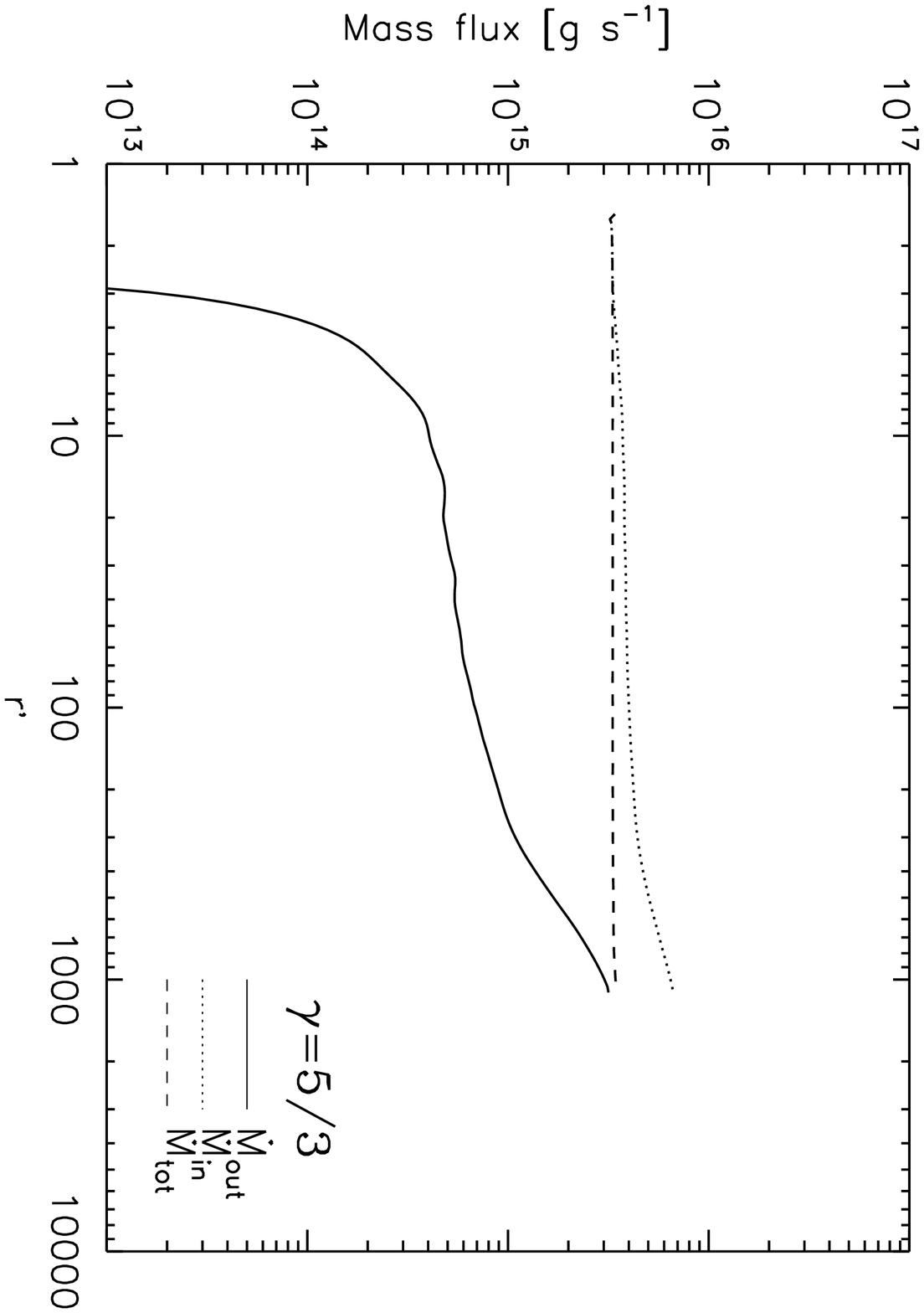}}

\put(400,300){\includegraphics{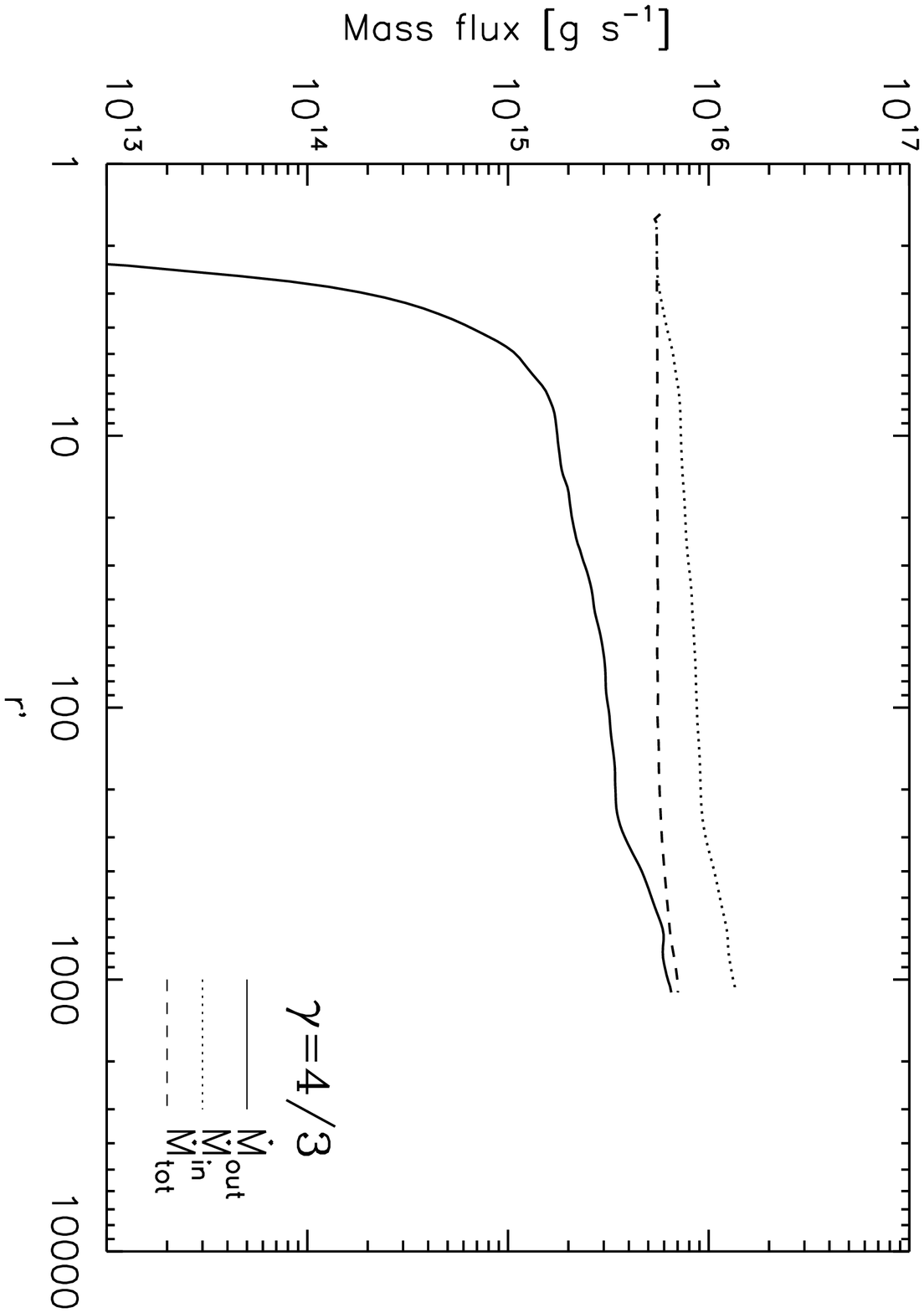}}

\put(100,100){\includegraphics{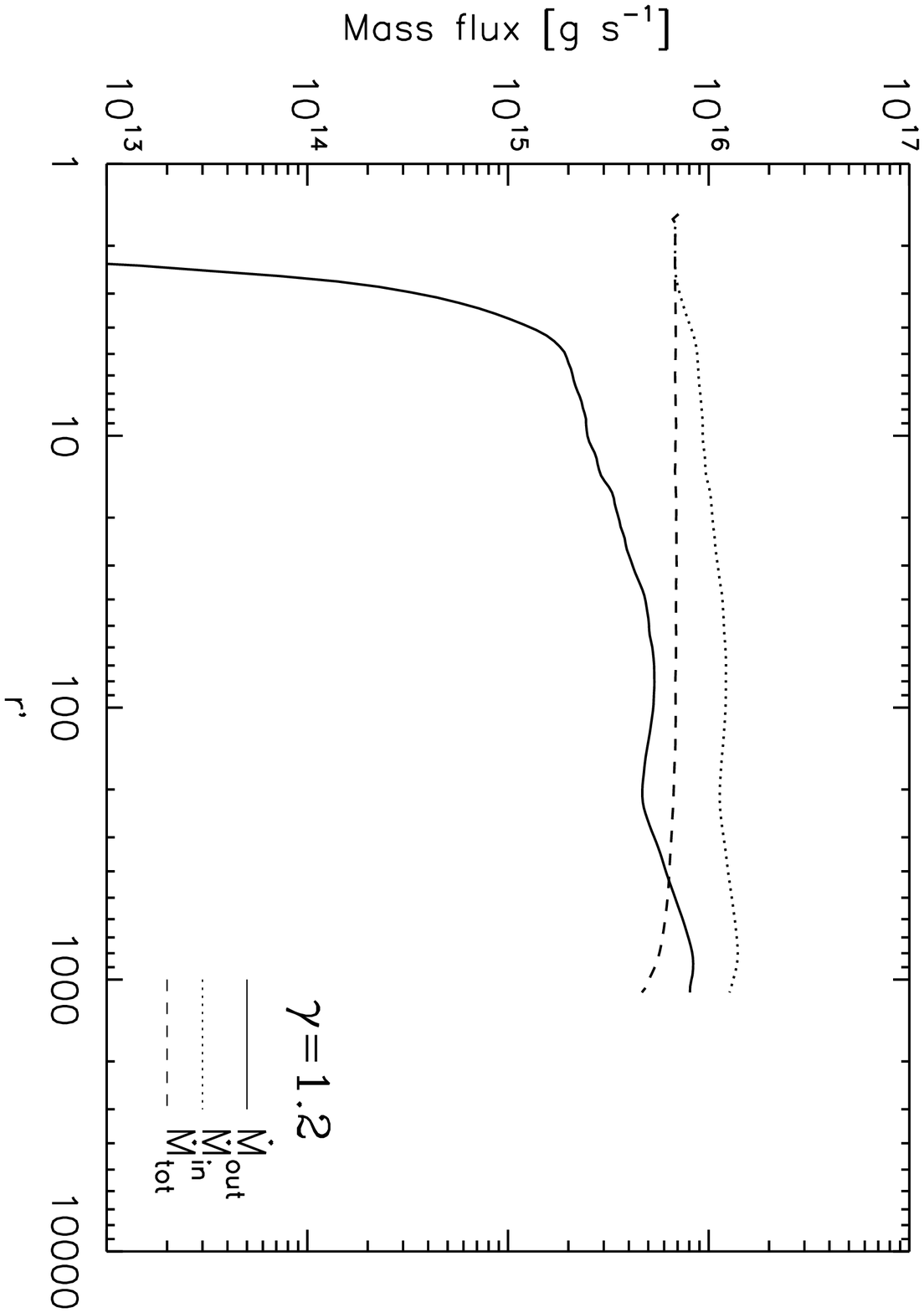}}

\put(400,100){\includegraphics{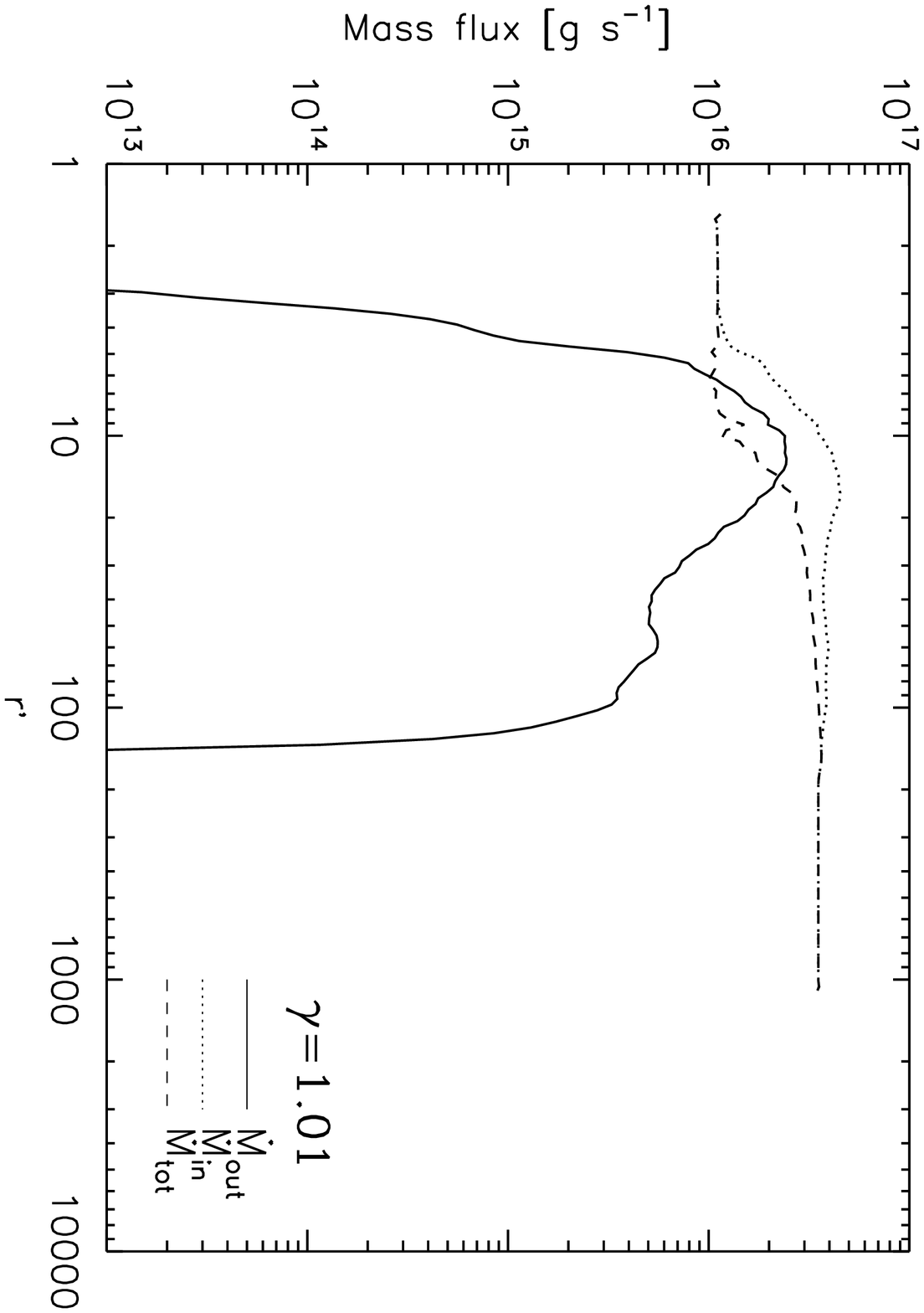}}

\end{picture}
\caption{Mass flux rates as functions of radius
for models with $R'_S=10^{-3}$ and different $\gamma$
time-averaged at the end the simulations. For
$\gamma$=5/3, 4/3, and 1.2, the average
is taken over $\Delta~t=1.75 \times 10^4$ - $t_f$ time period.
For $\gamma$=1.01, we average over $\Delta~t=2.5 \times 10^4$ - $t_f$.}
\label{fig:bilans}
\end{figure*}
\eject
\newpage

\begin{figure*}
\begin{picture}(0,600)

\put(100,300){\includegraphics{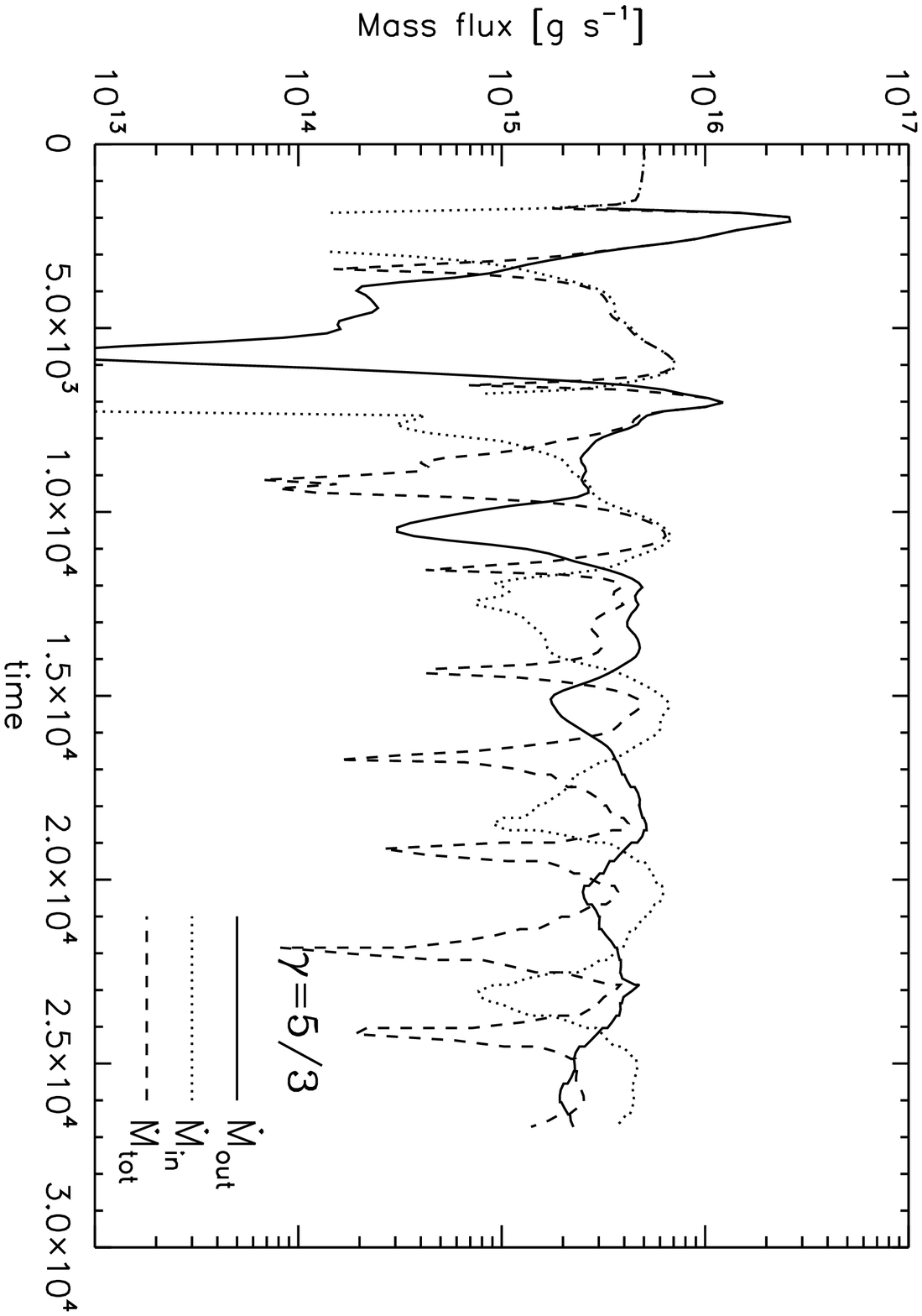}}

\put(400,300){\includegraphics{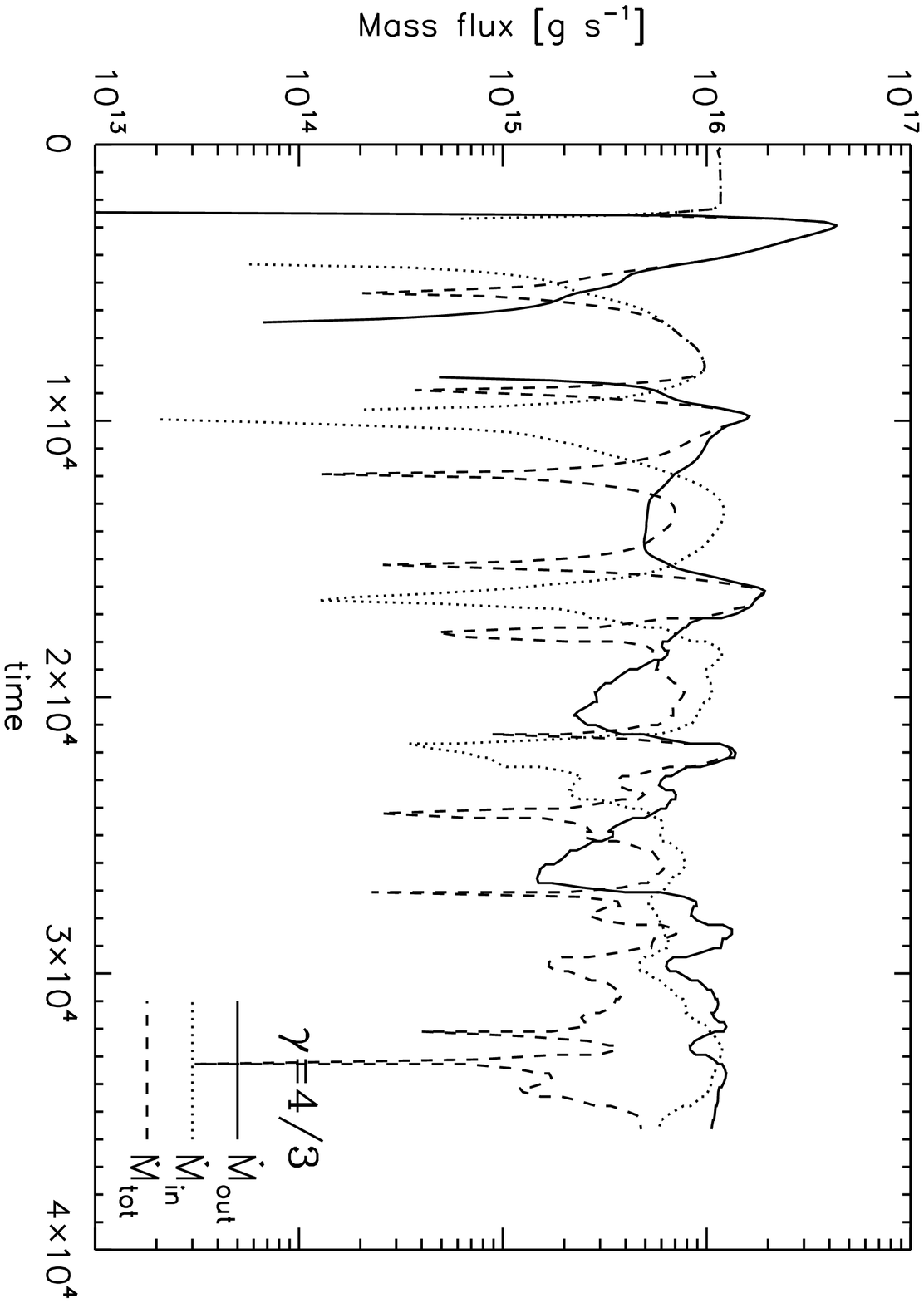}}

\put(100,100){\includegraphics{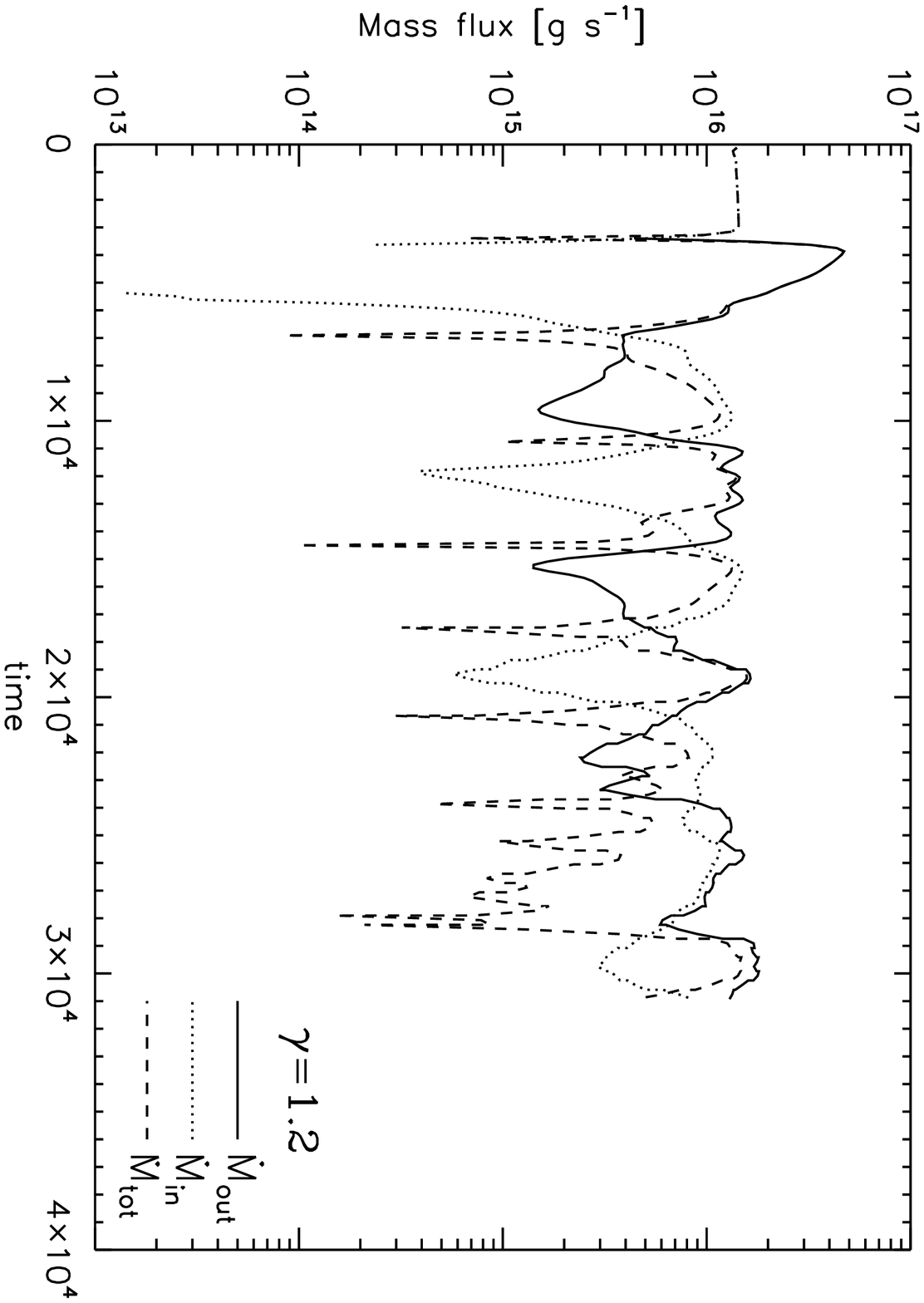}}

\put(400,100){\includegraphics{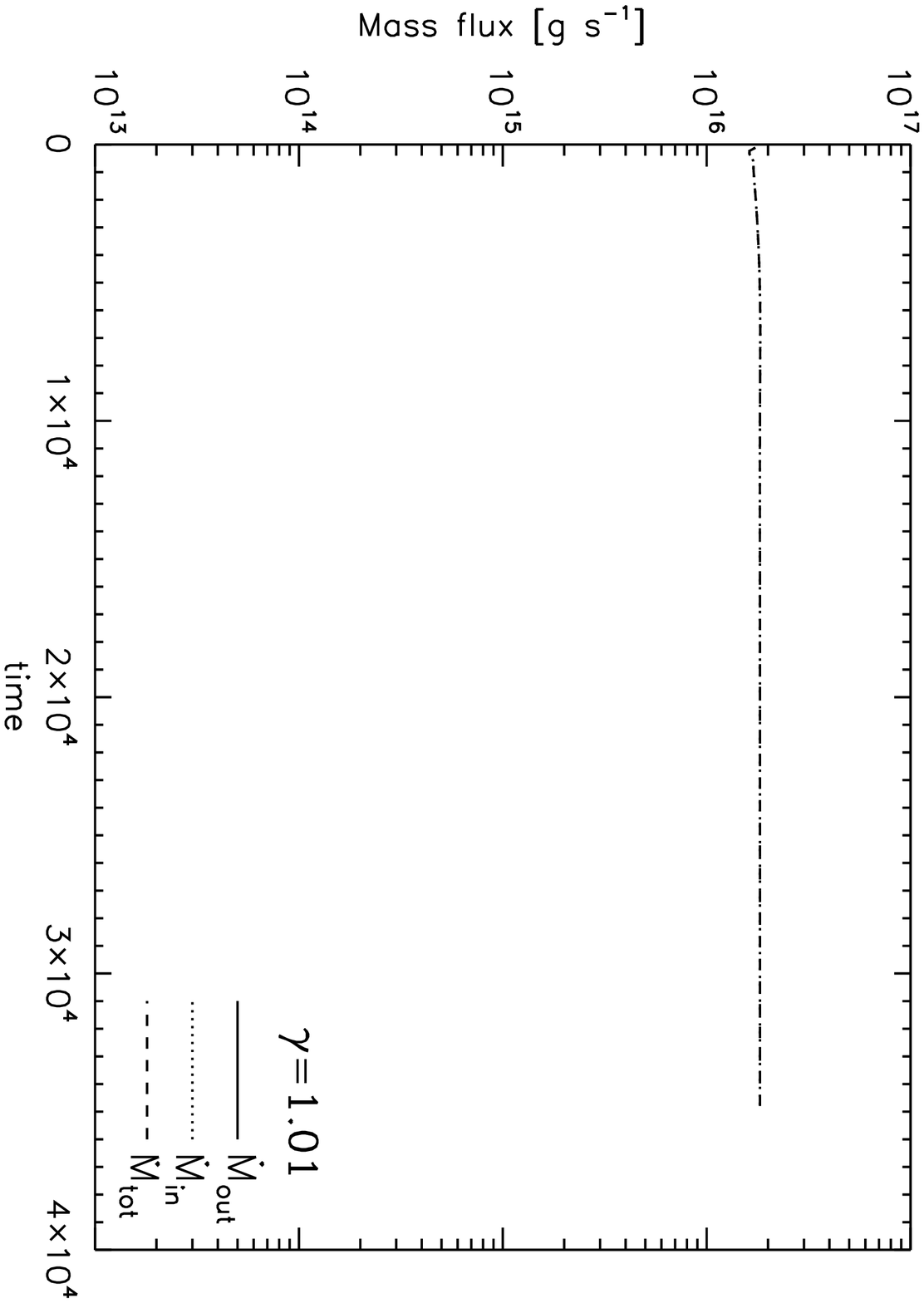}}

\end{picture}
\caption{Mass flow rates at the outer boundary as functions of time 
for models with $R'_S=10^{-3}$ and different $\gamma$.}
\label{fig:bilans_time}
\end{figure*}
\eject

\eject
\newpage

\begin{figure*}
\begin{picture}(0,550)

\put(320,390){\includegraphics{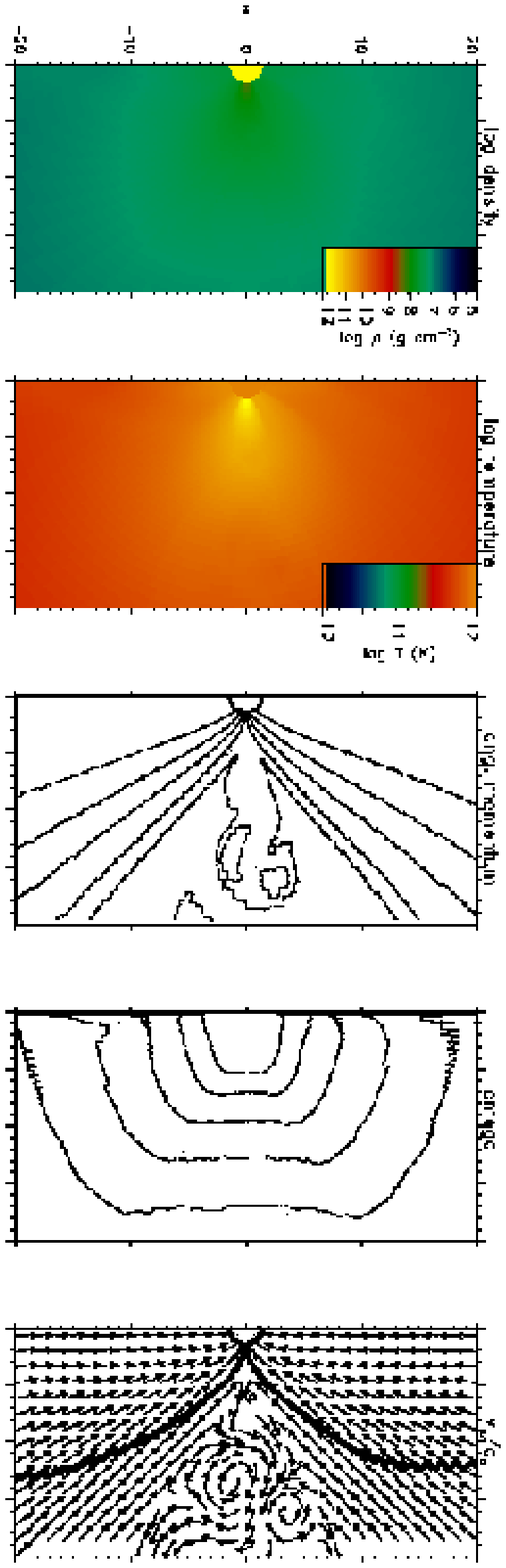}}

\put(320,260){\includegraphics{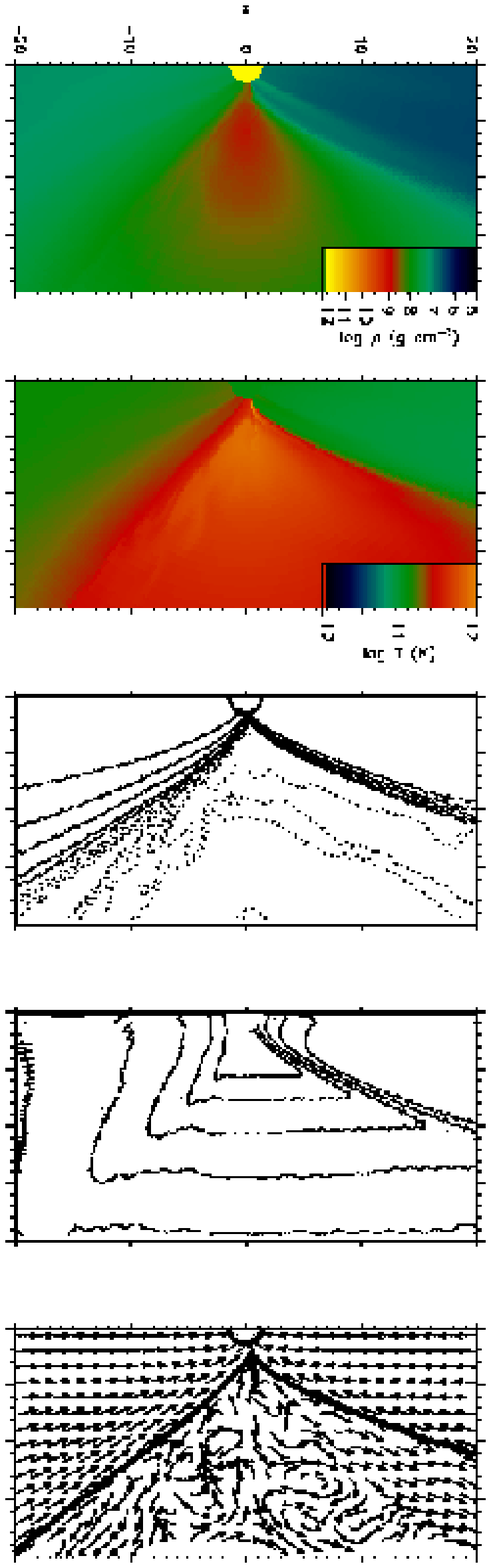}}

\put(320,130){\includegraphics{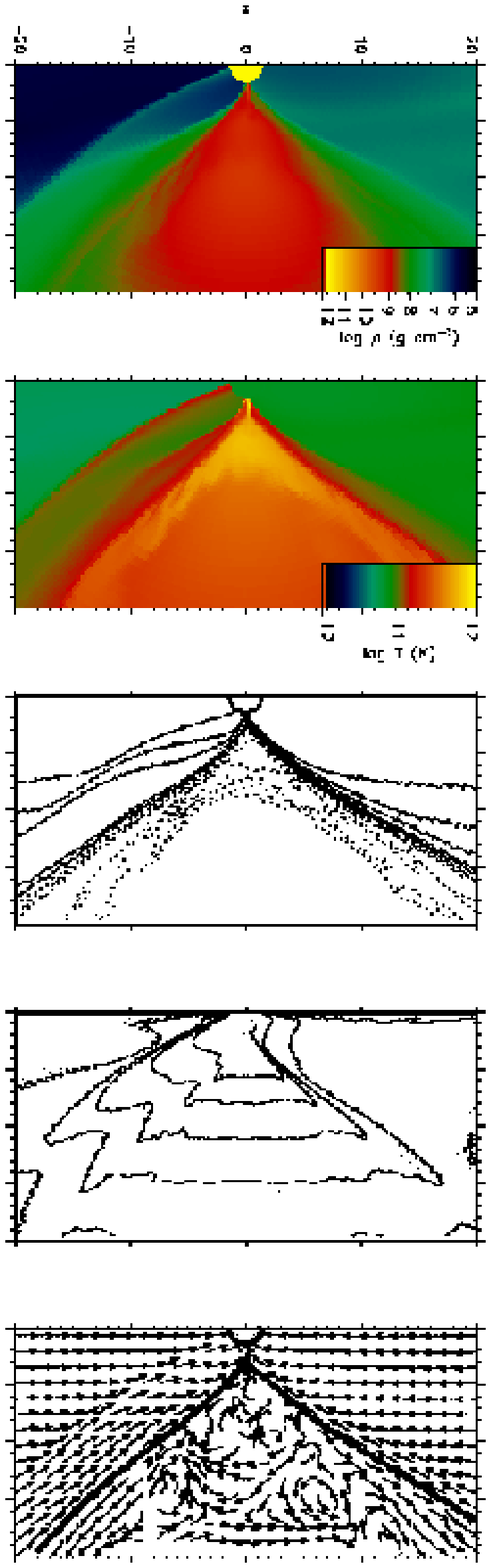}}

\put(320,0){\includegraphics{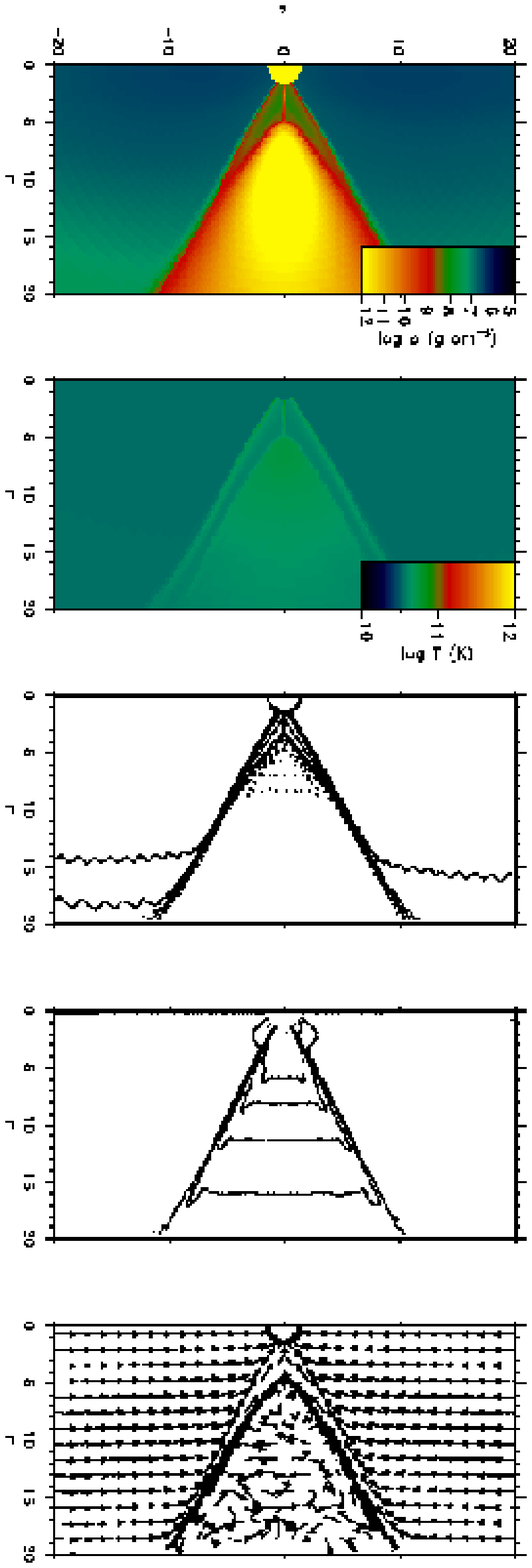}}

\end{picture}
\caption{ 
Two-dimensional structure of various quantities 
at the last time step for $\gamma$ (
$\gamma$ index 5/3, 4/3, 1.2, and 1.01 from top to bottom). 
The panels from left to right are snapshots of $\log~\rho$,
$\log~T$, $l$, $\log~\Omega$, and of the fluid poloidal velocity direction
overplot with sonic surface contour.
We do not draw labels for angular momentum and angular velocity because of strong compression
of contours.
The angular momentum contours are between 0.2 and 1.1 with the step of 0.1.
The contorus for angular velocity are between -2.5 and -1.5  with the step 
of 0.25. The scale of the figure is 20 $R_S$.
}
\label{fig:3a}
\end{figure*}

\eject
\newpage

\begin{figure*}
\begin{picture}(0,550)

\put(320,390){\includegraphics{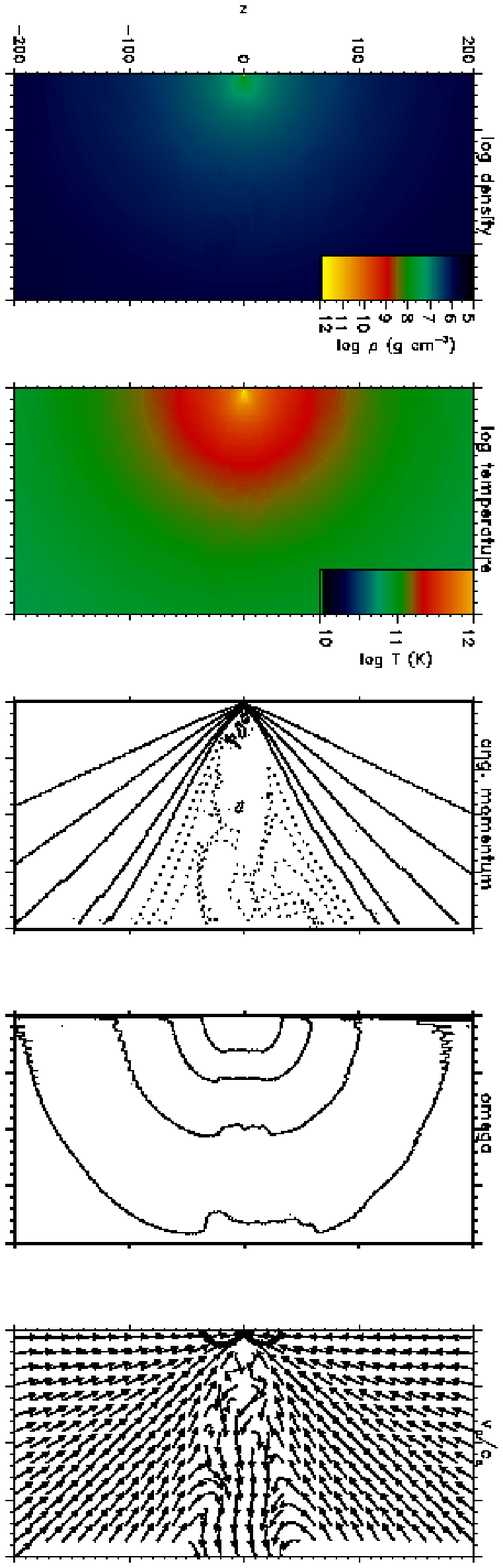}}

\put(320,260){\includegraphics{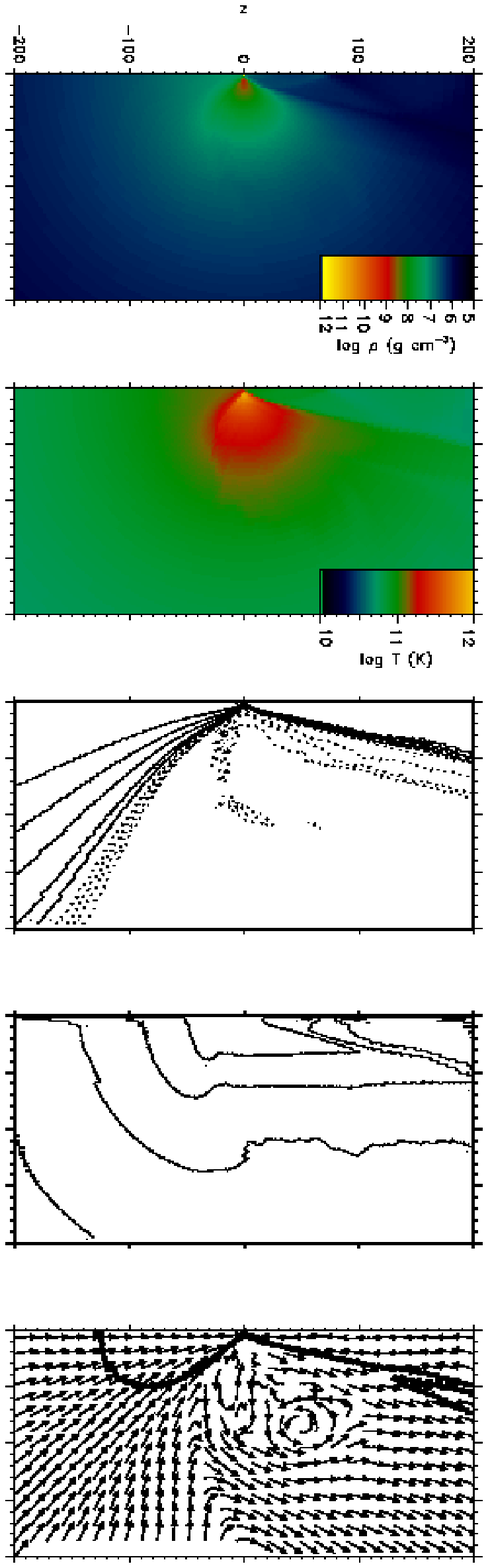}}

\put(320,130){\includegraphics{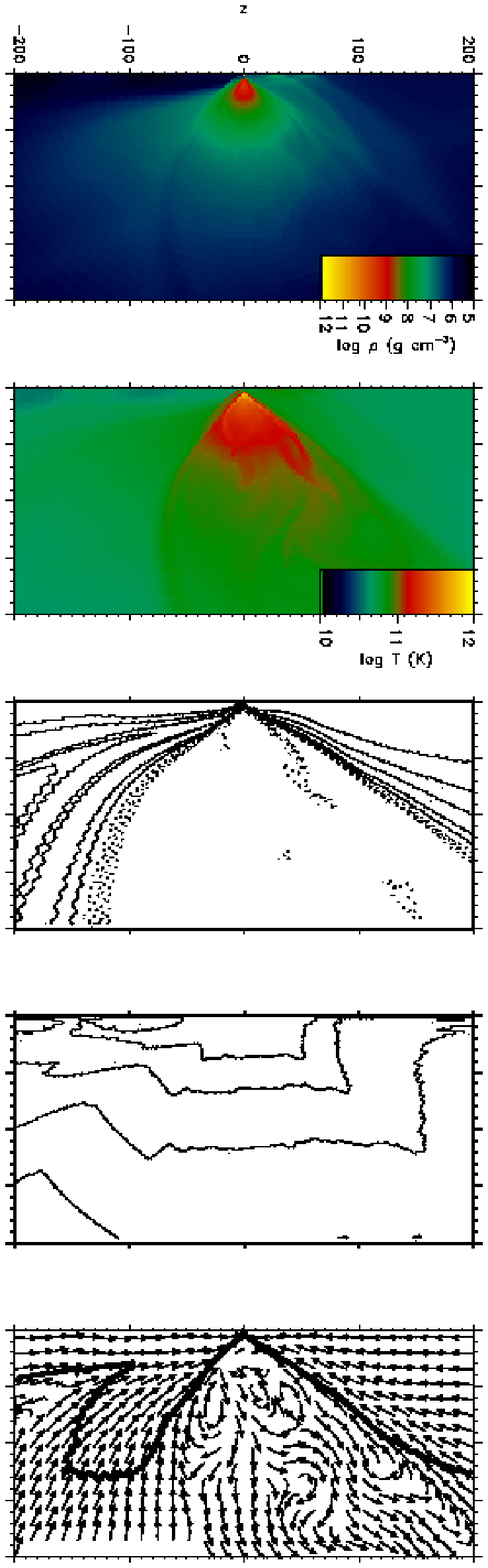}}

\put(320,0){\includegraphics{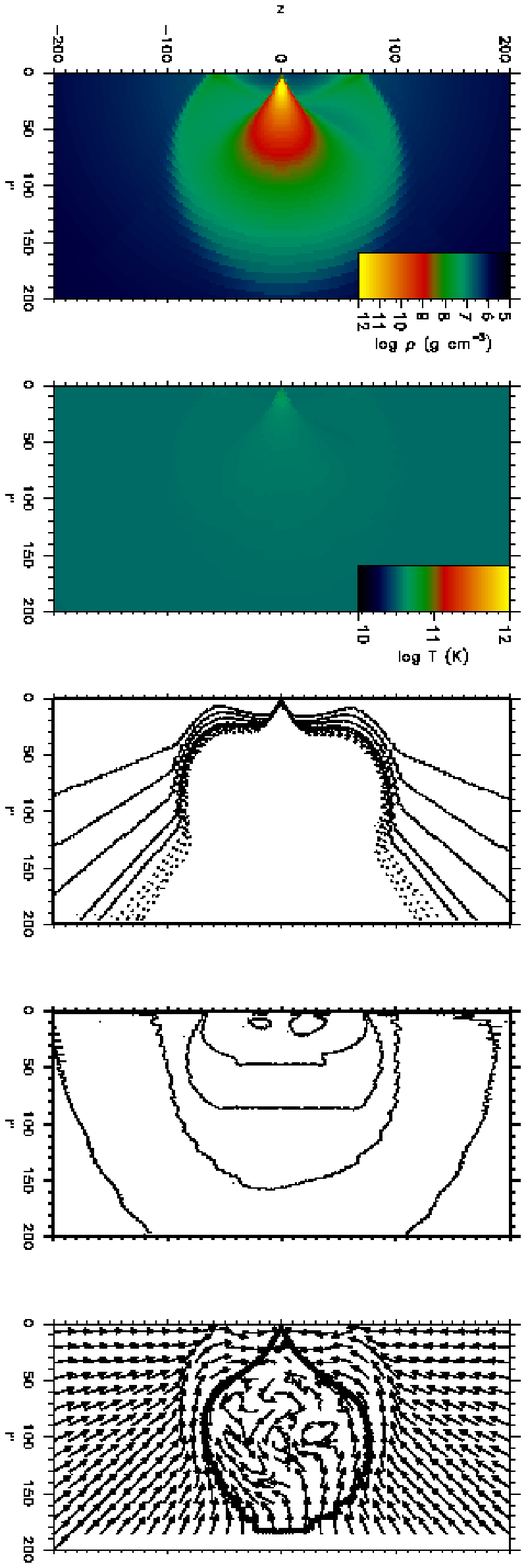}}

\end{picture}
\caption{ As in Fig. 8 but on the scale of 200 $R_S$.
The $l$ contours are between 0.2 and 1.1 with step of 0.1 while
the $\log~\Omega$ contours are between -5.5 and -3.0  with step of 0.5.
}\label{fig:3b}
\end{figure*}

\begin{figure*}
\begin{picture}(0,550)

\put(320,390){\includegraphics{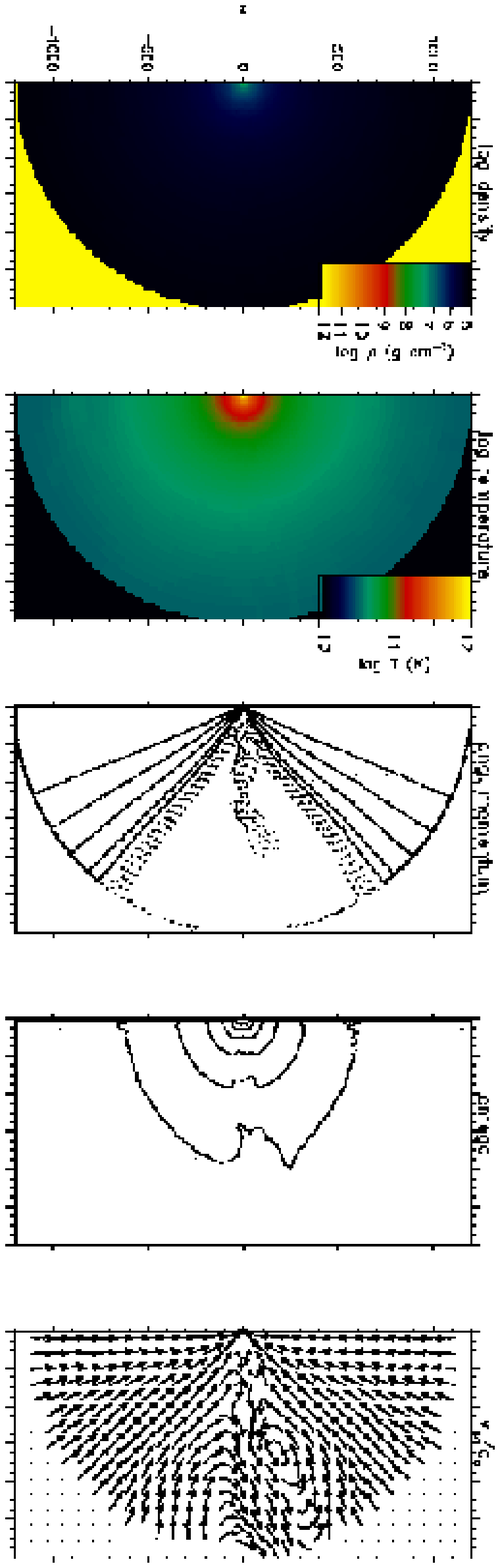}}

\put(320,260){\includegraphics{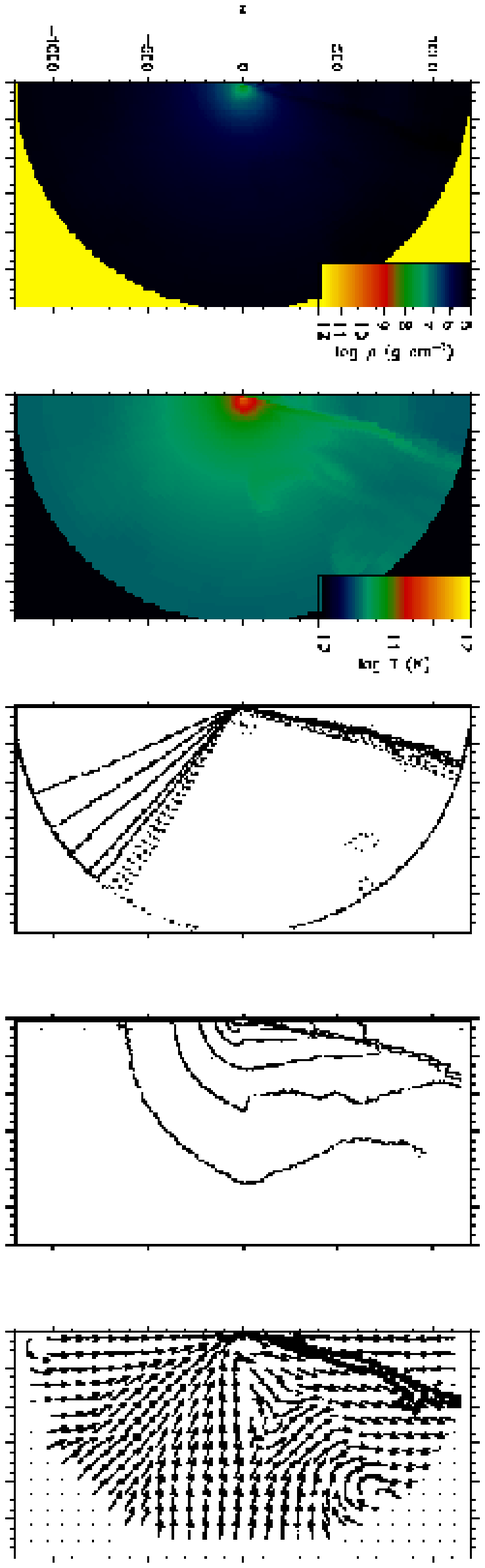}}

\put(320,130){\includegraphics{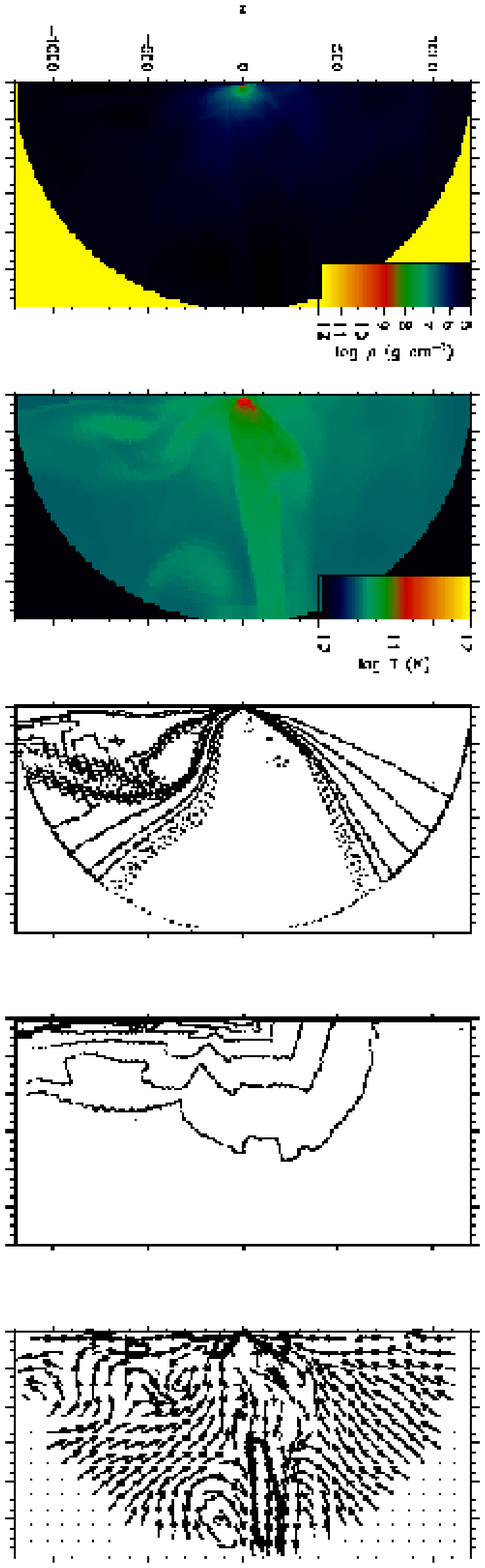}}

\put(320,0){\includegraphics{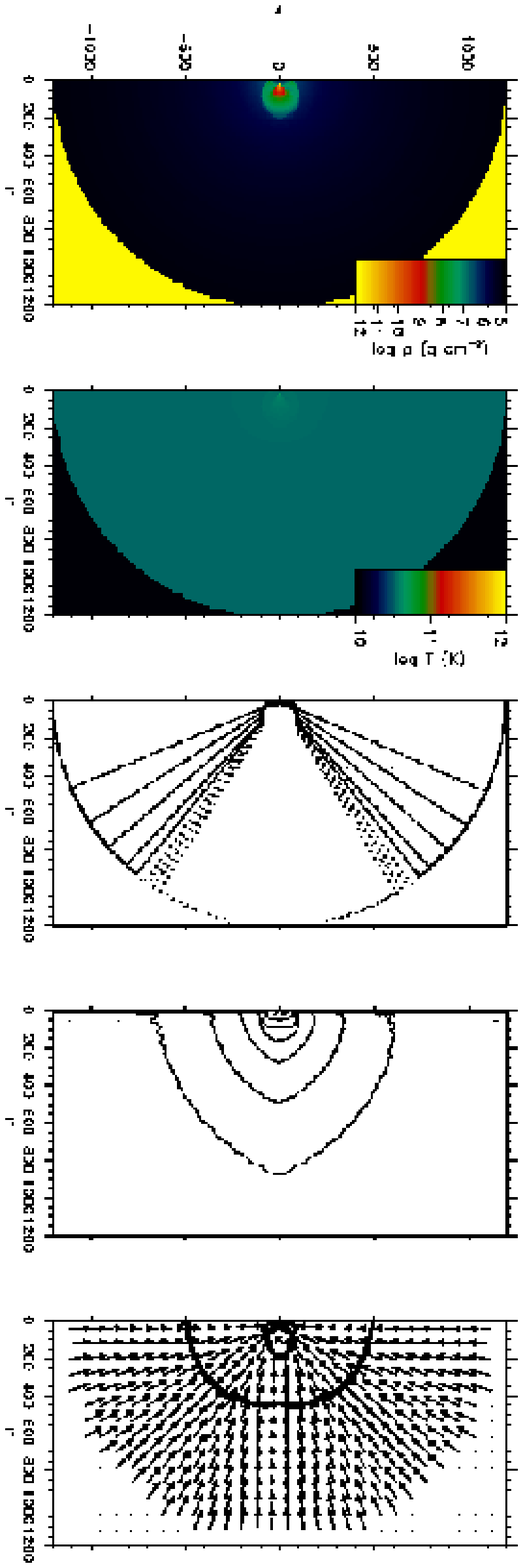}}

\end{picture}
\caption{ As in Fig. 8 but on the scale of 1200~$R_S$ 
(showing the whole computational domain).
The $l$ contours are between 0.2 and 1.1 with step of 0.1.
The $\log~\Omega$ contours are between -5.5 and -3.0  with step of 0.5.
}\label{fig:3c}
\end{figure*}

\eject

\newpage

\begin{figure*}
\begin{picture}(0,300)
\put(350,50){\includegraphics{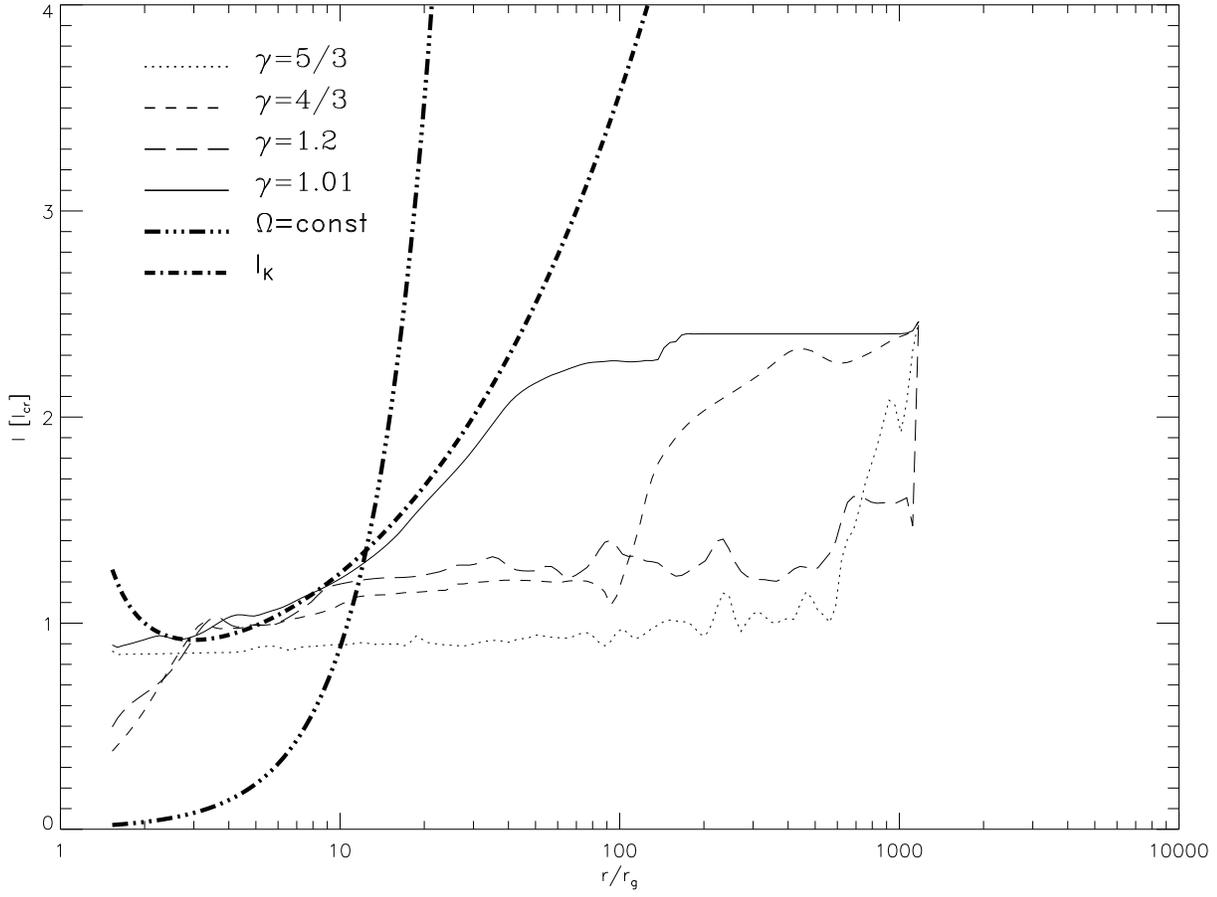}}
\end{picture}
\caption{Radial profile of angular momentum along the equatorial plane for 
different $\gamma$ at the end of simulations. For comparison Keplerian angular momentum and
angular momentum corresponding to constant angular velocity are shown, too.
\label{fig:5}}
\end{figure*}
\eject

\newpage

\begin{figure*}
\begin{picture}(0,550)

\put(50,300){\includegraphics{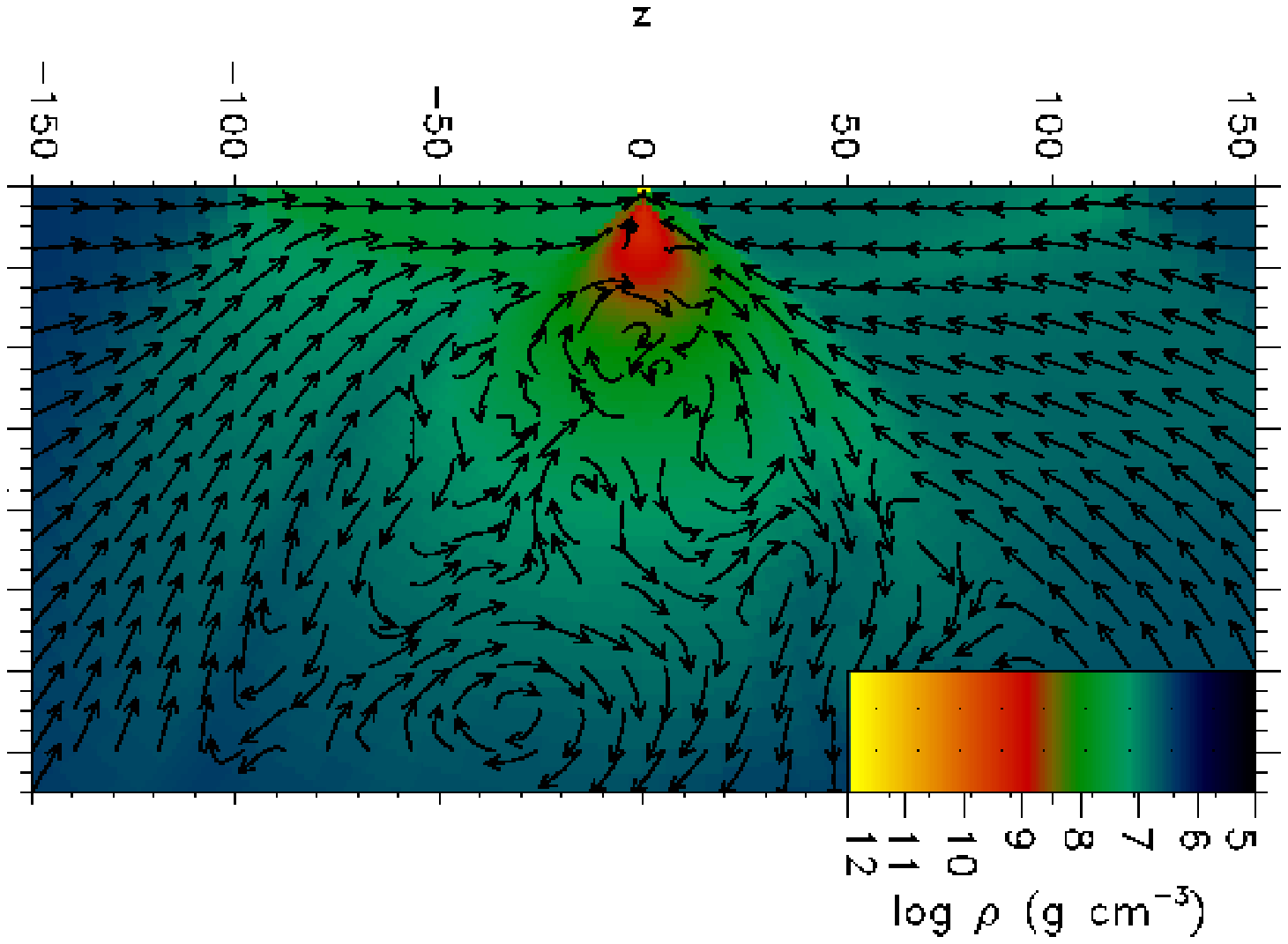}}

\put(189,300){\includegraphics{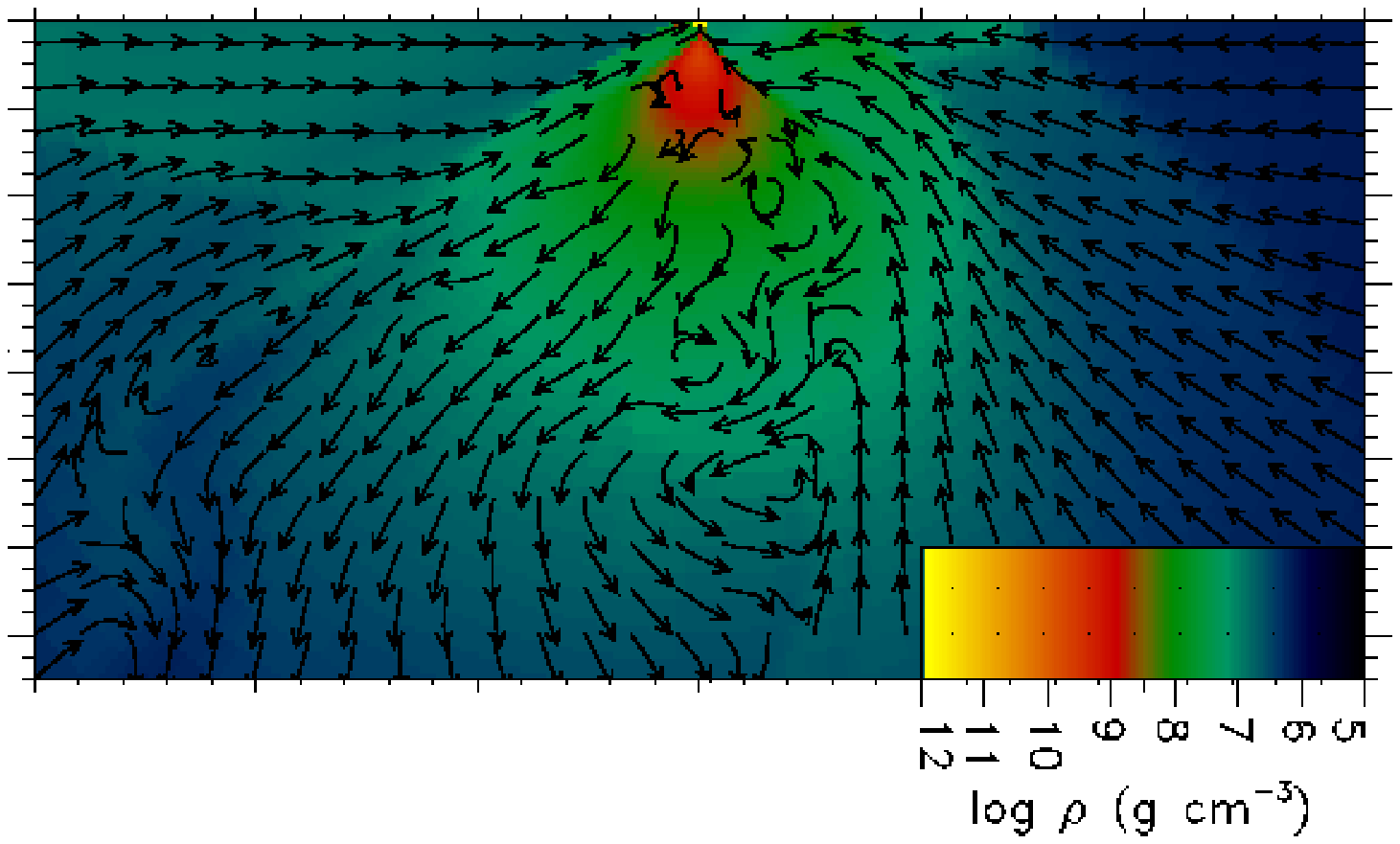}}

\put(328,300){\includegraphics{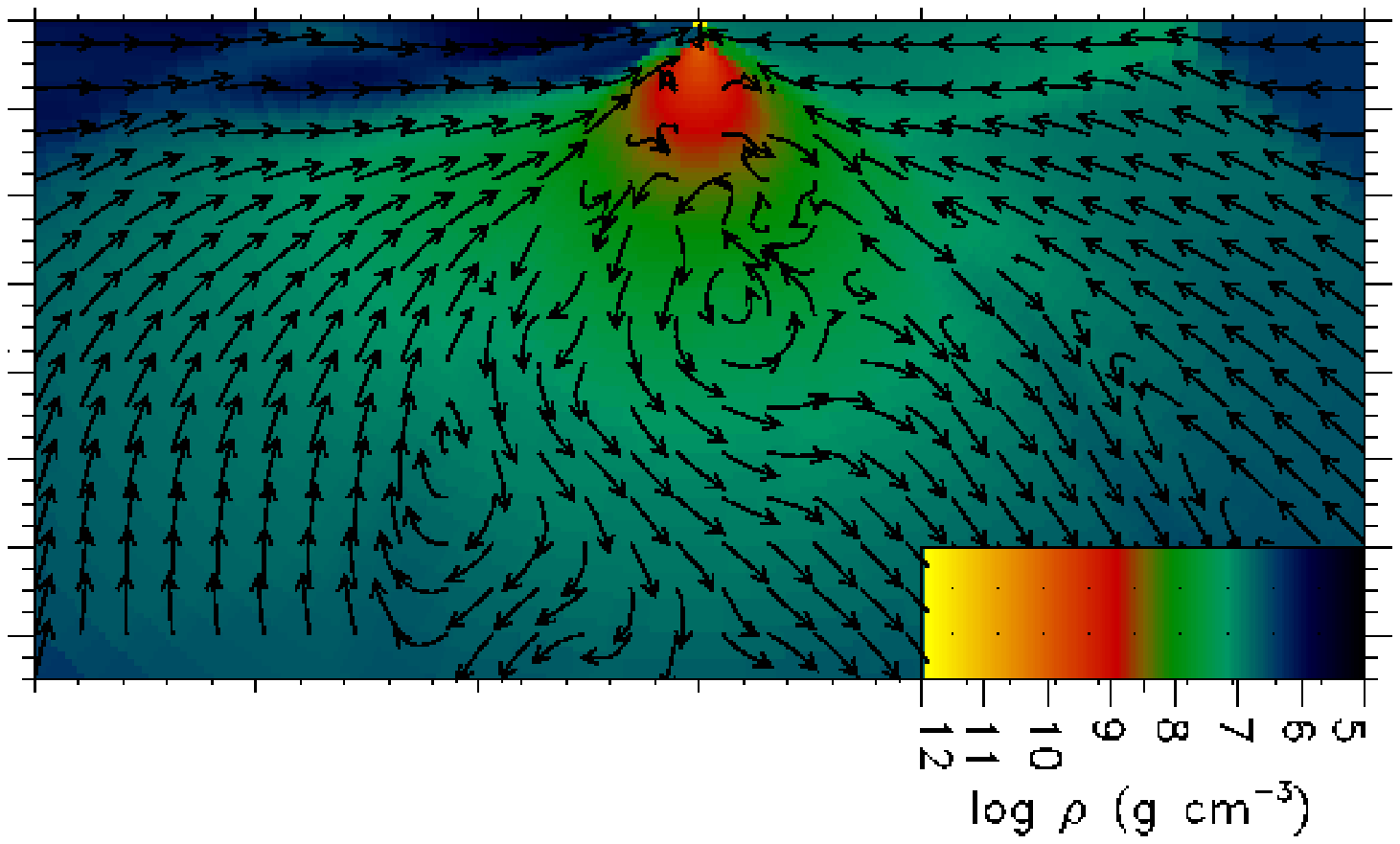}}

\put(50,0){\includegraphics{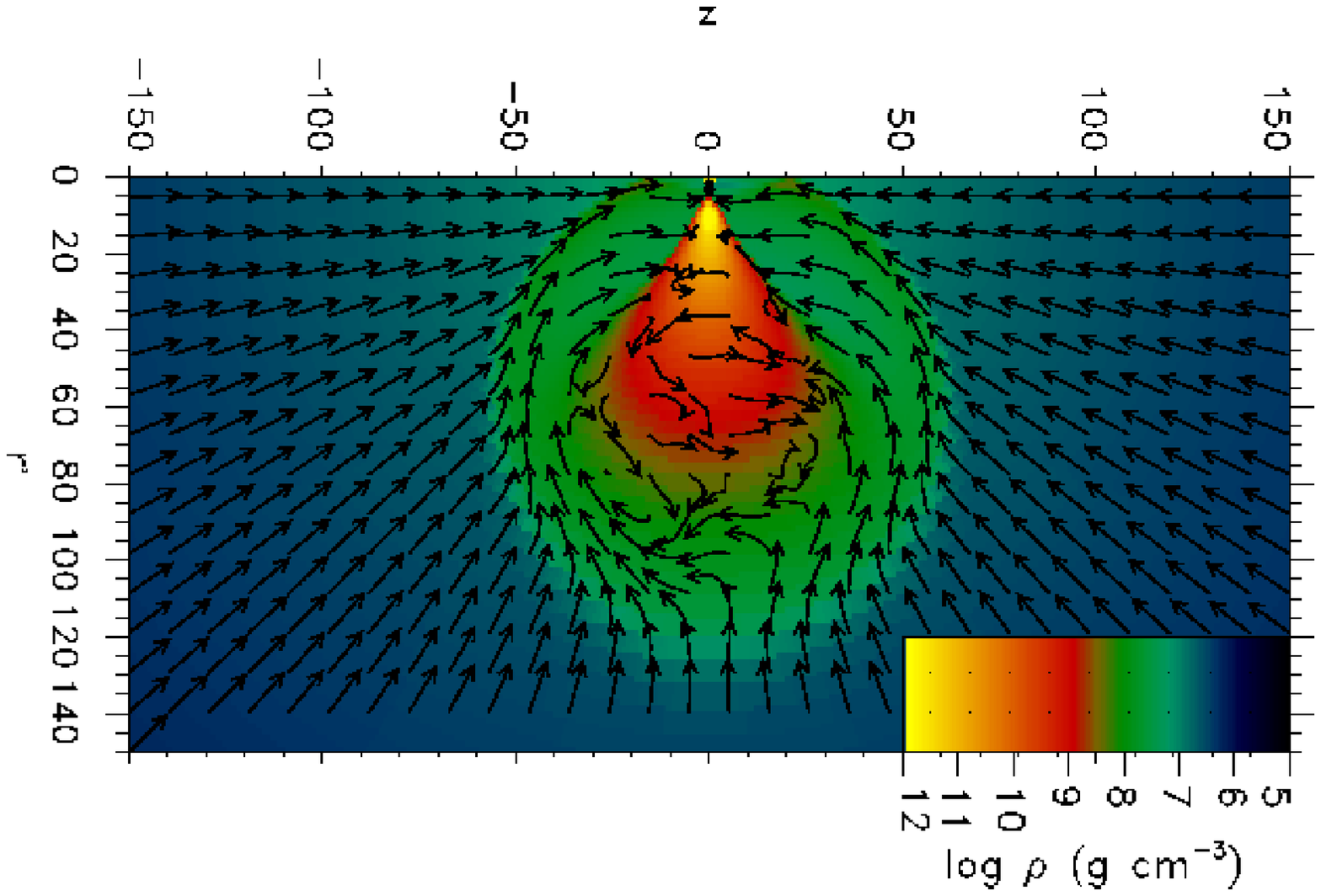}}

\put(189,0){\includegraphics{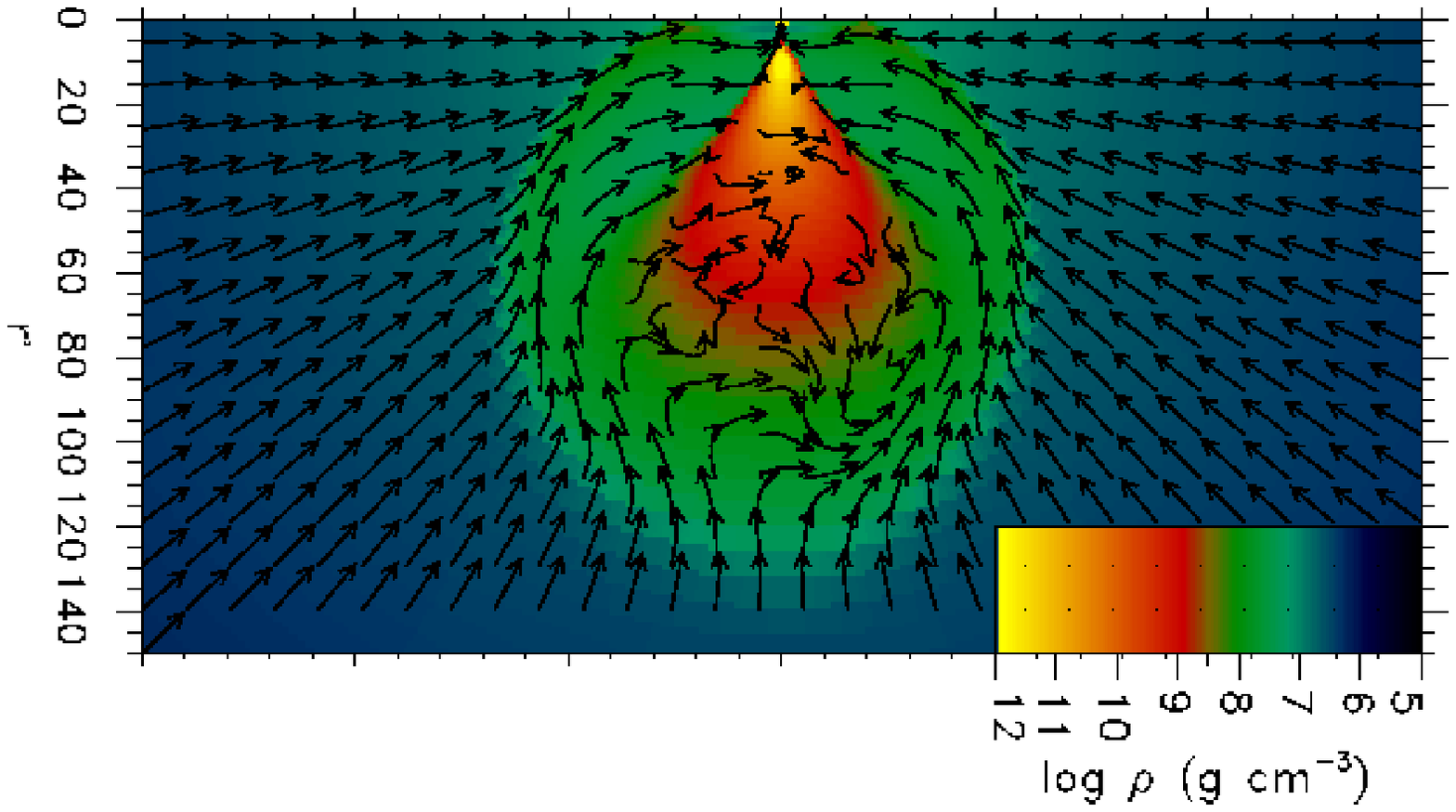}}

\put(328,0){\includegraphics{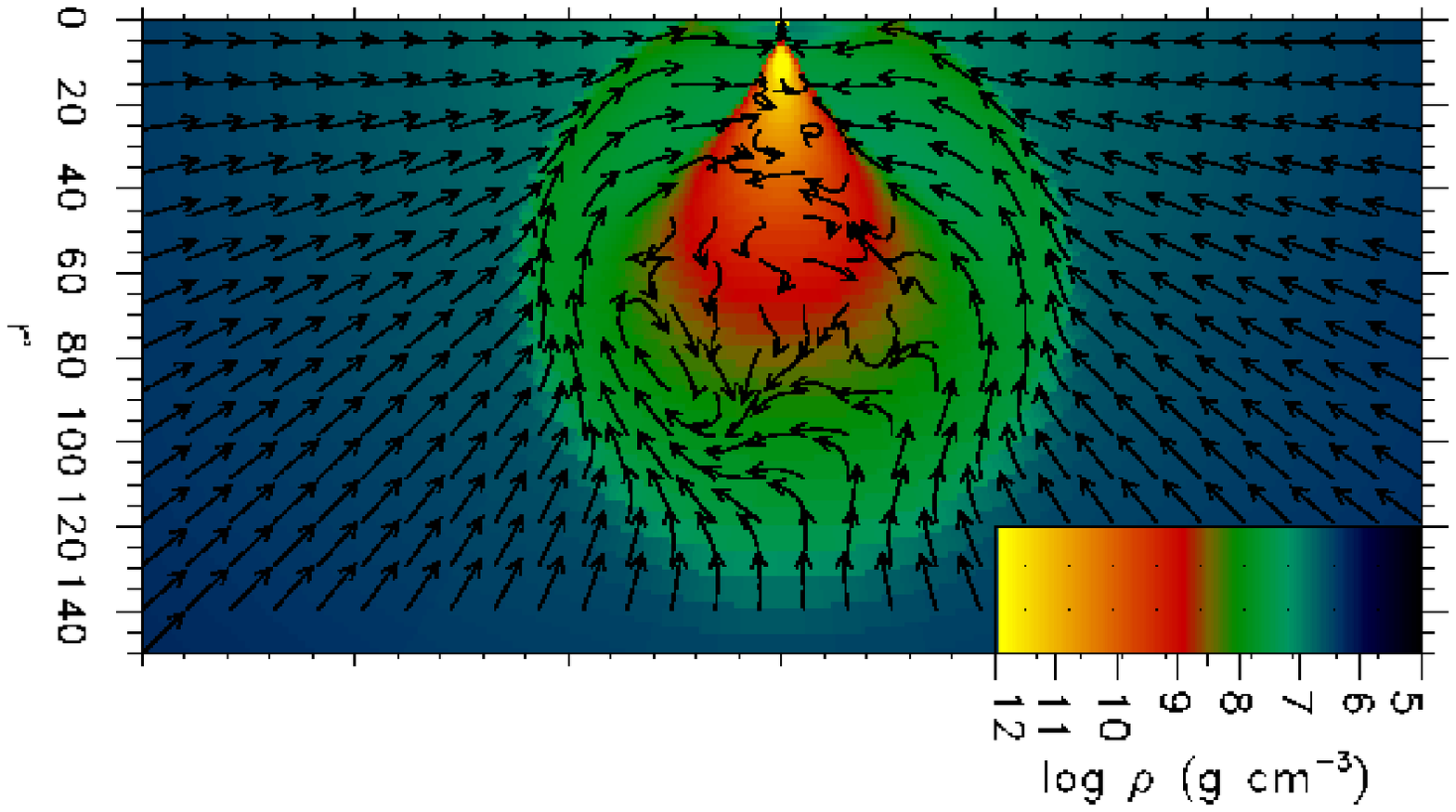}}

\end{picture}
\caption{Sequence of density maps overplotted by the direction
of the poloidal velocity for
$\gamma$=1.2 (upper panels) and $\gamma=1.01$ (lower panels) for
times t=$1.97 \times 10^4$, $2.1 \times 10^4$, and $2.8 \times 10^4$ (upper panels) and for
t=$9.7 \times 10^3$, $1.12 \times 10^4$, and $1.24 \times 10^4$ (lower panels).
These time sequences illustrate the evolution of oblique shocks forming
in the infalling gas in the polar funnel caused by the torus.
}\label{fig:4}
\end{figure*}

\eject
\newpage
\begin{figure*}
\begin{picture}(0,300)
\put(350,50){\includegraphics{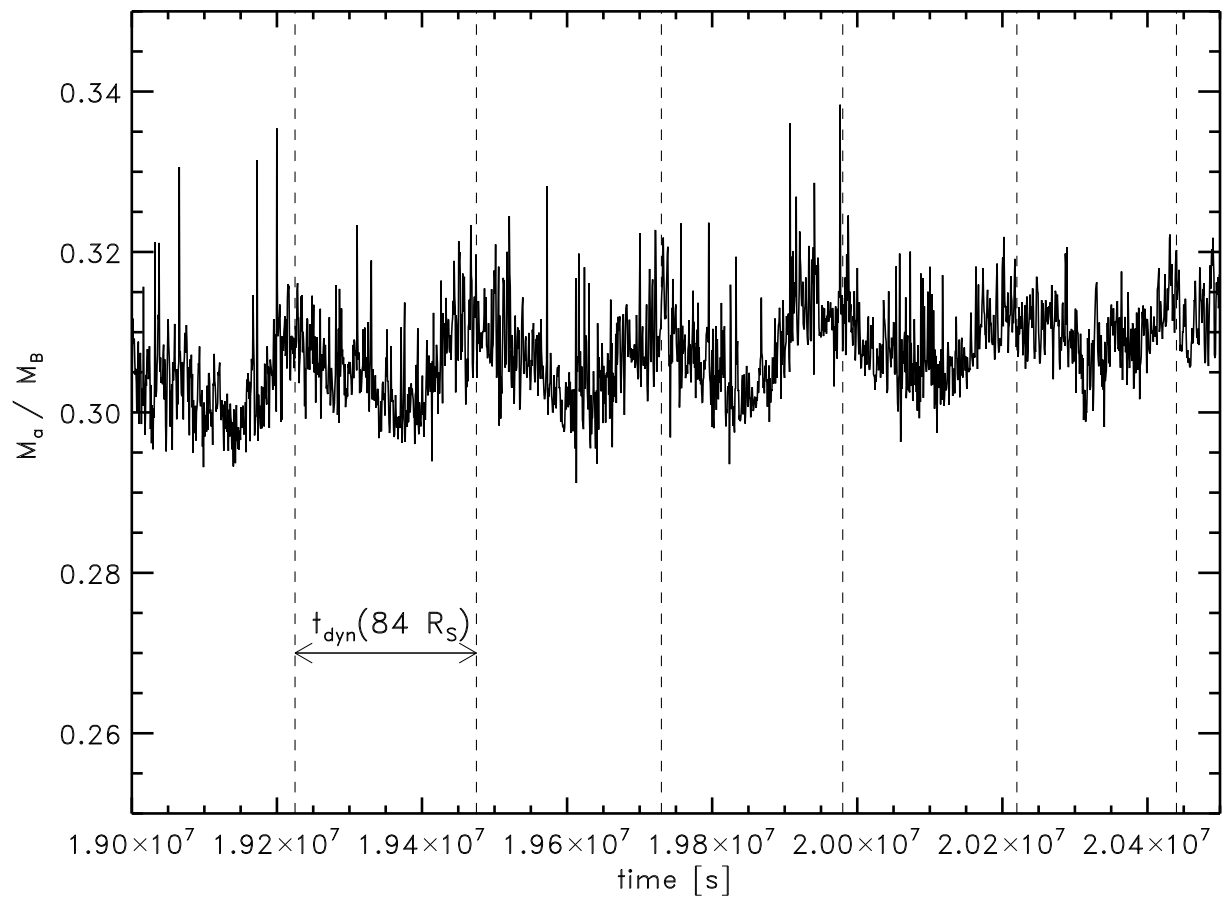}}
\end{picture}
\caption{Mass accretion rate curve for model G in the
later time of evolution.  The curve shows a periodic variability 
with
the period corresponding to the dynamical time
scale at radius of about 84 $R_S$, where we observe the strongest
oscillations the torus and asymmetry in the flow (see the lower panels
in Fig. 12.)}
\label{fig:var}
\end{figure*}
\eject

\eject
\newpage
\begin{figure*}
\begin{picture}(0,200)
\put(100,60){\includegraphics{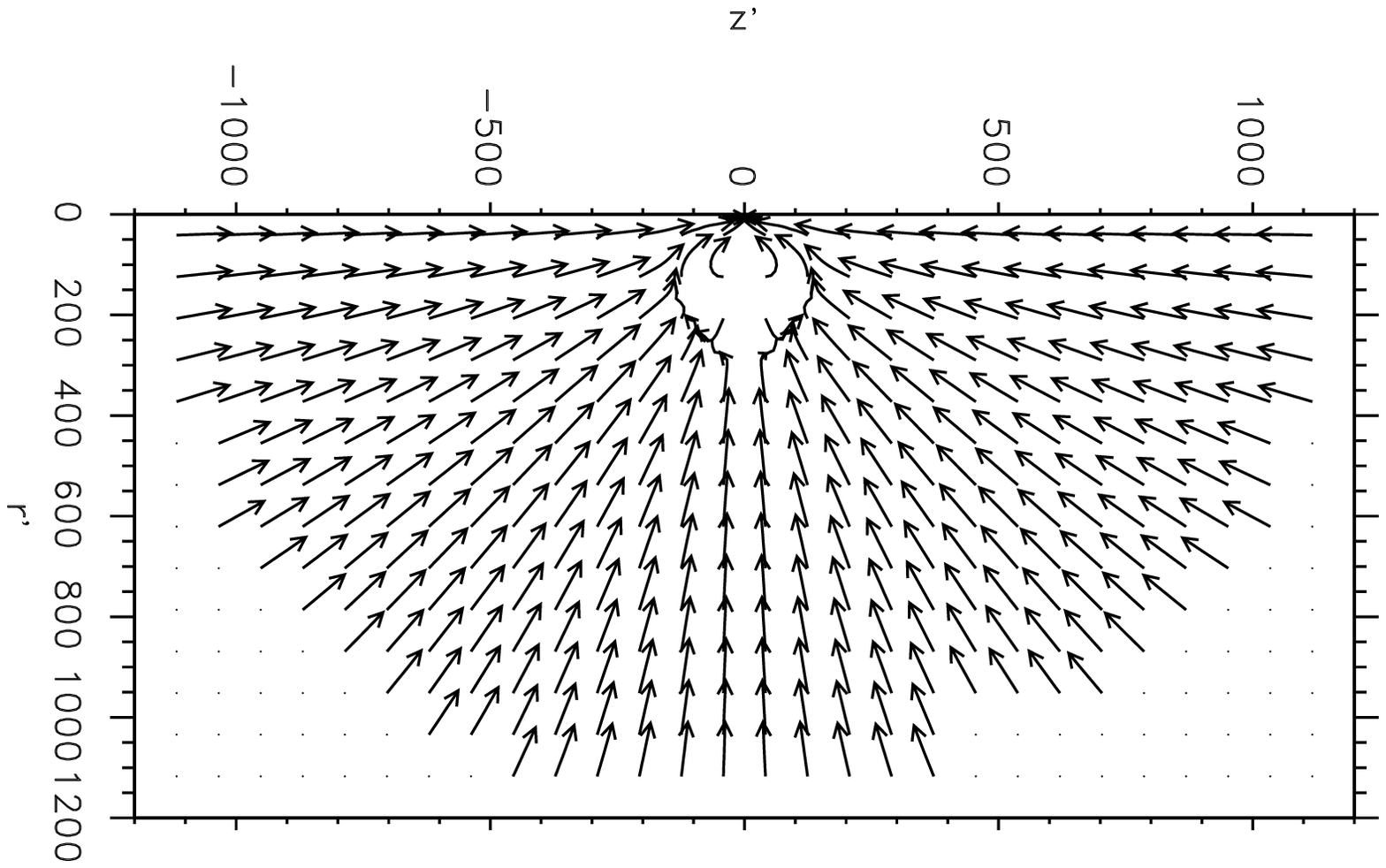}}
\put(200,60){\includegraphics{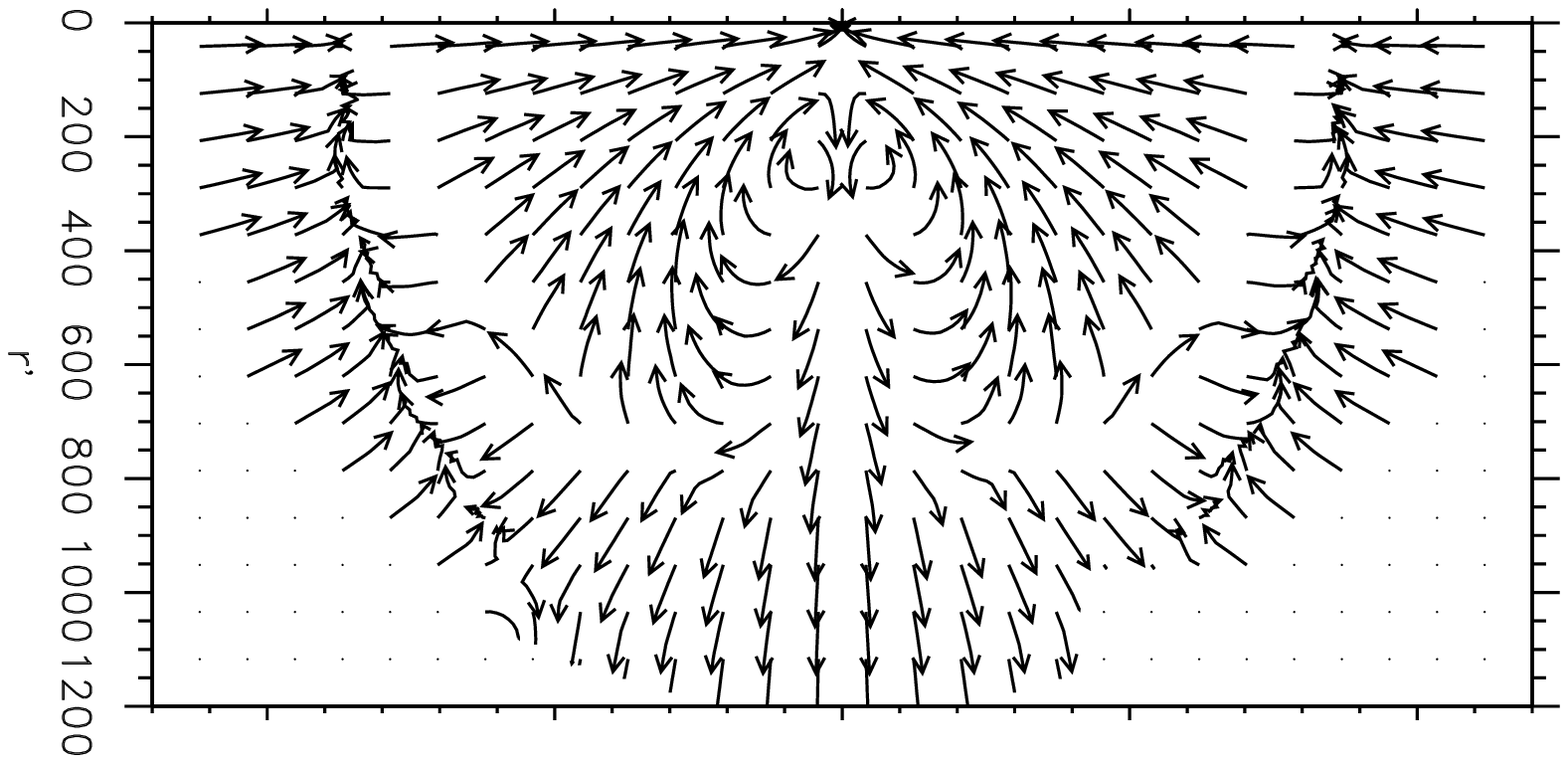}}
\put(300,60){\includegraphics{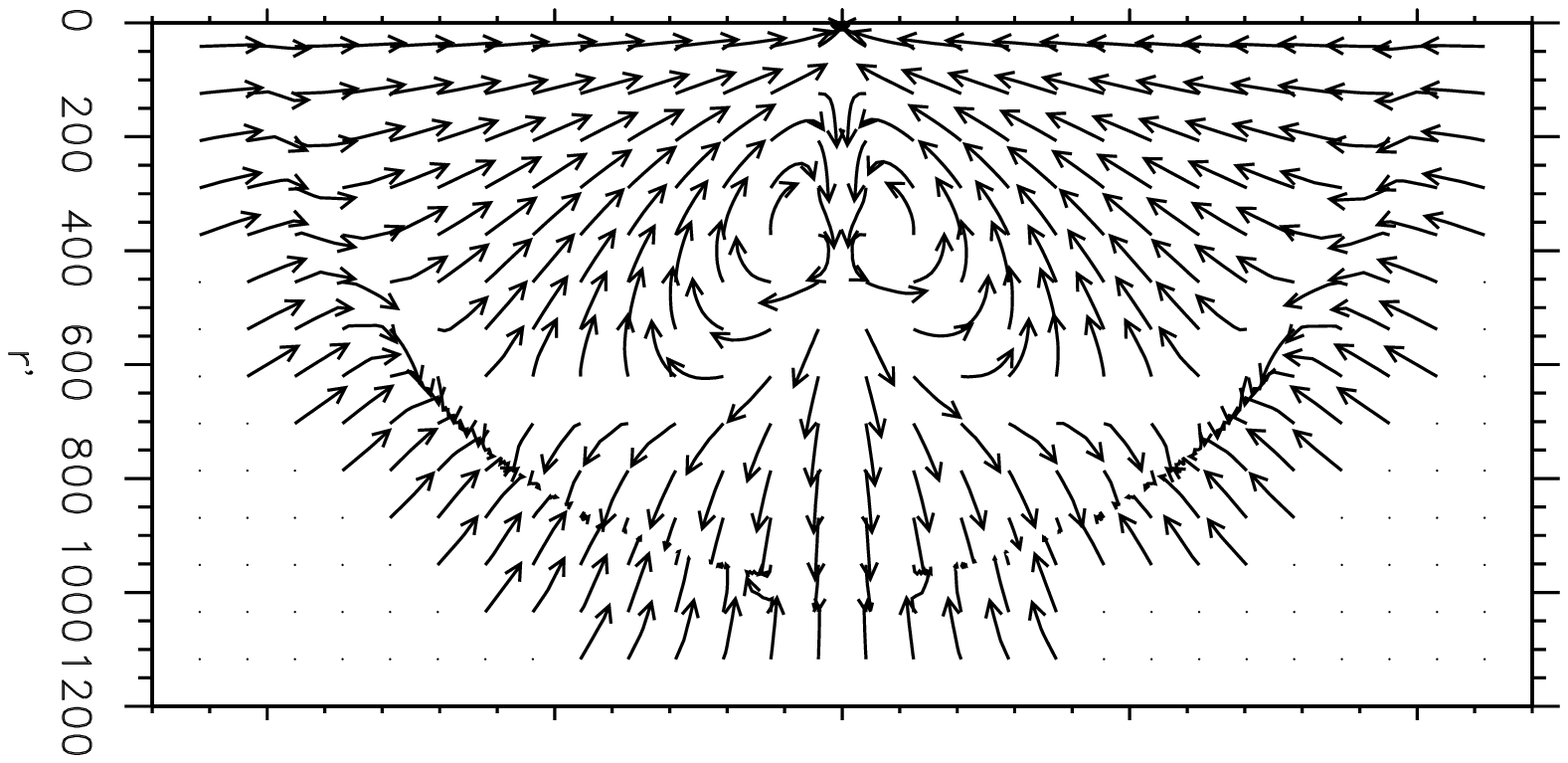}}
\put(400,60){\includegraphics{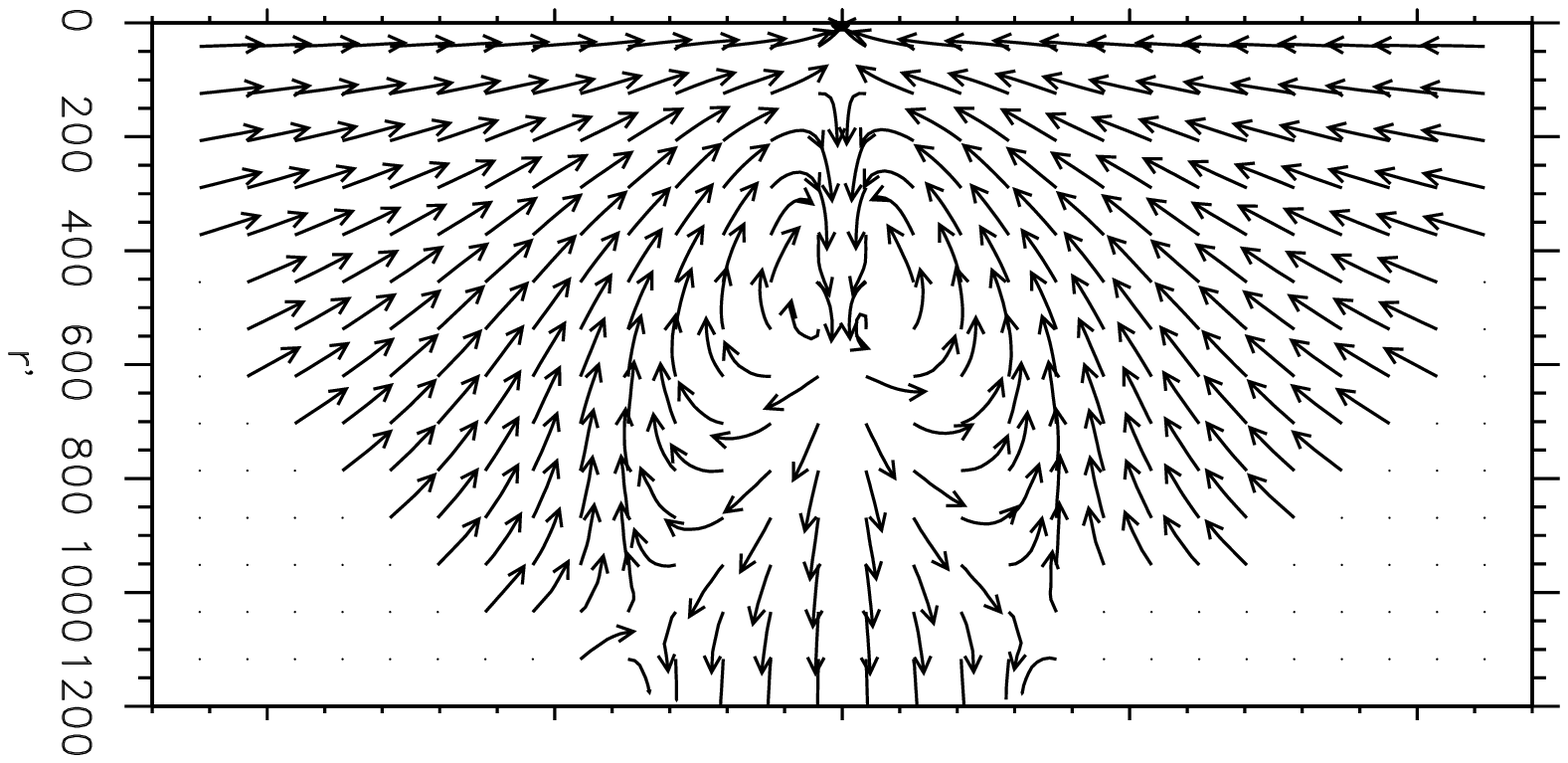}}
\put(500,60){\includegraphics{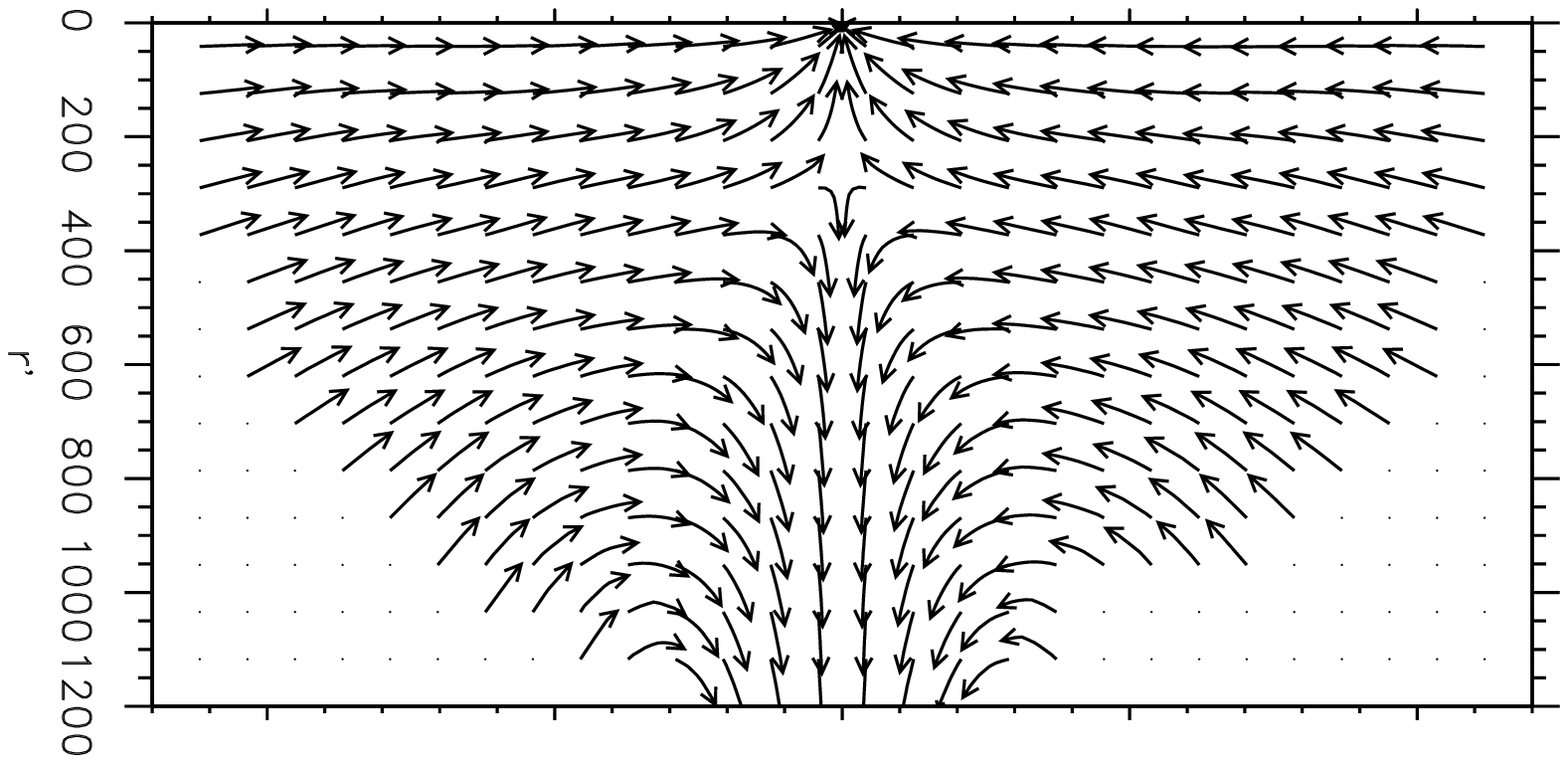}}
\end{picture}
\caption{Sequence of poloidal velocity fields
for five different time for with $R'_S=10^{-2}$. 
The computational domain  and initial condition are the same as in J model, 
except that the $c_{s,\infty}$ is much higher. 
}\label{fig:extra}
\end{figure*}

\end{document}